\PassOptionsToPackage{table, dvipsnames}{xcolor}
\documentclass[11pt, a4paper, logo, copyright, nonumbering]{map}
\usepackage[utf8]{inputenc} 
\usepackage[T1]{fontenc}    
\usepackage{caption}
\usepackage{CJKutf8}
\usepackage{xcolor}

\usepackage{booktabs}   
\usepackage{graphicx}   
\usepackage{array}      

\usepackage{tikz}
\usepackage{CJKutf8}
\usepackage{booktabs}
\usepackage{graphicx}
\usepackage{multirow}
\usepackage{multicol}
\usepackage{tabularx}
\usepackage{caption}
\usepackage{longtable}
\usepackage{subcaption}
\usepackage{afterpage}

\usepackage{wrapfig}

\usepackage{hyperref}
\usepackage{url}
\usepackage{makecell}
\usepackage{xspace}
\usepackage{listings}
\usepackage[export]{adjustbox}

\definecolor{boxcolorblue}{RGB}{141, 160, 203}
\definecolor{boxcolorgreen}{RGB}{103, 194, 165}  
\definecolor{boxcolororange}{RGB}{218, 137, 99}   
\definecolor{softblue}{rgb}{0.88, 0.95, 1.0} 
\definecolor{softred}{HTML}{FCE4D6}
\usepackage[colorlinks=true]{hyperref} 
\usepackage{tocloft}

\arrayrulecolor[HTML]{660000} 
\newcommand{\listmodelsname}{\normalsize{List of Model Score Tables}}
\newlistof{models}{csf}{\listmodelsname}

\usepackage{natbib}

\usepackage{CJKutf8}

\usepackage{xargs}  

\usepackage{todonotes}  

\usepackage{cleveref}

\usepackage{amsmath}
\usepackage{dsfont}

\usepackage{amsmath}
\usepackage{subcaption}
\usepackage{float}
\usepackage{svg}
\usepackage[breakable]{tcolorbox}
\usepackage{fontawesome}
\usepackage{xcolor}
\usepackage{colortbl}
\usepackage{lscape}
\usepackage[utf8]{inputenc}

\usepackage{multirow, hhline}
\usepackage{tabularx}
\usepackage{xcolor,colortbl}  
\definecolor{boxcolor}{HTML}{d92523} 
\definecolor{bulbcolor}{HTML}{e3b87f} 

\hypersetup{
    colorlinks=true,            
    linkcolor=blue,             
    filecolor=magenta,          
    urlcolor=cyan,              
    citecolor=purple,             
    pdftitle={Overleaf Example},
    pdfpagemode=FullScreen,
    }
 
\renewcommand{\today}{}
\newcommandx{\info}[2][1=]{\todo[linecolor=red,backgroundcolor=red!25,bordercolor=red,#1]{#2}}

\newcommand{\best}[1]{\textbf{#1}}
\newcommand{\secondbest}[1]{\fbox{#1}}
\title{YuE: Scaling Open Foundation Models for Long-Form Music Generation}

\author{
HKUST and MAP\\
{\small (alphabetical order)}\\
\small
GitHub: \href{https://github.com/multimodal-art-projection/YuE}{https://github.com/multimodal-art-projection/YuE}\\
Demo: \href{https://map-yue.github.io/}{https://map-yue.github.io/}
}

\tcbuselibrary{skins, breakable}  
\begin{document}
\begin{CJK*}{UTF8}{gbsn}

\begin{abstract}
We tackle the task of long-form music generation—particularly the challenging \textbf{lyrics-to-song} problem—by introducing \textbf{YuE (乐)}, a family of open foundation models based on the LLaMA2 architecture.
Specifically,
YuE scales to trillions of tokens and generates up to five minutes of music while maintaining lyrical alignment, coherent musical structure, and engaging vocal melodies with appropriate accompaniment. It achieves this through: (1) \textbf{track-decoupled next-token prediction} to overcome dense mixture signals, (2) \textbf{structural progressive conditioning} for long-context lyrical alignment, and (3) a \textbf{multitask, multiphase pre-training} recipe to converge and generalize. In addition, we redesign the \textbf{in-context learning} technique for music generation, enabling versatile style transfer (e.g., converting \textit{Japanese city pop} into an \textit{English rap} while preserving the original accompaniment) and bidirectional generation. Through extensive evaluation, we demonstrate that YuE matches or even surpasses some of the proprietary systems in musicality and vocal agility. In addition, fine-tuning YuE enables additional controls and enhanced support for tail languages. 
Furthermore,
beyond generation, we show that YuE's learned representations can perform competatively on music understanding tasks, 
where the results of YuE match or exceed state-of-the-art methods on the MARBLE benchmark.

\textbf{Keywords:} lyrics2song, song generation, long-form, foundation model, music generation 
\end{abstract}
\maketitle
\vspace{-1em} 

\linespread{0.9}\selectfont

\begin{figure}[b]
    \centering
    \includegraphics[width=0.7\linewidth]{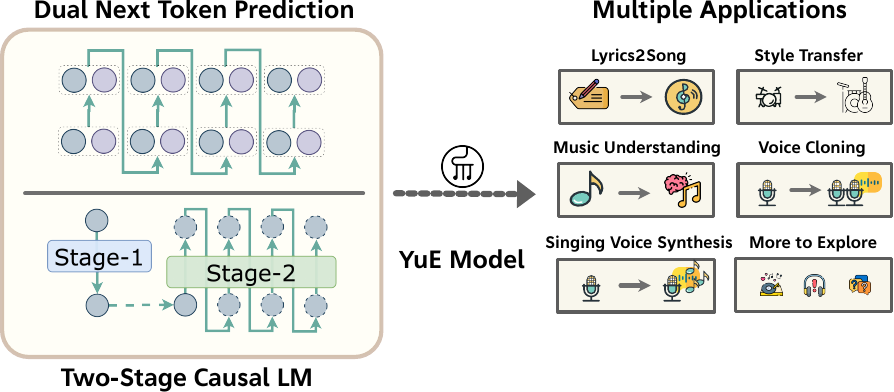}

    \caption{
    The General Application of YuE. The YuE model takes meta information and lyrics of the generated song in text and arbitrary audio as condition.
    The model can control outputs in multiple dimensions such as genre, emotion and languages. 
    }
    \label{fig:teaser}
\end{figure}
\vspace{1cm}

\newpage
\tableofcontents
\newpage

\section{Introduction}

Neural music generation represents a transformative intersection of technology and artistic creativity, offering profound commercial and cultural implications. By leveraging advanced algorithms, it is revolutionizing the music industry, enabling applications in entertainment, therapy, and personalized composition \citep{ma2024foundation}. Given the universal presence of music in human culture \citep{mehr2019universality}, these advances have the potential to democratize music creation, making it more accessible to a broader audience, while simultaneously reshaping traditional industry practices and fostering innovative approaches to musical expression.

Among various music generation tasks, \textbf{lyrics-to-song} audio generation, which involves creating full songs with vocals, accompaniment, from lyrics and control signals, is one of the most challenging. Despite its significance, no open-source system can achieve this at scale. While proprietary systems like Suno and Udio\footnote{\url{https://suno.com/}, \url{https://www.udio.com/}} have demonstrated impressive results, the lack of open-source alternatives limits accessibility, reproducibility, and innovation. Open-source ecosystems are crucial for advancing AI-driven music generation, enabling collaborative research and laying the foundation for AI models that can understand, compose, and innovate in the arts.

Most existing academic systems for AI-driven music audio generation are constrained to short, sub-30-second clips and treat singing voice synthesis \citep{liu2022diffsinger, chen2020hifisinger, zhang2022visinger} and instrumental generation \citep{musicgen,musicldm,agostinelli2023musiclm} separately \citep{li2024accompanied}. While recent efforts have started addressing full-song lyrics-to-music generation, they remain limited in effectiveness—producing short, low-quality output with poor musical coherence. The difficulty of this task arises from several key challenges: 1) \textit{Long-range dependencies}: Music exhibits complex temporal structures spanning several minutes, making it difficult for models to maintain coherence over extended durations. 2) \textit{Signal complexity}: Unlike speech or environmental sounds, music is inherently polyphonic, requiring precise coordination between multiple instrumental and vocal components. 3) \textit{Linguistic distortion}: Singing alters phonemes, durations, and prosody in ways that differ significantly from spoken language, complicating the alignment between lyrics and melody. 4) \textit{Data scarcity}: The lack of large-scale, high-quality paired datasets of lyrics, vocals, and accompaniment limits model training and generalization capabilities.

In this paper, we introduce \textbf{YuE}, the first\footnote{As of its release on Jan. 28, 2025, YuE familiy is the first publicly available, open-source lyrics-to-song model capable of full-song generation with quality on par with commercial systems.} family of open foundation models designed to push the boundaries of long-form lyrics-to-song generation. Built upon the LLaMA2~\citep{touvron2023llama, mapneo} architecture and trained on trillions of tokens, YuE generates \textbf{high-quality music up to five minutes long while maintaining lyrical alignment, musical coherence, and engaging vocal melodies}. 

By leveraging innovative pre-training and inference techniques, YuE addresses key challenges of lyrics-to-song generation and outperforms several proprietary systems in musicality, expressiveness, and controllability. We further examine subjective correlations with various automatic metrics. Interestingly, some traditional metrics (e.g., CLAP-score~\citep{clap}) fail to align with human preferences, while metrics like CLaMP3-score~\citep{wu2025clamp3universalmusic} and vocal range correlate strongly with subjective scores, suggesting the need for new, music-specific metrics that better capture listeners’ perceptual judgments (Section~\ref{sec:main_results}).

We further investigate potential \textbf{memorization effects} by thoroughly examining whether the model reproduces training data verbatim, and demonstrate that YuE largely avoids copying despite strong in-context conditioning (Section~\ref{sec:memorization}). 

Our main contributions include:
\begin{enumerate}[label=\arabic*),topsep=1pt,itemsep=2pt,leftmargin=20pt]
\item \textbf{Track-Decoupled Next-Token Prediction}: A dual-token strategy that separately models different audio tracks (vocals, accompaniment) at the frame level, resilient to challenging low vocal-to-accompaniment ratio scenarios like metal (Section~\ref{sec:dual-ntp}).
\item \textbf{Structural Progressive Conditioning}: A progressive conditioning strategy for long-form music generation, enabling song-level lyrics following and structure control (Section~\ref{sec:cot}).
\item \textbf{Redesigned In-Context Learning for Music}: A novel ICL framework enabling advanced style transfer, voice cloning, and bidirectional content creation (Section~\ref{sec:icl}).
\item \textbf{Multitask Multiphase Pre-training}: A training strategy that converges and generalizes on in-the-wild data (Section~\ref{sec:multitask_multiphase}).
\item \textbf{Strong Performance}: YuE demonstrates strong results in musicality, vocal agility, and generation duration compared to proprietary systems, supports multilingual lyrics following (Section~\ref{sec:multilingual}), while also excelling in music understanding tasks on representation learning benchmark MARBLE (Section~\ref{sec:marble}).
\end{enumerate}

\section{Related Work and Prelimenaries}
\textbf{Music Generation and Singing Voice Synthesis.} Early music generation approaches primarily focused on MIDI-based methods~\citep{huang2018musictransformer, MuseNet}, while recent models generate raw audio conditioned on tags or text~\citep{dhariwal2020jukebox, agostinelli2023musiclm, audioldm, huang2023noise2music, copet2023simple-musicgen, chen2024musicldm, evans2024stable}. However, most existing audio methods are limited to instrumental music with short durations (around 30 seconds) due to computational constraints. Although some efforts incorporate vocals, they typically lack coherent lyrical semantics~\citep{agostinelli2023musiclm, dhariwal2020jukebox}. Concurrently, deep learning has significantly advanced singing voice synthesis (SVS), leveraging techniques like GANs, diffusion models, and variational autoencoders for high-quality vocal synthesis~\citep{chen2020hifisinger, liu2022diffsinger, zhang2022visinger, hong2023unisinger}, and enabling nuanced control via language prompts or discrete tokens~\citep{singsong, wang2024prompt, wu2024toksing}. Nevertheless, these SVS systems mostly generate pure vocals with explicit melodic guidance. In contrast, our work proposes a novel approach capable of autonomously generating coherent and semantically meaningful vocals alongside instrumental accompaniments, supporting significantly extended song contexts of up to five minutes, thus substantially advancing automated music production.

\textbf{Song Generation.} 
Despite recent progress in music generation research, academic models still face significant limitations. Previous or concurrent work, such as Jukebox~\citep{jukebox}, MelodyLM~\citep{li2024accompanied}, SongCreator~\citep{lei2024songcreator}, SongGen~\citep{liu2025songgen} struggle to generate long-form music audio beyond 30 seconds while maintaining coherence and high-quality synthesis. These models often lack fully open-source implementations, making reproducibility and further improvements difficult. For instance, Jukebox utilizes a multi-scale VQ-VAE for raw audio modeling but suffers from noticeable artifacts and limited controllability. Similarly, SongCreator ~\citep{lei2025songcreator} and SongGen~\citep{liu2025songgen} introduce innovative transformer-based architectures for text-to-song generation, yet their performance is inferior to commercial counterparts. 
In contrast, industry-developed systems such as Tiangong Music (Kunlun Ltd.), Seed Music (ByteDance)\citep{bai2024seed}, Suno, Udio, and Hailuo Music (MiniMax) have demonstrated promising results in song-level audio generation, though their technical details remain undisclosed. Our work addresses these gaps by offering an open-source, song-level generative model with full technical transparency, achieving performance on par with leading proprietary systems.

\textbf{Audio Tokenizers.} Discrete modeling of audio often employs neural codec tokenizers, particularly Residual Vector Quantization GANs (RVQ-GANs) \citep{dac}, typically categorized into \textit{acoustic} and \textit{semantic} tokens \citep{defossez2024moshi, audiolm}. \textbf{Acoustic} tokens, optimized for reconstruction, encode fine acoustic details, causing significant token shifts even with minor acoustic variations. Prior studies \citep{musicgen} indicate these tokens require extensive training epochs; notably, we find acoustic tokens alone fail to converge efficiently on our dataset (Section~\ref{sec: tokenizer comparison}). Conversely, \textbf{semantic} tokens, derived from self-supervised learning encoders~\citep{schneider2019wav2vec, baevski2020wav2vec, chung2021w2v, baevski2022data2vec, ma2022mt4ssl}, produce semantically meaningful representations (e.g., phonemes, notes, genres) \citep{zhang2023speechtokenizer, yuan2024marble, wang2025spark}. Unlike previous work, we conduct extensive experiments on complex in-the-wild music datasets, perform qualitative comparisons, and report tokenizer convergence, demonstrating that fusing semantic information significantly enhance convergence.

\clearpage

\section{YuE}
\subsection{Overview}

\setlength{\intextsep}{5pt}  
\setlength{\columnsep}{5pt}  

\begin{wrapfigure}{r}{0.35\linewidth}
    \centering
    \includegraphics[width=0.9\linewidth]{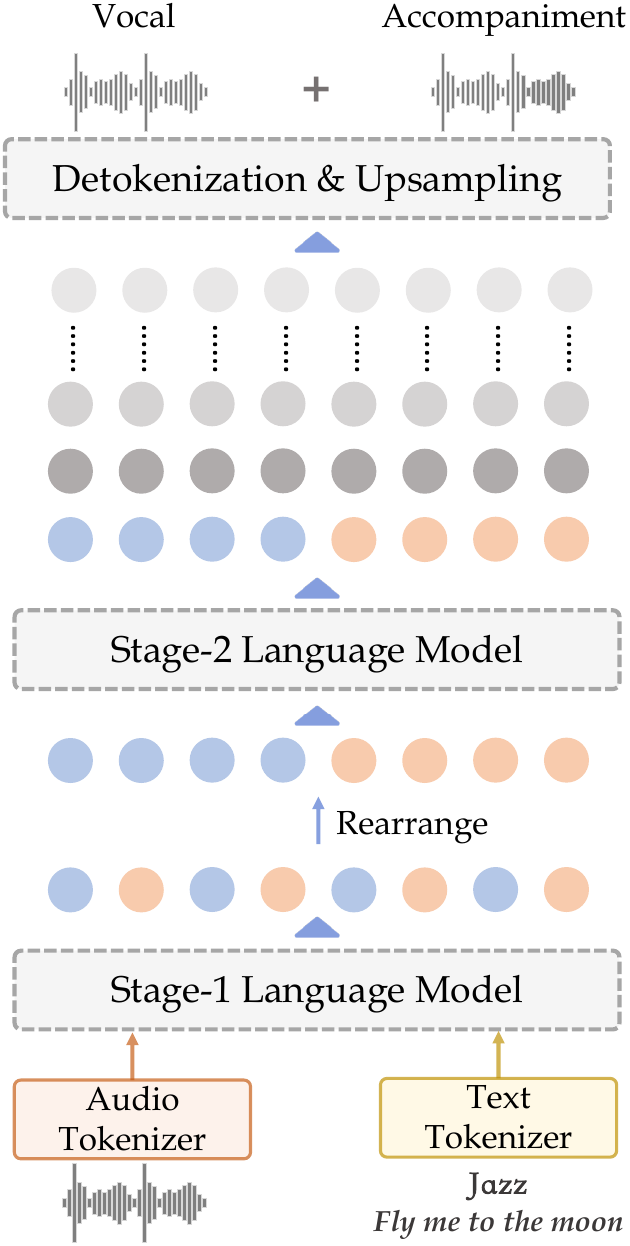}

    \caption{Overview of YuE framework: two-stage lyrics-to-song generation with audio/text tokenizers and two language models. Stage-1: music language modeling. Stage-2: residual modeling. \textcolor{blue}{Blue}: vocal tokens. \textcolor{orange}{Orange}: accompaniment tokens. \textcolor{gray}{Grey}: residual tokens.}
    \label{fig:overview}
\end{wrapfigure}

YuE is an autoregressive (AR) language model (LM)-based framework tailored for lyrics-to-song generation. As depicted in Figure~\ref{fig:overview}, YuE comprises four main components: an audio tokenizer (with a lightweight upsampler), a text tokenizer, and two language models (LMs). The audio tokenizer converts waveforms into discrete tokens using a semantic-acoustic fused approach. The Stage-1 LM is track-decoupled, trained on text tokens and semantic-rich bottom-level audio tokens (codebook-0 from residual VQ-VAE), modeling lyrics-to-song generation as an AR next-token prediction (NTP) task. In Stage-2, a smaller LM predicts residual tokens from codebook-0 tokens to reconstruct audio. Both LMs follow the widely-adopted LLaMA2 architecture \citep{llama, llama3}. Finally, a lightweight vocoder upsamples Stage-2's 16 kHz audio to 44.1 kHz output.


\subsection{Stage-1: Music Language Modeling}
Music language modeling stage (MuLM), illustrated in Figure~\ref{fig.stage-1}, enables music generation conditioned on diverse inputs (lyrics, tags, structures, reference audio). We introduce MuLM's core techniques: 1) track-decoupled next-token prediction (Section~\ref{sec:dual-ntp}), 2) structural progressive generation (Section~\ref{sec:cot}), and 3) music in-context learning (Section~\ref{sec:icl}).

\subsubsection{Track-Decoupled Next-Token Prediction}
\label{sec:dual-ntp}
\paragraph{Challenges of Standard NTP.} Popular LM-based approaches for modeling long RVQ sequences typically adopt a multi-stage design~\citep{vall-e, agostinelli2023musiclm, audiolm}, where the first stage commonly uses a single codebook-0 token to represent each audio frame.\footnote{We acknowledge single-stage methods such as MusicGen, which utilize delay or parallel decoding patterns to reduce sequence length. However, we observed that the parallel decoding pattern fails to converge on our dataset, while the delay pattern results in longer sequences compared to multi-stage approaches.} Let 
\(\mathbf{x}_{1:T} = (x_1, x_2, \dots, x_T)\) 
represent a sequence of audio tokens, where each $x_t$ corresponds to one frame. In a standard NTP framework, we factorize the joint probability of $\mathbf{x}_{1:T}$ as:
\begin{MiddleEquation}
\begin{equation}
    p(\mathbf{x}_{1:T}) 
    \;=\; \prod_{t=1}^{T} p\bigl(x_{t}\mid x_{<t}; \,\theta\bigr),
    \label{eq:standard-ntp}
\end{equation}
\end{MiddleEquation}
where $\theta$ is the model parameter. During inference (generation), the model predicts the next token $\hat{x}_{t}$ which maximizes the conditional distribution:
\begin{MiddleEquation}
\begin{equation}
    \hat{x}_{t} 
    \;=\; \arg\max_{x_{t}}\; p\bigl(x_{t}\mid x_{<t}; \,\theta\bigr).
    \label{eq:standard-inference}
\end{equation}
\end{MiddleEquation}

This approach works well for tokens $\mathbf{x}_{1:T}$ representing purely vocal (text-to-speech, TTS) or instrumental (text-to-music, TTM) signals but struggles when encoding both vocals and accompaniment simultaneously due to differing dynamics, as in lyrics-to-song tasks combining TTS and TTM.

\begin{figure}
    \vspace{-10pt}
    \centering
    \captionsetup{skip=0pt} 
    \includegraphics[width=0.4\linewidth]{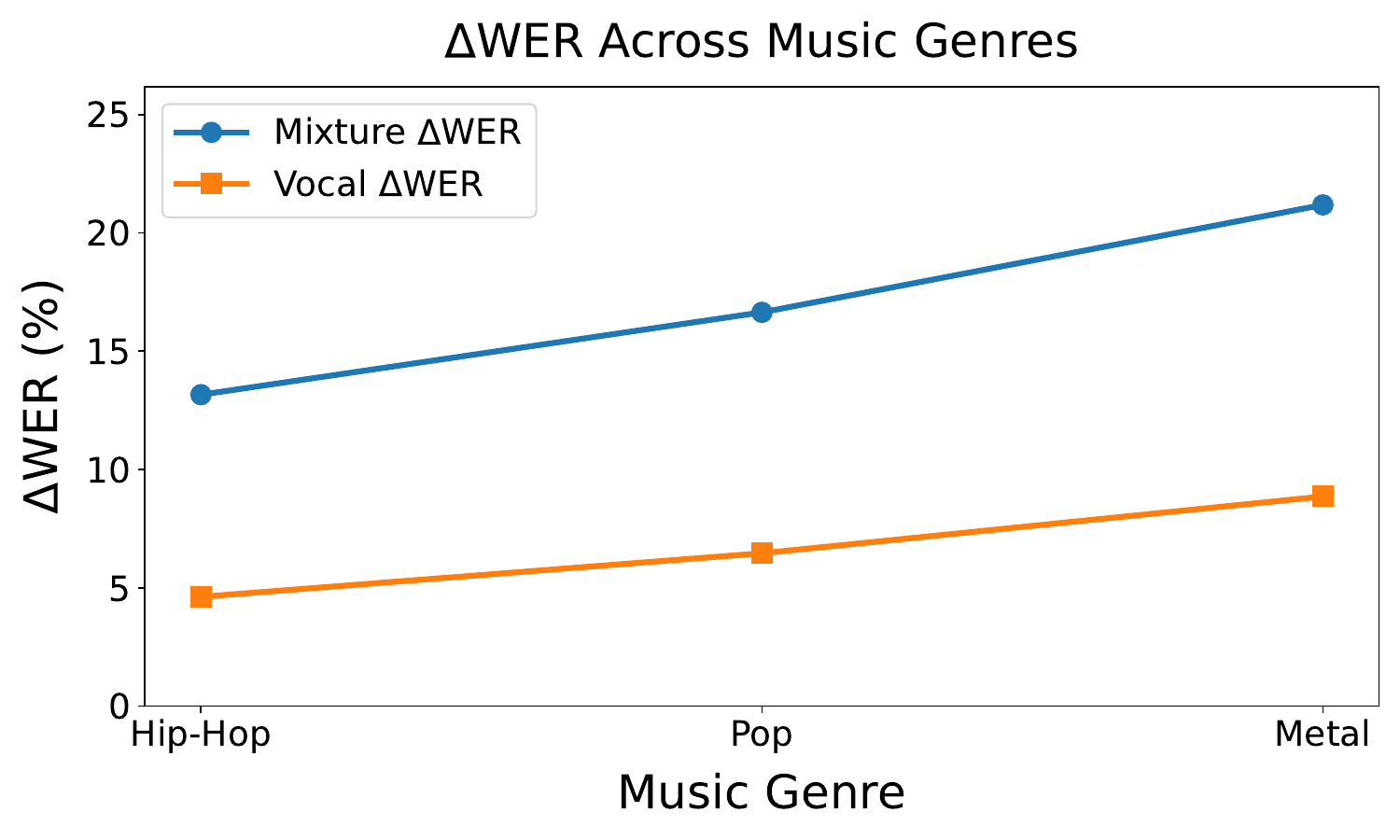}
    
    \caption{$\Delta\text{WER}$ across different music genres for mixture / vocal-only tracks. $\Delta\text{WER}\propto\text{LLAT}$.}
    \label{fig:delta_wer}
    \vspace{-5pt}
\end{figure}

We quantify \textbf{L}inguistic information \textbf{L}oss \textbf{A}fter \textbf{T}okenization (\textbf{LLAT}) using delta \textbf{W}ord \textbf{E}rror \textbf{R}ate ($\Delta$WER), defined as $\Delta\text{WER} = \text{WER}_{\text{recon}} - \text{WER}_{\text{ori}}$, where $\text{WER}_{\text{recon}}$ and $\text{WER}_{\text{ori}}$ are estimated by a fine-tuned Whisper\footnote{A Whisper V3 checkpoint fine-tuned on an internal song dataset with manual transcription.} model on tokenizer-reconstructed\footnote{We use \texttt{X-Codec} as our tokenizer. See more discussion in Section~\ref{sec:tokenizer} and ~\ref{sec: tokenizer comparison}.} and original mixture audio, respectively. Figure~\ref{fig:delta_wer} illustrates the relationship between $\Delta$WER and music genre (hip-hop, pop, metal) using 1k sampled tracks. An upward trend is evident, with metal exhibiting the highest LLAT followed by pop and hip-hop, indicating greater modeling difficulty in acoustically dense genres. Vocal-only tracks consistently achieve lower $\Delta$WER compared to mixtures, indicating lower LLAT after source separation.

\begin{figure}[t!]
\centering
\graphicspath{{figures/Yue_main/}}
\includegraphics[width=0.9\linewidth]{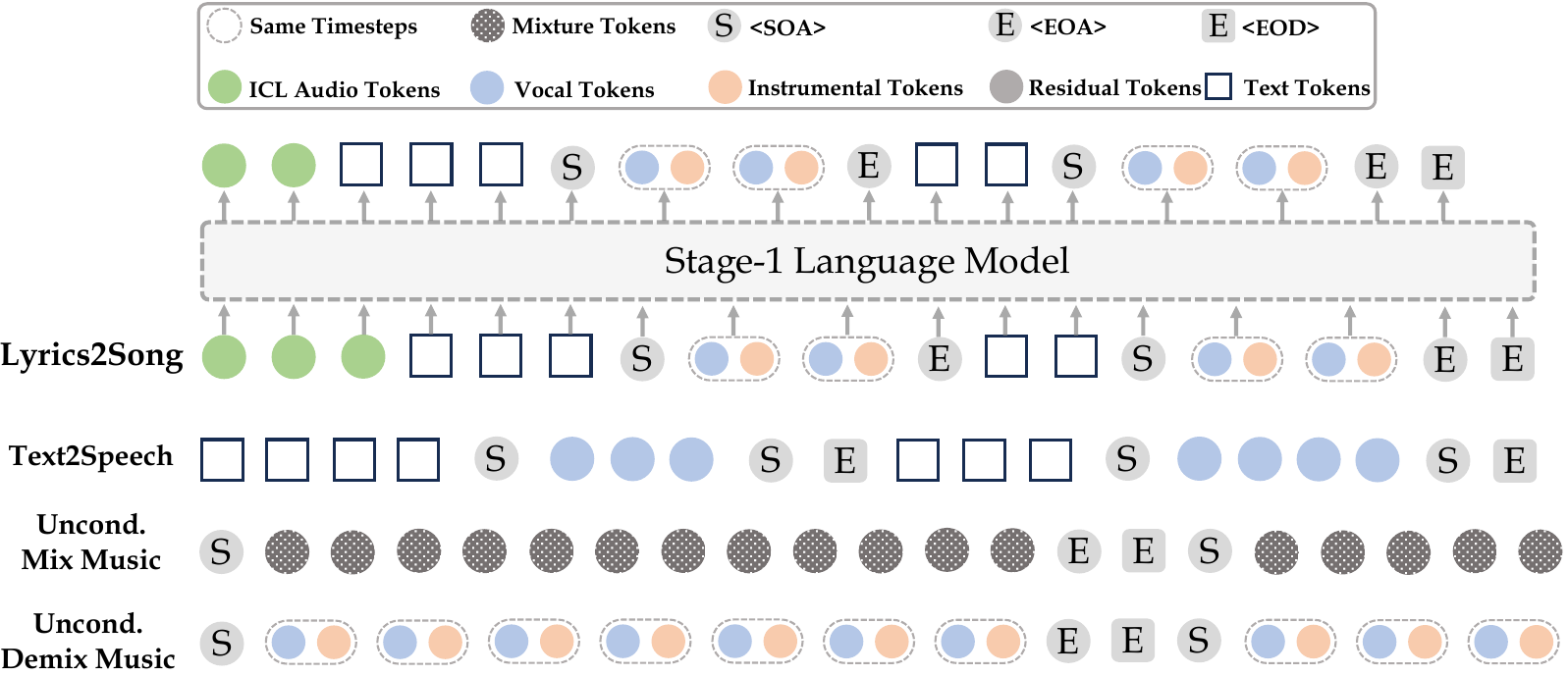}
\caption{The Stage-1 Framework of YuE. Dotted lines: Dual-NTP (Section~\ref{sec:dual-ntp}). Text interleave: CoT (Section~\ref{sec:cot}). Green tokens: ICL (Section~\ref{sec:icl}). Multitask learning (Section~\ref{sec:multitask_multiphase}).}
\label{fig.stage-1}
\end{figure}

\paragraph{Track-Decoupled Next-Token Prediction (Dual-NTP).} The above observation suggests that the issue arises from forcing a single token \(\,x_t\) to represent two distinct signals: vocal and music.  
Accompaniment can overshadow the vocal track, degrading lyric intelligibility.  
To overcome these shortcomings, we propose a method that explicitly incorporates a source separation prior, splitting each time step into two tokens: one for \textbf{vocal} and one for \textbf{accompaniment} (see dotted token pairs in Figure~\ref{fig.stage-1}). 

In the proposed method, each time step \(t\) outputs two tokens: \(v_t\) (vocal token) and \(a_t\) (accompaniment token).
The model’s sequence of tokens thus becomes:
\begin{MiddleEquation}
\begin{equation}
\bigl(\underbrace{v_1}_{\text{vocal}}, \underbrace{a_1}_{\text{accomp.}},\, 
\underbrace{v_2}_{\text{vocal}}, \underbrace{a_2}_{\text{accomp.}},\dots, 
\underbrace{v_T}_{\text{vocal}}, \underbrace{a_T}_{\text{accomp.}}\bigr).
\end{equation}
\end{MiddleEquation}

To formally define this, let \(\mathbf{v}_{1:T} = (v_1, v_2, \dots, v_T)\) and \(\mathbf{a}_{1:T} = (a_1, a_2, \dots, a_T)\).  
We factorize their joint probability as:
\begin{MiddleEquation}
\begin{equation}
    p\bigl(\mathbf{v}_{1:T},\, \mathbf{a}_{1:T}\bigr)
    \;=\;
    \prod_{t=1}^{T}\,
    p\Bigl(v_{t},\,a_{t}\;\Big|\;v_{<t},\,a_{<t};\,\theta\Bigr).
    \label{eq:dualNTP}
\end{equation}
\end{MiddleEquation}
At inference time, the next pair \(\bigl(\hat{v}_t,\,\hat{a}_t\bigr)\) is chosen to maximize this joint conditional:
\begin{equation}
    \bigl(\hat{v}_t,\,\hat{a}_t\bigr)
    \;=\;
    \arg\max_{(v_{t},\,a_{t})}\;
    p\Bigl(v_{t},\,a_{t}\;\Big|\;v_{<t},\,a_{<t};\,\theta\Bigr).
    \label{eq:dualNTPinfer}
\end{equation}
Although this probability is written in joint form, it can be decomposed as:
\begin{equation}
    p\Bigl(v_{t},\,a_{t}\;\Big|\;v_{<t},\,a_{<t};\,\theta\Bigr)
    \;=\;
    p\Bigl(v_{t}\;\Big|\;v_{<t},\,a_{<t};\,\theta\Bigr)
    \,\times\,
    p\Bigl(a_{t}\;\Big|\;v_{\le t},\,a_{<t};\,\theta\Bigr),
    \label{eq:twoFactor}
\end{equation}
making it straightforward to implement in standard AR decoding frameworks.

\paragraph{Discussion.} Existing work has explored modeling dual tracks using various approaches \citep{lei2024songcreator, li2024accompanied}, often requiring large modifications to the LM architecture or modeling the tracks sequentially. In contrast, our proposed method offers a more effective solution with the following advantages:
\begin{enumerate}[label=\arabic*),topsep=1pt,itemsep=2pt,leftmargin=20pt]
    \item \textbf{Scalability:} 
    By preserving the existing LM architecture, we leverage well-established pre-training infrastructures and enable straightforward scalability.
    \item \textbf{Convergence:}
    Empirically, Dual-NTP converges to lower training loss compared to standard NTP. Notably, it demonstrates robust lyric adherence even within challenging minority genres (e.g., metal music)\footnote{We encourage the readers to listen to our demo page \url{https://map-yue.github.io/}.}, illustrating its adaptability to heterogeneous data distributions.
    \item \textbf{Joint Modeling of Tracks:}
    Our approach jointly contextualizes both tracks in a single forward pass, avoiding track synchronization issues, and allowing coherent and natural musical planning.
    \item \textbf{Granular Modeling \& Processing:}
    The explicit segregation of vocal and accompaniment tokens enables independent modeling, allowing for the capture of finer nuances, particularly in instrumentally-dense segments. This also facilitates separate post-processing and mastering for each track.
\end{enumerate}


\subsubsection{Structural Progressive Conditioning}
\label{sec:cot}
\paragraph{Challenges of Full-song Generation.} 
While typical TTS and TTM systems operate on less than 30 seconds of context \citep{audioldm, audioldm2, vall-e, audiolm, musicgen}, full-song modeling requires handling minutes-long contexts. Although some  proprietary systems like Suno.ai and Udio achieved this, their methodologies remain undisclosed. We find that extending the LM context to full-song modeling is non-trivial. Simply scaling up the LM context length does not yield effective song-level lyrics-following capabilities and demands substantial computational resources.

A key challenge is the \textbf{long-term decay property} of commonly adopted Rotary Position Embedding (RoPE) \citep{su2024roformer}. In autoregressive TTS and TTM systems, text conditioning applied at the start degrades as audio tokens extend further. Empirically, this degradation begins around 3K tokens and leads to complete failure beyond 6K tokens, even with 16K-token pre-trained contexts. Our mitigation attempts, such as increasing the RoPE base (10K to 100K) or curriculum learning with gradually increasing audio lengths, have been ineffective. See ablation in Section~\ref{sec:cot_ablation} for more details.

\paragraph{Structural Progressive Conditioning (CoT).}\hspace{-1em}\footnote{We named it ``CoT'' to pay tribute to the concept of Chain-of-Thought prompting \citep{wei2022chain}, as we adopt similar instructions and leverage intermediate conditioning tokens as guidance. However, our approach fundamentally differs from the original Chain-of-Thought prompting in implementation and application. We also acknowledge that there is CoT-like work proposed for audio language models~\citep{du2024cot, ma2025audio, wang2025spark}.}\hspace{0.6em}
To address the long-term decay property issue, we propose an elegant solution that leverages the inherent structural priors of music. Songs are typically composed of distinct segments, such as intro, verse, chorus, bridge, and outro \citep{nieto2020audio, bruderer2009perception, lerdahl1996generative}. We use all-in-one \citep{kim2023all} to automatically segment songs into musical sections, with most of the sections shorter than 30 seconds. A song is on average segmented into 14 sessions. Within each structure section, text form segment labels, lyrics and audio are paired together. From a full song perspective, structured text and audio tokens are \textbf{interleaved} (see lyrics2song token arrangement in Figure~\ref{fig.stage-1}). Special tokens are incorporated to indicate the start and end of the audio. 

We describe a single training example as a concatenation of several components. In our setup, a song is constructed as
\[
\begin{array}{rcl}
\mathcal{D}_{\mathrm{cot}} & = & \underbrace{\text{Instruct} \;\circ\; \text{Tag} \;\circ\; \text{Lyrics}}_{\text{Prompt}} \;\circ\; \left( \bigcirc_{i=1}^{N} s_i \right) \;\circ\; \textsc{<EOD>}.
\end{array}
\]
Specifically,
$\circ$ denotes sequence concatenation.
\(\text{``Instruct''}\) is the instruction,
a task prefix as follows:
  \[
    \texttt{``Generate music from the given lyrics segment by segment.''}
  \]
  \(\text{``Tag''}\) denotes the musical tags, which is the style control string. An example of \(\text{Tag}\)  is as follows:
  \[
    \texttt{[Genre]\ jazz male deep vocal romantic big band}.
  \]
\(\text{``Lyrics''}\) represents the raw lyric text provided before any segmented annotations.
\(\textsc{<EOD>}\) is an end-of-document token.



In addition, each segment \(s_i\) is structured as follow:
{\small
\[
s_i = \textsc{[start\_of\_segment]} \;\circ\; \tau_i \;\circ\; \ell_i \;\circ\; \textsc{<SOA>} \;\circ\; \psi_i \;\circ\; \textsc{<EOA>} \;\circ\; \textsc{[end\_of\_segment]}.
\]}
\(\tau_i \in \{\texttt{[intro]},\ \texttt{[verse]},\ \texttt{[chorus]},\ \texttt{[bridge]},\ \texttt{[outro]}\}\) is a structure label, \(\ell_i\) representing the segment's lyric content\footnote{Interestingly, replacing lyrics string with \texttt{\textbackslash n} can enable instrumental music generation.},
and \(\psi_i\) denoting a sequence of Dual-NTP audio tokens\footnote{We prepend tokenizer type special token, e.g. \texttt{<xcodec>}, at the beginning of audio token sequence.}.

In summary, each document in CoT begins with an instruction, metadata, and raw lyrics, followed by a series of annotated segments, and ends with the \(\textsc{<EOD>}\) token.

\subsubsection{Music In-Context Learning}
\label{sec:icl}
\paragraph{Deficiencies of Speech ICL.}
Previous work in TTS~\citep{vall-e, du2024cosyvoice} often defines speech ICL via a continuation-based approach. The sequence is constructed as:
\[
\underbrace{T_{\mathrm{ref}}}_{\text{reference text}} \;\circ\;
\underbrace{T_{\mathrm{input}}}_{\text{input text}} \;\circ\;
\underbrace{A_{\mathrm{ref}}}_{\text{reference audio}} \;\circ\;
\underbrace{A_{\mathrm{gen}}}_{\text{generated audio}}
\]
While this framework can be suitable for speech-based tasks, there are three major issues when directly applying it to music:

\begin{enumerate}[label=\arabic*),topsep=1pt,itemsep=2pt,leftmargin=20pt]
    \item \textbf{Necessity of reference text.}\,
    Requiring a text transcript for the reference audio can be redundant in a musical context, and lyrics may be unavailable or challenging to obtain.

    \item \textbf{Unidirectional assumption.}\,
    Continuation is unidirectional and restricts the task generalization in scenarios requiring bidirectional creativity, e.g., writing an entire piece from a short chorus snippet.

    \item \textbf{Entanglement.}\,
    Continuation imposes strong constraints on the style and content of the generated audio. Given that music often features structural repetition, the model may simply replicate the reference melody or even entire segments, raising copyright concerns. Moreover, this tight coupling between reference and generated segments diminishes the effectiveness of control prompts or tags designed to steer the creative process.
\end{enumerate}

\paragraph{Re-designing ICL for Music.} The aforementioned issues necessitate a novel approach to ICL for music. We propose a revised formulation of music ICL in two modes: single-track and dual-track. In single-track mode, the reference audio can be an accompaniment, vocal, or full mixture track. In dual-track mode, we incorporate both the separated vocal and accompaniment tracks in a token-level interleaved manner, akin to Dual-NTP.

Extending the ICL format from CoT data, we randomly sample a 20--40s segment from the reference track(s) and prepend its token sequence to the CoT data:
\[
\mathcal{D}_{\mathrm{icl}} = A_{\mathrm{ref}} \;\circ\; \mathcal{D}_{\mathrm{cot}}.
\]

We find that this form of ICL can be effectively activated with minimal computational overhead (\textasciitilde2\% of the total pre-training cost). However, ICL constitutes a strong conditioning signal and can be considered as ``easy'' data. Our preliminary experiments reveal that incorporating ICL data too early encourages \textbf{shortcut learning}~\citep{geirhos2020shortcut}, where the model tends to directly copy the reference audio rather than composing novel music. This strong content entanglement even disrupts lyrical control. Once shortcut learning occurs, the model's creative capabilities cannot be easily restored. Removing ICL data and continuing training on CoT alone fails to resolve the issue—without reference audio, the model struggles to generate meaningful outputs, exhibiting poor musicality.

To address this, we introduce a \textbf{delayed activation strategy}. We introduce a small amount of ICL data (\textasciitilde10B tokens) only during the annealing phase, ensuring no ICL data is used beforehand. This strategy facilitates \textbf{disentangled} control between text and reference audio. For instance, using a Japanese city pop track with a female vocal as reference, the model can transform the lyrics into English while preserving the same vocalist and genre, or even generate a male English rap version of the city pop track.

\subsection{Stage-2: Residual Modeling}

\begin{figure}[t!]
\centering
\graphicspath{{figures/Yue_main/}}
\includegraphics[width=0.9\linewidth]{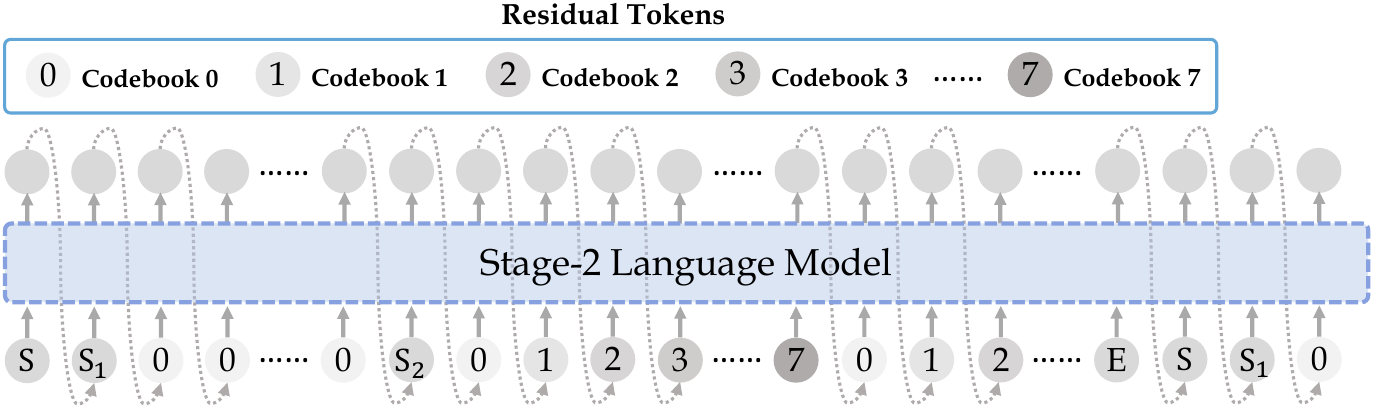}
\vspace{1em}
\caption{Stage-2 Framework of YuE. $S$: \texttt{<SOA>}, $S$=\texttt{<SOA>}, $E$=\texttt{<EOA>}, $S_i$=\texttt{<stage\_i>}.}
\label{fig:stage-2}
\end{figure}

As shown in Figure~\ref{fig:stage-2}, after Stage-1 yields coarse semantic tokens (codebook-0), Stage-2 refines the audio with additional codebooks \(1,2,\dots,7\). Denote the total number of codebooks by \(K=8\) (indexed from 0 to 7). Although codebook-0 is already produced by Stage-1, we train Stage-2 to predict \emph{all} codebooks \(\{0,1,\dots,7\}\) jointly in a single autoregressive framework. This design ensures that the model has a unified view of both the high-level structure (codebook-0) and the residual details (codebooks \(1\)--\(7\)).

\textbf{Architecture Overview.}
Let \(\mathbf{x}^{(0)}_{1:T} = (x^{(0)}_1, \dots, x^{(0)}_T)\) be the Stage-1 codebook-0 tokens for \(T\) frames. In Stage-2, we introduce additional codebooks, collectively denoted by 
\[
  \mathbf{x}^{(1:7)}_{1:T}
  \;=\;
  \bigl(x^{(1)}_1, \dots, x^{(7)}_1; \; \dots; \; x^{(1)}_T, \dots, x^{(7)}_T\bigr).
\]
For training, we treat the output space as \(\mathbf{x}^{(0:7)}_{1:T}\), i.e., each timestep \(t\) has a tuple
\[
  \mathbf{x}^{(0:7)}_t
  =
  \bigl(x^{(0)}_t, x^{(1)}_t, \dots, x^{(7)}_t\bigr).
\]
Although codebook-0 tokens are the same as those from Stage-1, they are included in the training target so the model learns to predict them as well, thus capturing complete frame-level dependencies across all codebooks.

\textbf{Aligned Autoregressive Factorization.}
We maintain a strictly time-aligned factorization:
\begin{align}
  p\Bigl(\mathbf{x}^{(0:7)}_{1:T}\Bigr)
  \;=\;
  \prod_{t=1}^T
  p\Bigl(\mathbf{x}^{(0:7)}_t
    \,\Big\lvert\, \mathbf{x}^{(0:7)}_{<t}
  \Bigr).
  \label{eq:stage2-factorization}
\end{align}
This ensures that at each frame \(t\), the model conditions on all previously generated tokens across \emph{all} codebooks, while still maintaining frame alignment with codebook-0.

\textbf{Cross-Conditioning.}  
During training, we organize the sequence as:
\[
  \bigl[
    \underbrace{x^{(0)}_1, \dots, x^{(0)}_T}_{\text{all codebook-0 first}},
    \underbrace{x^{(0)}_1, x^{(1)}_1, \dots, x^{(7)}_1, \;
                x^{(0)}_2, x^{(1)}_2, \dots, x^{(7)}_2,\;\dots,\;
                x^{(0)}_T, x^{(1)}_T, \dots, x^{(7)}_T}_{\text{blocks of }0\text{-}7\text{ per frame}}
  \bigr].
\]
That is, the first segment is only the codebook-0 tokens, followed by repeated 8-token blocks \(\{0,1,\dots,7\}\) for each frame. We apply standard teacher forcing on this extended sequence and minimize
\[
  \mathcal{L}_{\mathrm{Stage2}}
  \;=\;
  -\sum_{t=1}^T
    \log p\Bigl(
        \mathbf{x}^{(0:7)}_t \,\Big\lvert\,
        \mathbf{x}^{(0:7)}_{<t}
    \Bigr).
\]
By placing all codebook-0 tokens at the beginning, the model is guaranteed to “see” the entire semantic structure before it encounters any mixed (0–7) blocks. This allows the model to plan the later residuals by attending to a complete semantic outline from Stage-1.

\textbf{Inference.}  
At test time, codebook-0 tokens \(\mathbf{x}^{(0)}_{1:T}\) come from Stage-1 and are treated as fixed (i.e., clamped). Even though the model is trained to predict codebook-0 as part of the joint sequence, during inference we replace any predicted codebook-0 tokens with the Stage-1 output. Consequently, the only “free” outputs in the autoregressive generation are the residual codebooks \(\mathbf{x}^{(1:7)}_{1:T}\). This ensures the sequence alignment.

\textbf{Implementation.}
Our model is a 1B-parameter Transformer with an 8K-token context window, trained on consecutive 6-second single-track segments. It employs a shared acoustic codebook space to model various audio types, including speech, vocals, instrumentals, and mixtures.

\subsection{Tokenization and Audio Reconstruction}
\label{sec:tokenizer}
Following the design space of \citet{audiolm, wang2023neural-valle}, the stage-1 LM models text tokens and semantic-rich codebook-0 tokens. 
After investigation, we realized that the vanilla text-to-speech (TTS) / text-to-music (TTM) method performs poorly on our task, where musicality and song-level lyrics-following capability are the two key challenges.

\begin{wraptable}{l}{0.5\textwidth}
    \vspace{-0.5em}
    \centering
    \captionsetup{skip=2pt} 
    \caption{Special tokens and their descriptions.}
    \label{tab:special_tokens}
    \begin{footnotesize}
    \begin{tabular}{ll}
        \toprule
        \textbf{Token} & \textbf{Description}\\
        \midrule
        \texttt{<EOD>} & End of document \\
        \texttt{<SOA>} & Start of audio \\
        \texttt{<EOA>} & End of audio \\
        \texttt{<stage\_1>} & Start of Stage~1 \\
        \texttt{<stage\_2>} & Start of Stage~2 \\
        \texttt{<encodec32k>} & Tokenizer type (Encodec 32k) \\
        \texttt{<xcodec>} & Tokenizer type (XCodec) \\
        \texttt{<semanticodec>} & Tokenizer type (SemantiCodec) \\
        \texttt{<hificodec>} & Tokenizer type (HiFiCodec) \\
        \bottomrule
    \end{tabular}
    \end{footnotesize}
    \vspace{-1em}
\end{wraptable}

\paragraph{Text Tokenizer.} In this work, the vocabulary of the LMs contains two sections: text and audio. For the text part, we reuse LLaMA tokenizer with a size of 32000 unique BPE tokens. Instructions, genres, lyrics, structure annotations, and structure segment boundary signals are represented with text format and tokenized with BPE.

\paragraph{Semantic-Acoustic Fused Codec.} For the audio vocabulary, we experimented with several open-source music and universal neural codecs. Detailed ablations are provided in Section \ref{sec: exp}. Ultimately, we adopted a semantic-acoustic fused strategy \citep{defossez2024moshi, zhang2023speechtokenizer, liu2024semanticodec, ye2024codec}. Specifically, we utilized \texttt{X-Codec} \citep{ye2024codec} as our off-the-shelf audio tokenizer. We employed a general-purpose version of \texttt{X-Codec}, trained on a mixture of 200k hours of 16 kHz audio with a ratio of music : speech : audio effects = 1 : 1 : 0.05. 

The \texttt{X-Codec} tokenizer fuses a 100M-parameter HuBERT-based universal semantic representation into the codec latent space. It has a 50Hz frame rate, consists of 12 RVQ layers, each with a codebook size of 1024. For this study, we used only the first 8 layers, as including more layers did not yield noticeable quality improvements. Notably, codebook-0 alone captures rich semantic information such as melody and vocal content, which are critical for our task.

\paragraph{Vocabulary Expansion and Special Tokens.} We expand the SentencePiece tokenizer vocabulary to support multiple audio tokenizers and special tokens. Specifically, we include \texttt{Encodec-32khz-music} \citep{encodec,musicgen}, \texttt{HiFi-Codec-universal} \citep{yang2023hifi,yang2023uniaudio}, \texttt{X-Codec-general} \citep{ye2024codec}, and \texttt{Semanticodec-100tps} \citep{liu2024semanticodec}. 

For special tokens, we introduce the following: \texttt{<EOD>} represents the end of a document, \texttt{<SOA>} denotes the start of audio, and \texttt{<EOA>} signifies the end of audio. Additionally, stage indicators, \texttt{<stage\_1>} and \texttt{<stage\_2>}, mark the beginning of Stage~1 and Stage~2 tokens, respectively. Tokenizer type indicators specify the corresponding tokenizer types, which are inserted between \texttt{<SOA>} and the actual audio token IDs. Note that stage indicators are only used in residual modeling and positioned between \texttt{<SOA>} and the tokenizer type indicator.

\textbf{Light-weight Upsampling Module.} To achieve better perceptual audio quality, we upsample the reconstructed 16kHz audio to 44.1kHz. For this, we utilize a light-weight upsampling vocoder adapting Vocos \citep{siuzdak2023vocos} to predict the higher-frequency components. To enhance the robustness of the upsampler, we apply codebook dropout randomly and introduce a small amount of Gaussian noise during training.

\section{Training and Inference Strategies}

\subsection{Scaling Up Stage-1 Pre-Training}
\label{sec:multitask_multiphase}

\subsubsection{Multitask Learning} 

\begin{wraptable}{r}{0.45\textwidth}
  \centering
  \begin{footnotesize} 
    
    \caption{Decomposition of Lyrics-to-Song Generation Capabilities}
    \vspace{1em}
  \begin{tabular}{@{}l@{}}
    \toprule
    \textbf{Essential Capabilities} \\ \midrule
    1) Modeling of Human Vocal \\
    2) Modeling of Instrumental \\
    3) Joint Modeling of Vocal and Instrumental \\
    4) Aligning Cross-Modal/Same-Modal Controls \\
    \quad (lyrics, style, structure, in-context learning) \\ \bottomrule
  \end{tabular}
  \label{tab:capabilities}
  \end{footnotesize}
\end{wraptable}

Conditional lyrics-to-song data are inherently scarce, as most available music data exist in an \textit{unconditional} format. Our preliminary experiments show that \textit{large models tend to overfit to dominant learning signal}\footnote{See more discussion in Section~\ref{lab:unsuccessful_attempts}.}, making it difficult for them to adhere to control signals when pre-training is predominantly driven by unconditional data.

To address this, we propose a \textit{multitask pre-training} approach that facilitates \textbf{knowledge transfer} from auxiliary tasks to enhance lyrics-to-song generation. We decompose the essential capabilities required for this task into the four key components listed in Table~\ref{tab:capabilities}. These components serve as guiding principles for our multitask setup, which includes:

\textbf{Text-to-Speech.} Establishing alignment between linguistic control and human vocalization necessitates the use of speech data paired with text transcripts. 
This task is essential for enabling lyric-following capabilities, as discussed in capabilities 4) and 1). Omitting this task results in ineffective lyric control when training on in-the-wild data.

TTS data primarily consists of short-form speech, typically under 20 seconds, which is significantly shorter than music tracks. To mitigate this sequence length mismatch, text-speech pairs are sequentially concatenated to form full-context sequences. Additionally, the task instruction \texttt{Generate speech:} is prepended to transcripts with a dropout rate of 50\% to enhance robustness.

While this task is beneficial, the proportion of TTS data used is crucial. Excessive TTS training biases the generated token space towards speech, effectively modeling rap music but degrading performance on other genres require singing\footnote{Overfitting TTS data turns the model into a rap machine.}. Conversely, insufficient TTS training leads to poor adherence to lyrics. Striking an optimal balance is essential for achieving effective lyric control across diverse musical styles.

\textbf{Music Generation.} The majority of our dataset consists of unconditional music.
We annotate all tracks using Qwen2-Audio \citep{chu2024qwen2} to obtain open-vocabulary tags. Tags are in the style of MTG-Jamendo \citep{bogdanov2019mtg}, categorized by genre, instrument, and mood. The input to Qwen2-Audio is a 30-second clip sampled from each track. Prompt is shown in appendix~\ref{box:music tagging prompt}.

Furthermore, 40\% of the tracks are separated into vocal-instrumental dual-track format using UVR\footnote{\url{https://github.com/Anjok07/ultimatevocalremovergui}}. We employ ensemble predictions\footnote{We use the minimal signal of each track.} from three models: \texttt{htdemucs\_ft}, \texttt{Kim\_Vocal\_1}, and \texttt{UVR-MDX-NET-Inst\_3}.

The processed tracks are tokenized and arranged into either tag-conditioned NTP or Dual-NTP formats. Text instructions are prepended before the audio sequences to distinguish the two prediction objectives: \texttt{Generate music based on the given tags} or \texttt{Generate music in dual-track format based on the given tags.} 
The tag condition consists of a \texttt{[genre]} string followed by shuffled tags separated by spaces, inserted between the instruction and the audio sequence.

While this is a relatively challenging task, training on it improves musicality, facilitates the development of capabilities 1), 2), and 3), while enabling style control within capability 4).

\textbf{Lyrics-to-Song.} Obtaining high-quality paired lyrics-audio data is challenging, as sources from web searches and platform-provided transcripts often contain noise, irrelevant text, misaligned timestamps, and version discrepancies. To address this, we implement heuristic filtering to remove irrelevant content and exclude overly short lyrics (less than 10 sentences), retaining only approximately 10\% of matched tracks. Despite filtering, some inconsistencies remain.

The CoT design addresses these issues by leveraging segment-level rather than sentence-level lyrics-audio alignment, thus reducing reliance on precise matches. Additionally, incorporating a TTS auxiliary task further enhances model robustness against imperfect alignment. Manual quality inspection on over one hundred segments confirmed an approximate 80\% match rate, defined by the audible presence of the majority of text in the audio.

For ICL, we support single-track and dual-track modes. Reference token sequences (20s–40s) are randomly selected from the mixed, vocal, accompaniment tracks, or combinations thereof, and are prepended directly to corresponding CoT samples. We introduce vocal tags during ICL, prompt shown in appendix~\ref{box:vocal tagging prompt}.

\subsubsection{Multiphase Training}
\label{sec:multiphase_training}


\textbf{Phase-1: Warm Up.} 
In the first phase, we warm up the model with a linear learning rate schedule from $lr=0$ to $lr=3\times10^{-4}$ over 280B tokens. Only English and Chinese data are used, as manual verification showed that these two languages dominate the dataset and exhibit relatively high quality. To save computational costs, we use a context length of 8192 (approximately 163s for mix music data and 81s for dual-track data) and a global batch size of 768 (around 6.29M tokens). This phase rapidly establishes basic musical generation capabilities.

\textbf{Phase-2: Constant Learning Rate.} 
In this phase, we maintain a constant learning rate of $3\times10^{-4}$ and introduce additional in-the-wild, lower-quality datasets, including multilingual data. The total processed tokens reach 1T. When incorporating new data, we maintain a 2:1 ratio of old to new data to prevent excessive distribution shifts.

\textbf{Phase-3: Context Extension.} 
We retain the learning rate at $3\times10^{-4}$ and extend the context length. Since music inherently involves long sequences, we simply increase the maximum positional embedding and sequence length to 16384 without modifying the data composition. We remove the single-track unconditional data during this phase. This phase continues training for an additional 750B tokens, further enhancing the model’s ability to handle long-context dependencies across multiple languages.

\textbf{Phase-4: Annealing with Control Injection.}
This is the final phase of the Stage-1 LM training. The learning rate follows a cosine schedule, gradually annealing to $3\times10^{-5}$. At this stage, we completely remove speech and unconditional music data while introducing stronger control signals. The control signals include reference audio (ICL), gender tags, vocal timbre tags, and BPM control. However, BPM control was later removed due to its coupling with lyrics length, which degrades lyrics following. 

To improve training data quality, we constructed quality signals and selected approximately 20K hours of high-quality data. The quality signals include playback count, likes, comments, and dataset source quality ratings (based on manual inspection pass rates). We performed annealing experiments across multiple languages, including English, Chinese, Japanese, and Korean. During annealing, we applied a CoT to ICL ratio of 2:1 to prevent excessive reliance on reference songs. Remarkably, with only 40B tokens (\textasciitilde2\% of the total compute budget), the model successfully enabled all control signals introduced in this stage.

\subsection{Stage-2 Pre-training.}
We train a Stage-2 LM with a context length of 8192. This phase incorporates all speech, demixed music, and mixed music datasets. The compute budget is set to 2T tokens. The learning rate follows a linear warm-up and cosine annealing schedule with a maximum learning rate of $3\times10^{-4}$. We find in preliminary experiments that scaling the Stage-2 LM from 0.5B to 1B parameters and increasing the dataset size leads to improvements in audio quality; therefore, we adopt a 1B-parameter model for this stage.

\subsection{Test-time Strategies}
\textbf{Forced Decoding.} In stage-1 LM decoding, only vocabulary tokens within the audio range are permitted until the \texttt{<EOA>} token is predicted. Subsequently, the prompt for the next segment is forcibly provided based on user input. In stage-2 LM, the codebook-0 tokens, predicted by the previous stage, are enforced at each frame. When decoding the corresponding residual token, only the vocabulary of the respective codebook is allowed.

\textbf{Sampling and Classifier-Free Guidance.} The sampling parameters are set as follows: top-\(k = 50\), repetition penalty \(= 1.1\), top-\(p = 0.93\), temperature \(= 1\), and maximum new tokens \(= 3000\). Classifier-free guidance is applied with a scale of \(s = 1.5\) for the first segment and \(s = 1.2\) for subsequent segments\footnote{A lower guidance scale in later segments promotes diversity.} to improve the good-case rate. Given the conditional log-probability \(\ell_c(k) = \log p_\theta(k \mid x)\) for token \(k\) given prompt \(x\) and the unconditional log-probability \(\ell_u(k) = \log p_\theta(k \mid \varnothing)\), the CFG-adjusted log-score is computed as:
\[
\ell_{\mathrm{cfg}}(k) = s\bigl[\ell_c(k) - \ell_u(k)\bigr] + \ell_u(k).
\]

\textbf{Music In-Context Learning.} Using a song's chorus section for in-context learning significantly enhances musicality and stability. Moreover, we find that dual-track ICL enables better audio quality than single-track ICL mode. Consequently, dual-track ICL mode is enabled by default unless specified otherwise.

\section{Experiments}
\label{sec: exp}
\subsection{Data \& Training Setup}

\textbf{Data Setup.} 
For conditional speech data (TTS), we leverage three widely used English and Chinese TTS datasets—WeNetSpeech (zh), LibriHeavy (en), and GigaSpeech (en)—comprising a total of 70k hours of data. For unconditional music data (music generation), we mine 650K hours of in-the-wild music recordings from the Internet. 10\% of the music data has corresponding lyrics after filtering. 

After tokenization, Stage-1 comprises 13B conditional speech tokens, over 200B unconditional music tokens (both mixed and demixed), and 28B CoT music tokens. During annealing, a high-quality subset of 10B CoT tokens is sampled and expanded fourfold, creating a 40B ICL dataset. This dataset includes variants such as vocal-ICL, accompaniment-ICL, mix-ICL, and dual-ICL. Prior to annealing, the data mixture is set at \textit{Conditional : Unconditional = 3 : 1} and \textit{Music : Speech = 10 : 1}. During annealing, only CoT and ICL data are used, maintaining a ratio of \textit{CoT : ICL = 2 : 1}.

\textbf{Training Setup.}  
Our codebase is built upon Megatron-LM~\citep{shoeybi2019megatron}, following the LLaMA2 architecture~\citep{llama, mapneo}. Most of our Stage-1 experiments use a 0.5B-scale model trained on 16 NVIDIA H800 GPUs, with a typical token budget of 100B tokens. Under this budget, models usually produce valid outputs, show preliminary lyric-following capabilities, and exhibit basic musical discernment. For scaling experiments, we increase the token budget to 500B and scale models to 0.5B, 2B, and 7B parameters, trained respectively on 32, 96, and 512 NVIDIA H800 GPUs. We further train the 7B LM with additional data, scaling up to a total of 1.75T tokens before starting an annealing phase, during which we apply a 40B-token annealing process. We maintain a global batch size of 768 when possible by adjusting micro-batch size, gradient accumulation steps, and tensor parallelism; when computational resources are constrained, we reduce the global batch size to 512 or 256. For optimization, we use the Adam optimizer with gradient clipping set to 1.0, weight decay 0.1, $\beta_1=0.9$, $\beta_2=0.95$, $\epsilon=10^{-8}$, and parameter initialization with standard deviation 0.02. Detailed training procedures are described in Section~\ref{sec:multiphase_training}.

\subsection{Evaluation Protocol}
\paragraph{Baselines.} As of the writing of this paper, apart from YuE, no academic or open-source system provides usable long-form song generation capabilities, and known prior works exhibit limited performance~\citep{jukebox}. Therefore, we selected five popular closed-source systems for benchmarking: Suno V4\footnote{\url{https://suno.com}}, Udio\footnote{\url{https://www.udio.com}}, Hailuo\footnote{\url{https://hailuoai.com/music}}, and Tiangong\footnote{\url{https://www.tiangong.cn/music}}.
It is important to note that due to the black-box nature of these closed-source models, our evaluation conducted in January 2025 reflects the comparative performance between YuE and these systems at that specific point in time. 

All systems support lyric-based inputs; however, their support for style control inputs varies significantly. Specifically, Tiangong does not support textual style prompts, so we used our own reference audio as style control. Hailuo provides 18 predefined style tags, thus we selected the tag closest to our desired style prompt and used the system’s default built-in reference audio, as uploading custom references is not supported. 

\paragraph{Human Evaluation.} We conducted a human evaluation involving 40 researchers, including 12 experts in Speech/Music AI\footnote{Worked on text-to-speech, text-to-music, singing voice synthesis.} and 7 trained musicians. None of the evaluators participated in model training, ensuring objectivity. Following prior studies~\citep{singsong, qu2024mupt, yuan2024chatmusician}, we adopted an A/B test format.

In the main experiments in Section~\ref{sec:main_results}, each model generated 42 full-length songs based on a diverse set of English prompts specifying genre, instruments, emotion, lyrics, and tempo. These prompts utilized real lyrics that were rewritten by GPT and paired with corresponding 30s chorus segments as reference audio. Similarly, for the multilingual experiment, we used 10 Chinese prompts and 10 Japanese/Korean prompts. Some multilingual prompts contained sentences with more than one language, e.g., EN-JA-KR mixes. For evaluation involving non-English multilingual samples, we invited native speakers or language-major students proficient in the respective languages to conduct assessments.

Evaluators blindly compared pairs of music pieces produced by two different systems according to several criteria: Overall Musicality, Vocal Quality (VocalQual), Accompaniment Quality (AccompQual), Music Arrangement (MusicArr), Melodic Attractiveness (MelodicAttrac), Vocal-Accompaniment Compatibility (VocalAccompComp), Song Structure Clarity (SongStruct), Lyrics Following Accuracy (LyricFollow)\footnote{We observe that Whisper transcription accuracy is insufficiently robust for reliable automated lyrics-following evaluation. Therefore, lyrics alignment with input prompts is manually evaluated by human raters in the main experiments.}, Genre Controllability (GenCtrl), Instrument and Vocal Configuration Controllability (InstrCtrl), Emotional Expressiveness (EmoCtrl), and Tempo and Rhythm Control (Tempo/RhyCtrl).

For ablation studies in Section~\ref{sec:ablation}, unless otherwise specified, we utilize a set of 15 GPT-generated English prompts (see Appendix~\ref{gpt_15_prompts}). Each study undergoes small-scale A/B testing, with inference performed twice per prompt, resulting in a total of 30 samples per setting.

\paragraph{Automatic Evaluation.} We also report automatic evaluation metrics, including Kullback–Leibler (KL) divergence for measuring distributional differences in generated audio features using \texttt{audioldm\_eval}\footnote{\url{https://github.com/haoheliu/audioldm_eval}}, Frechet Audio Distance (FAD)~\citep{kilgour2019frechet} for assessing audio quality and realism (also via \texttt{audioldm\_eval}), Audiobox-Aesthetic~\citep{audioboxaesthetics} for capturing perceived musical aesthetics (Production Quality (PQ), Production Complexity (PC), Content Enjoyment (CE), and Content Usefulness (CU)) using a neural audio embedding model, CLAP score\footnote{\url{https://github.com/Stability-AI/stable-audio-metrics}} and CLaMP 3 score~\citep{wu2025clamp3universalmusic}\footnote{\url{https://github.com/sanderwood/clamp3}} to measure semantic alignment between text prompts and audio outputs, vocal agility quantifying song-level vocal range and flexibility (pitch estimated with RMVPE\footnote{\url{https://github.com/yxlllc/RMVPE}}, applying 40ms note filtering and human verification), and generation duration as a practical measure of song-level audio modeling capability.

\begin{figure}[t!]
    \centering
    \begin{subfigure}[t]{0.48\linewidth}
        \centering
        \includegraphics[width=\linewidth]{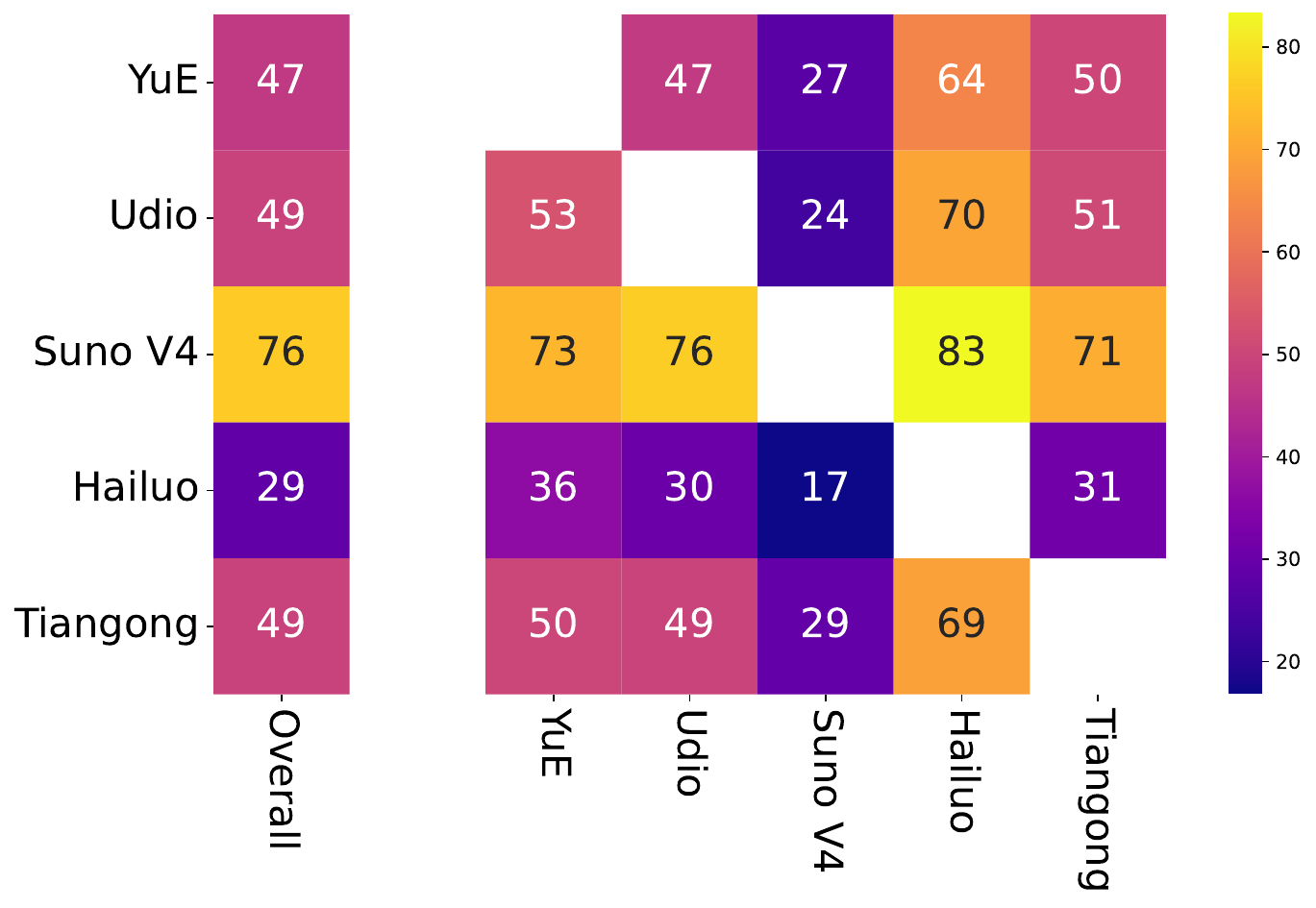}
        \label{fig:heatmap_avg}
    \end{subfigure}
    \hfill
    \begin{subfigure}[t]{0.48\linewidth}
        \centering
        \includegraphics[width=\linewidth]{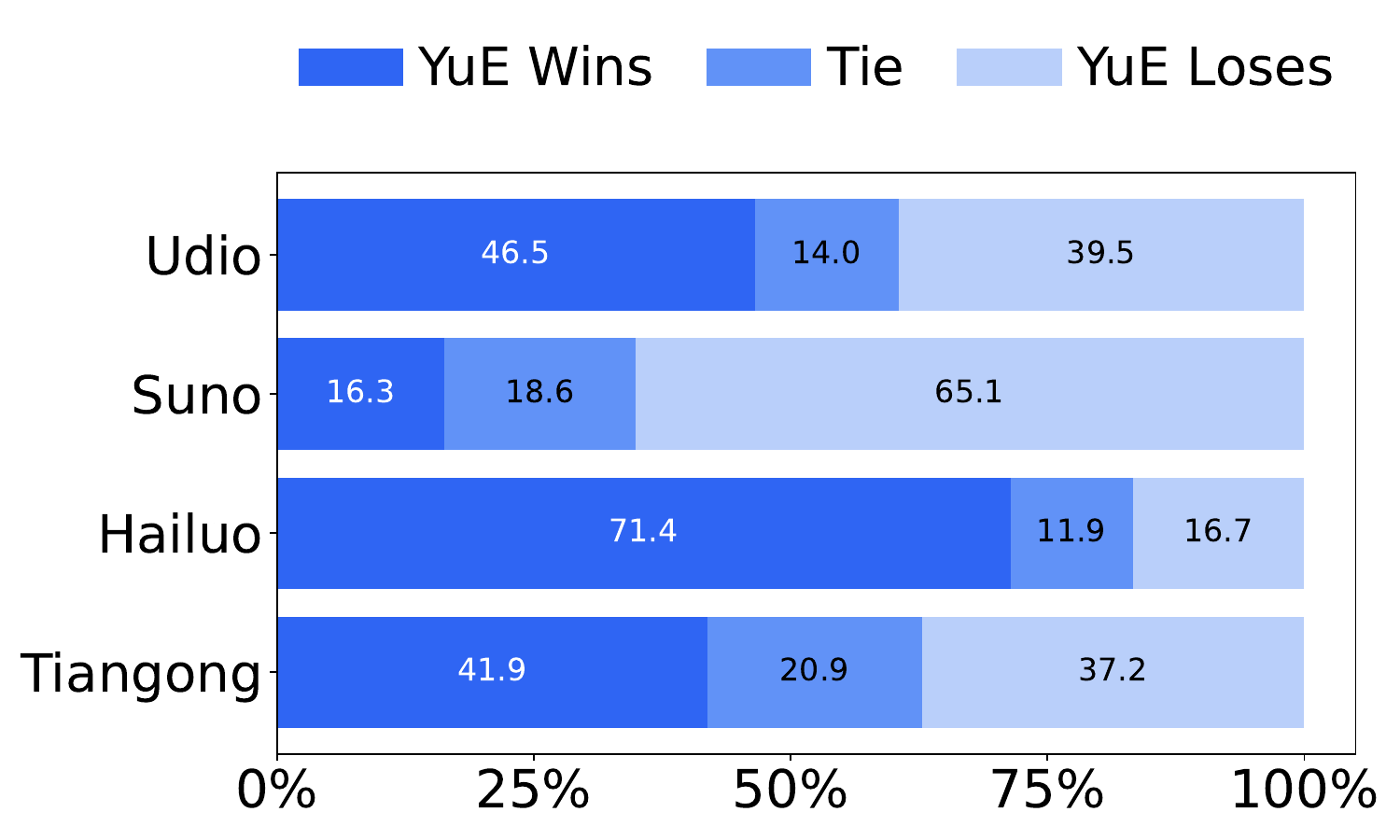}
        \label{fig:wintieloss_musicality}
    \end{subfigure}

    \caption{Human evaluation comparing YuE to 4 proprietary systems. YuE matches two of it (Tiangong, Udio) and outperforms one (Hailuo). Left: Average human preference on all aspects (warmer colors / larger numbers indicate higher preference); Right: win-tie-loss on musicality.}
    \label{fig:overall_human}
\end{figure}

\section{Main Results}
\label{sec:main_results}
\subsection{Human Evaluation}
\label{en_human_eval}
We report the generation result on English in section~\ref{en_human_eval}.
\subsubsection{Overall Comparison with Proprietary Systems.}
In human evaluation \autoref{fig:overall_human}, our model, YuE, demonstrates competitive performance relative to four proprietary systems in both average human preference\footnote{Obtained by averaging win rate over all aspects.} and musicality. Specifically, YuE outperforms Hailuo by a clear margin, achieves comparable results to Tiangong and Udio, but still trails behind Suno V4, which remains the state-of-the-art system. In detailed musicality comparisons, YuE shows balanced win–loss ratios against Tiangong and Udio, decisively outperforms Hailuo, but underperforms compared to Suno V4. These results indicate that while proprietary products still lead in the best quality, YuE represents a promising step toward high-quality open-source music generation.

\subsubsection{Detailed Comparison with Proprietary Systems.}
\begin{figure}[t!]
    \centering
    \begin{subfigure}[b]{0.465\linewidth}
        \centering
        \includegraphics[width=\linewidth]{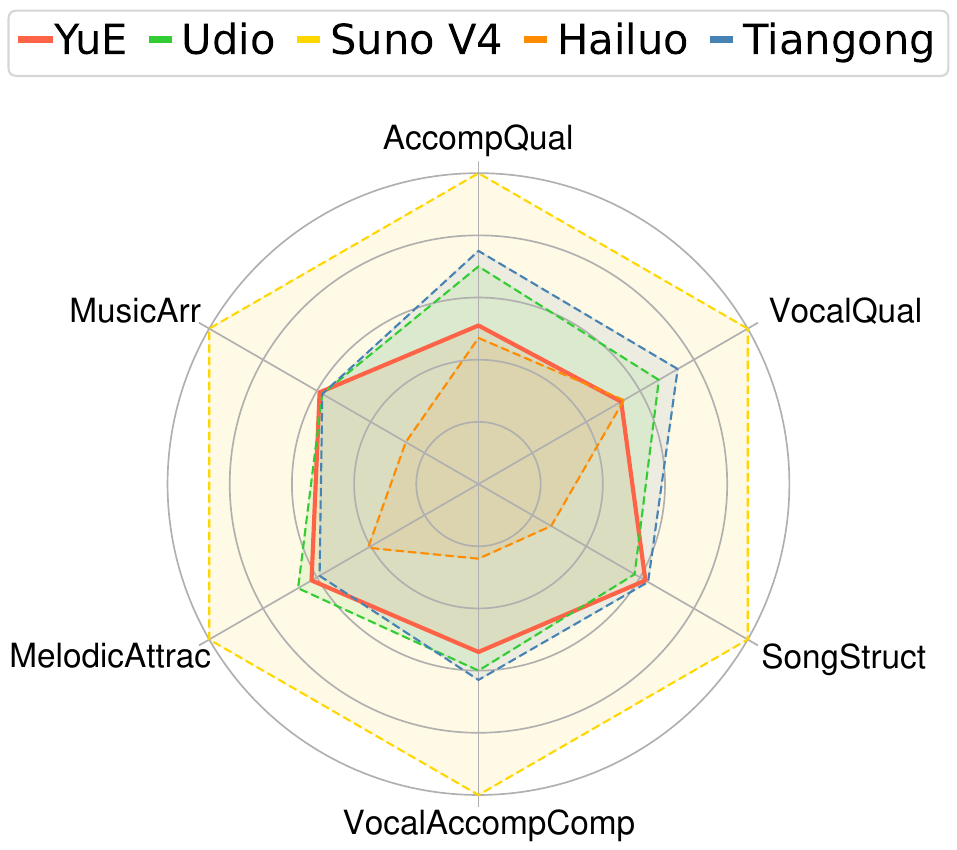}
        \label{fig:subjective-radar}
    \end{subfigure}
    \hfill
    \begin{subfigure}[b]{0.48\linewidth}
        \centering
        \includegraphics[width=\linewidth]{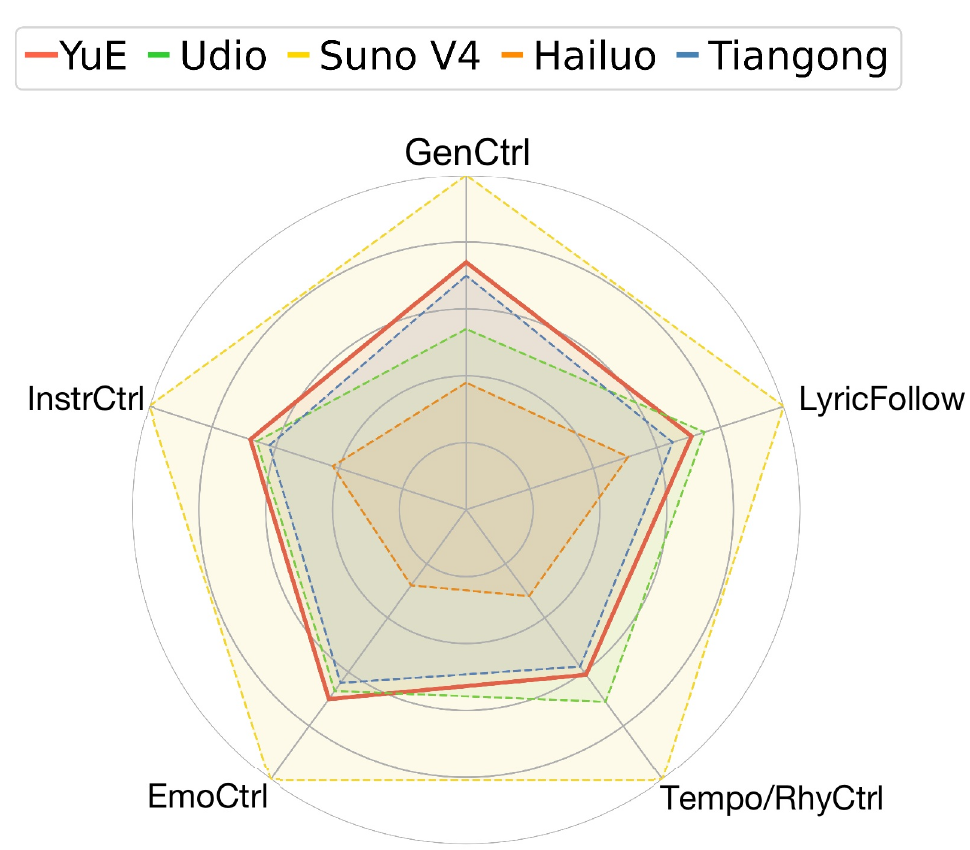}
        \label{fig:objective-radar}
    \end{subfigure}
    \caption{Normalized human preference on different music aspects. Left: scores across 6 musical aspects; Right: performance on 5 types of control.}
    \label{fig:combined-radar}
\end{figure}

\paragraph{Aspects of Musicality and Acoustic Quality.} To evaluate the subjective musical qualities of YuE and comparative models, we conducted a detailed A/B test on six dimensions: \textit{vocal (acoustic) quality}, \textit{accompaniment (acoustic) quality}, \textit{music arrangement}, \textit{melodic attractiveness}, \textit{vocal-backtrack matching}, and \textit{song structure}. We visualize the win rate with radar plot in \autoref{fig:combined-radar}(L). Suno V4 consistently outperforms all other models across these aspects, thus we normalized the win rate by Suno to improve visual clarity. Among other models, YuE excels notably in music structure and music arrangement, highlighting its capability for coherent long-form composition capability. However, YuE shows clear deficiencies in vocal and accompaniment acoustic quality, likely due to limitations of its current audio tokenization method. While YuE achieves decent musicality and convergence, the semantic-fused tokenizer requires improvements in acoustic detail via an enhanced decoder or a super-resolution backend.
\paragraph{Controllability.} Similarly, we evaluated the controllability of YuE and comparative models through A/B testing on five dimensions: \textit{genre control}, \textit{instrument/vocal control}, \textit{emotion control}, \textit{tempo/rhythm control}, and \textit{lyrics following}. Given limitations of existing classifiers and transcription systems, user preference win rate was our primary evaluation metric, with results presented in \autoref{fig:combined-radar}(R). Suno v4 consistently outperforms all models across controllability metrics. 
Among other models, YuE performs strongest in genre adherence, instrument/vocal consistency, and emotion, highlighting its effectiveness in generating stylistically coherent music aligned with textual prompts. YuE demonstrates moderate performance in emotion and tempo control, indicating the need for improved lyric alignment and tempo tagging systems due to considerable noise observed in the pseudo label on the training corpus provided by Qwen2Audio. Overall, these results affirm YuE’s robust controllability capabilities.

\subsection{Automatic Evaluation}
\subsubsection{Vocal Agility}
\begin{figure}[htbp]
    \centering
    \includegraphics[width=0.7\linewidth]{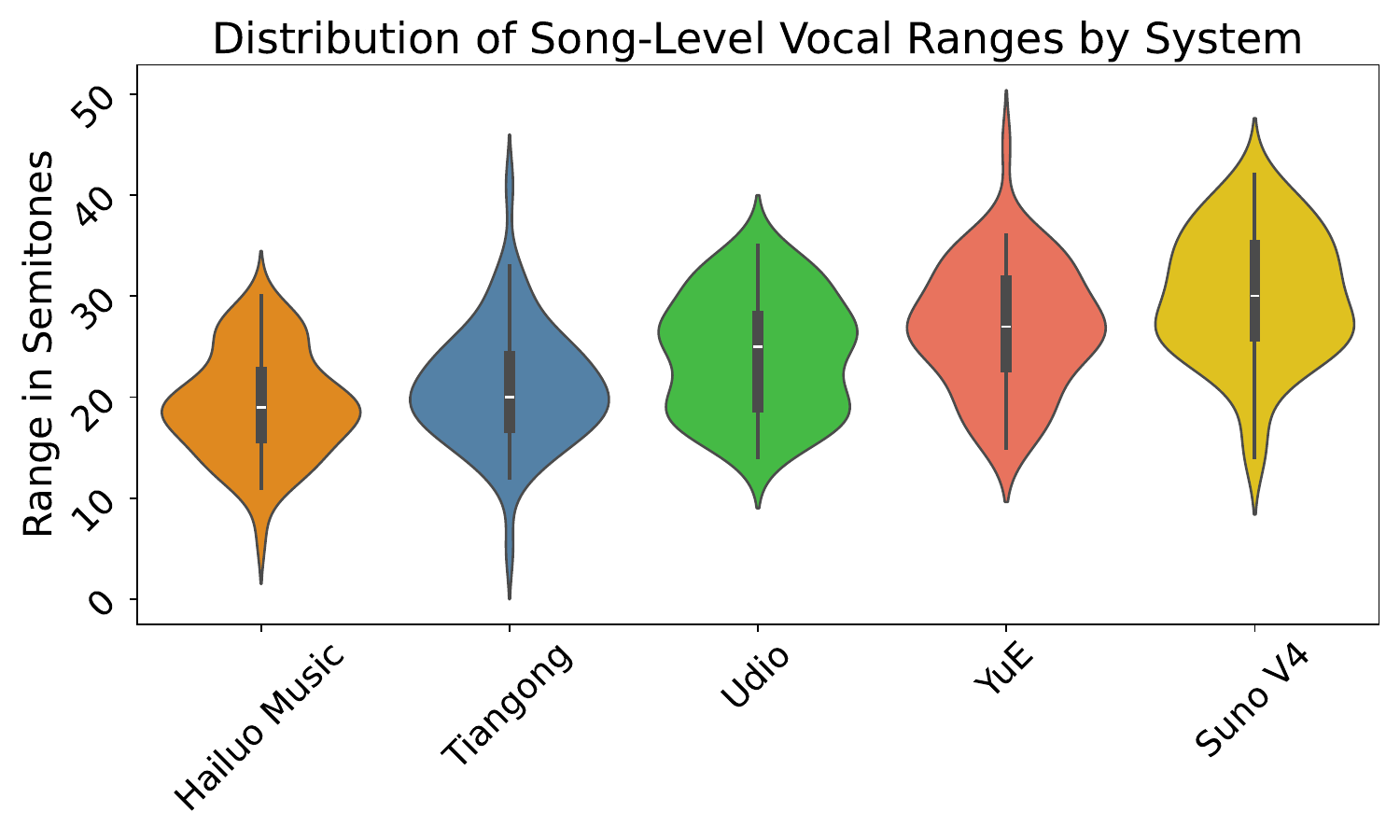}

    \caption{Song-level vocal range on different systems. Higher values indicates better vocal agility, e.g. range=12 means the vocal only span through an octave in a given song. YuE's vocal range is among the top close-source systems.}
    \label{fig:objective-vocal-range}
\end{figure}
As shown in \autoref{fig:objective-vocal-range}, the distribution of song-level vocal ranges across different systems reveals notable variations in vocal agility. Higher values indicate greater vocal expressiveness. Among the models, YuE demonstrates one of the widest vocal ranges (medium $\sim=27$ semitones), closely matching top-performing closed-source systems like Suno V4. This suggests that YuE is capable of generating diverse and dynamic vocal performances. In contrast, models like Hailuo and Tiangong show a more constrained vocal range (medium number around 20 semitones), indicating potential limitations in expressiveness. These findings highlight YuE’s strength in producing vocally rich and varied song compositions.

\subsubsection{Duration}
\begin{figure}[htbp]
    \centering
    \includegraphics[width=0.7\linewidth]{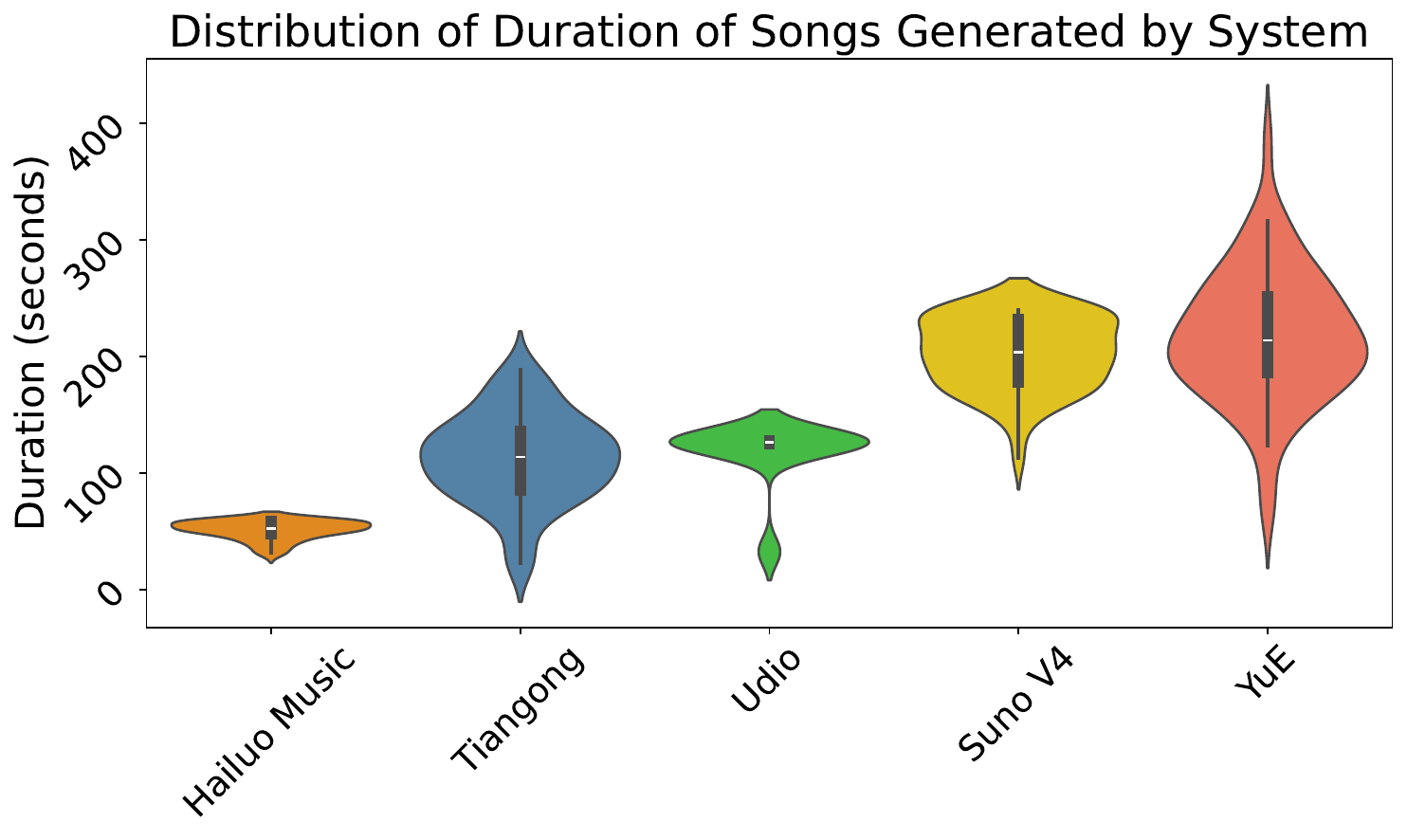}
    \caption{Duration range on different systems. YuE generates the longest audio.}
    \label{fig:objective-duration-range}
    \vspace{1em}
\end{figure}
The distribution of generated song durations across different systems reveals substantial variation in length constraints as demonstrated by \autoref{fig:objective-duration-range}. YuE produces the longest audio, with a significantly wider duration range compared to all other models, demonstrating its ability to generate full-length songs beyond typical AI-generated clips. SunoV4 and Tiangong also generate relatively long audio. In contrast, Hailuo Music show the most restricted durations, suggesting limitations in modeling long-term musical structure. These results highlight YuE’s advantage in handling extended temporal dependencies, making it more suitable for full-song generation.

\subsubsection{Model Based Evaluation}
Table~\ref{tab:autometric} illustrates model based automatic evaluation results, including distribution metrics KL and FAD, aesthetics metrics proposed by meta, and audio-text alignment score such as CLAP score~\citep{clap} and CLaMP 3 score~\citep{wu2025clamp3universalmusic}. \textbf{Note that not all metrics align well with human perception}. We will further discuss the correlation between each metric and human evaluation in Section~\ref{sec:auto_human_corr}.

\begin{table*}[t]
    \centering
    \small
    \renewcommand{\arraystretch}{1.2} 
    \caption{Comparison of various music generation models across multiple metrics.}
    \begin{tabular}{l cc cccc cc}
    \toprule[1.5pt]
    \textbf{Metric} & \multicolumn{2}{c}{\textbf{Distrib. Match}} & \multicolumn{4}{c}{\textbf{Content Based}} & \multicolumn{2}{c}{\textbf{Alignment}} \\
    \cmidrule(lr){2-3} \cmidrule(lr){4-7} \cmidrule(lr){8-9}
     & KL\(\downarrow\) & FAD\(\downarrow\) & CE\(\uparrow\) & CU\(\uparrow\) & PC\(\uparrow\) & PQ\(\uparrow\) & CLAP\(\uparrow\) & CLaMP 3\(\uparrow\) \\
    \midrule
    Hailuo   & 0.756 & 2.080 & 7.350 & 7.737 & \textbf{6.793} & 8.132 & 0.265 & 0.106 \\
    SunoV4   & 0.620 & 1.544 & \textbf{7.474} & \textbf{7.813} & 6.601 & 8.120 & 0.265 & 0.160 \\
    Tiangong & 0.708 & 2.547 & 7.421 & 7.766 & 6.060 & \textbf{8.220} & 0.244 & 0.114 \\
    Udio     & 0.503 & \textbf{1.222} & 7.112 & 7.520 & 6.626 & 7.803 & \textbf{0.310} & 0.156 \\
    YuE      & \textbf{0.372} & 1.624 & 7.115 & 7.543 & 6.280 & 7.894 & 0.118 & \textbf{0.240} \\
    \bottomrule[1.5pt]
    \end{tabular}
    \label{tab:autometric}
\end{table*}

\paragraph{Distribution Matching Metrics.} We report KL and FAD to evaluate how well generated audio matches the target distribution. YuE achieves the best performance on KL divergence (0.372), significantly outperforming others such as Udio (0.503) and SunoV4 (0.620). While Udio attains the lowest FAD (1.222), YuE remains competitive (1.624), demonstrating effective audio quality and distribution matching capabilities. Although distribution-based metrics can suffer from sample size biases, they remain valuable for comparative purposes, particularly when evaluating against closed-source systems where large-scale sampling is impractical. We refrain from adopting the traditional MusicCaps-based evaluation scheme since MusicCaps contains a large amount of purely instrumental content, rendering it unsuitable as a reference set for song generation tasks.

\paragraph{Content Based Metrics.} Scores above 7 across audiobox-aesthetic dimensions indicate a strong overall performance. Specifically, YuE achieves competitive scores—PQ (7.894), PC (6.280), CE (7.115), and CU (7.543)—which closely match state-of-the-art closed-source systems such as SunoV4 (CE 7.474, CU 7.813) and Tiangong (PQ 8.220). These results suggest YuE performs comparably in terms of perceived audio aesthetics and usability.

\paragraph{Alignment Metrics.} YuE attains the highest alignment score according to CLaMP 3 (0.240)~\citep{wu2025clamp3universalmusic}, closely aligning with the human-evaluated ``control'' indicators from the previous section. However, we observe a notably lower alignment for YuE according to the CLAP score (0.118), which not only diverges from human evaluation trends but also directly contradicts the findings from CLaMP 3. These discrepancies highlight potential limitations of the CLAP score in accurately capturing human perceptions of controllability, possibly due to differences in pretraining data and modeling strategies. In contrast, CLaMP 3 appears to benefit from recent methodological improvements and broader, web-scale training resources, resulting in more reliable evaluation outcomes.\footnote{We use CLaMP3 as the CLaMP 3 score backend, which is a more recent model compared to CLAP, showing improved results in representation quality and music retrieval tasks due to extensive web-scale pretraining. In contrast, CLAP may suffer from limited exposure to singing/musical content during its training, which could lead to discrepancies in evaluating certain music types.}


\subsubsection{Correlation Between Automatic Metrics and Human Evaluation}
\label{sec:auto_human_corr}

\begin{table*}[htbp]
\centering
\vspace{2em}
\caption{Pearson correlation between subjective metrics (\textit{Musicality}, \textit{Average}) and automatic metrics. \textbf{Vocal Range} strongly impacts Musicality and Average ratings.}
\label{tab:pearson_corr_vocal}
\resizebox{\textwidth}{!}{
\begin{tabular}{lccccccccc}
\toprule
 & KL & FAD & CE & CU & PC & PQ & CLAP & CLaMP 3 & VocalRange \\
\midrule
Musicality & -0.232 & -0.249 & 0.368 & 0.320 & -0.268 & 0.112 & -0.072 & 0.333 & \textbf{0.857} \\
Average    & -0.199 & -0.351 & 0.357 & 0.303 & -0.128 & 0.054 &  0.086 & 0.264 & \textbf{0.858} \\
\bottomrule
\end{tabular}}
\end{table*}

\paragraph{Correlation with Musicality \& Average Preference} When considering musicality and average human preference (Table~\ref{tab:pearson_corr_vocal}), the \textbf{Vocal Range} metric stands out, correlating most strongly (above 0.85) with both subjective ratings. This highlights the crucial role of vocal expressiveness and melodic diversity\footnote{One possible explanation relates to AR music generation behavior. Such models often favor high-probability tokens, biasing melodies toward conservative choices like tonic, chord tones, or previously generated notes. Poor optimization (e.g., overfitting) or overly conservative sampling exacerbates this issue, reducing melodic diversity.} in listeners’ overall impressions of generated music. We find vocal range to be a practical proxy for musicality and recommend its adoption. 

\begin{table}[htbp]
\centering
\vspace{1em}
\caption{Pearson correlation of alignment metrics vs. human preference on controllability.}
\begin{tabular}{lccccc}
\toprule
 & LyricFollow & GenCtrl & InstrCtrl & EmoCtrl & Tempo/RhyCtrl \\
\midrule
CLAP\(\uparrow\) & -0.25 & 0.01 & -0.07 & 0.14 & 0.09 \\
CLaMP 3\(\uparrow\) & \textbf{0.42} & \textbf{0.37} & \textbf{0.44} & \textbf{0.33} & \textbf{0.36} \\
\bottomrule
\end{tabular}
\label{tab:clap-clamp3-correlation}
\end{table}

\paragraph{Alignment Metrics.} The correlation results in Table~\ref{tab:clap-clamp3-correlation} demonstrate that CLaMP 3 scores consistently correlate better with human evaluations of controllability compared to CLAP scores. This is particularly evident in tasks such as LyricFollow (0.42 vs. -0.25) and InstrCtrl (0.44 vs. -0.07). Interestingly, the genre-following capability measured by the CLaMP 3 backend~\citep{wu2025clamp3universalmusic} appears to be closely related to lyric-following performance, even though lyrics are not explicitly included in the computation of the CLaMP 3 score. This indicates a correlation between genre controllability and lyric adherence in music generation models. Conversely, the weaker correlations observed with CLAP suggest limitations in its capacity to capture nuanced perceptual aspects, likely due to insufficient exposure to singing and music-specific content during pre-training.

\begin{table}[htbp]
\centering
\caption{Pearson correlation of KL and FAD on acoustic quality preference metrics.}
\label{tab:kl_fad_accomp_vocal}
\begin{tabular}{lcc}
\toprule
 & AccompQual & VocalQual \\
\midrule
KL & 0.14 & 0.23 \\
FAD & \textbf{-0.15} & \textbf{-0.11} \\
\bottomrule
\end{tabular}
\end{table}

\paragraph{Distribution Matching Metrics.} We employed the more advanced PaSST~\citep{passt} backbone instead of the conventional VGGish~\citep{vggish} to evaluate distribution matching metrics. Despite its sophistication, the AudioSet pre-trained backbone may inherently suffer from out-of-distribution (OOD) issues when dealing with generative music, particularly with singing or vocal elements. Additionally, sample size bias may contribute significantly, as limited availability of extensive audio samples from closed-source generative systems hinders accurate distribution estimations.

As shown in Table~\ref{tab:kl_fad_accomp_vocal}, both \textbf{KL} and \textbf{FAD} exhibit weak correlations with \emph{accompaniment (acoustic) quality} (AccompQual) and \emph{vocal (acoustic) quality} (VocalQual), suggesting that distribution-level metrics may not fully capture subtle subjective perceptions of acoustic fidelity in our case. However, as indicated in Table~\ref{tab:pearson_corr_vocal}, these same metrics correlate more strongly with \emph{musicality} and overall human preference\footnote{Both KL and FAD are negatively correlated, since lower values indicate better alignment.}. This implies that while distribution matching may not always reflect finer acoustic details, they sometimes reflect qualities relevant to perceived musicality and listener satisfaction.

\begin{table}[htbp]
\centering
\caption{Pearson correlation of content-based metrics vs. related preference metrics.}
\label{tab:ce_cu_pc_pq_all}
\begin{tabular}{lcccccc}
\toprule
 & AccompQual & VocalQual & SongStruct & VAComp & MelAttrac & MusicArr \\
\midrule
CE & \textbf{0.56} & \textbf{0.66} & \textbf{0.33} & \textbf{0.35} & \textbf{0.30} & \textbf{0.31} \\
CU & 0.50 & 0.61 & 0.27 & 0.29 & 0.25 & 0.26 \\
PC & -0.09 & 0.00 & -0.24 & -0.20 & 0.00 & -0.16 \\
PQ & 0.27 & 0.36 & 0.05 & 0.06 & -0.03 & 0.02 \\
\bottomrule
\end{tabular}
\end{table}

\paragraph{Content-Based Metrics.} In Table~\ref{tab:ce_cu_pc_pq_all}, \textbf{CE} exhibits the strongest correlations, particularly with subjective acoustic quality measures such as \textit{VocalQual} (0.66) and \textit{AccompQual} (0.56). This indicates that CE might be especially sensitive to acoustic characteristics perceived by listeners. By contrast, correlations with musicality-related aspects—such as \textit{SongStruct} (0.33), \textit{VAComp} (0.35), \textit{MelAttrac} (0.30), and \textit{MusicArr} (0.31)—are relatively lower, suggesting a lesser sensitivity of CE to detailed musical attributes. Meanwhile, both \textbf{PC} and \textbf{PQ} show notably weaker or inconsistent correlations across these subjective metrics, implying limitations in their ability to capture musicality related perceptual elements.

\section{Fine-tuning To More Languages}
\label{sec:multilingual}
Our results (detailed in Appendix~\ref{appendix:multilingual_subjective}) demonstrate YuE's strong adaptability and effectiveness through fine-tuning to multiple languages (Chinese, Korean, Japanese) within a 40B-token budget\footnote{Fine-tuning was conducted by re-annealing from the last constant learning rate checkpoint using an enhanced mixture of target language data.}. As shown in Table~\ref{tab:multilingual_results}, YuE notably achieves the highest lyrics-following performance in Japanese (70\%). In Chinese lyrics-following, YuE secures second-best performance (60\%) behind Suno (73\%), while in Korean lyrics-following, it ranks third (55\%). These results highlight YuE's robust adaptability and suggest potential for further improvement with targeted fine-tuning.

YuE also demonstrates competitive musicality, placing second in Chinese (62\%) and Korean (55\%), which indicates effective cross-lingual transfer of musical features. However, its gap relative to Suno in Chinese musicality highlights the need for more culturally-specific training. Overall, these findings underscore YuE's promising multilingual capability and the importance of addressing linguistic and cultural nuances in fine-tuning approaches.

\begin{table}[htbp]
    \centering
    \caption{Human preference rate for lyrics following and musicality across languages. \textbf{Bold} indicates the best-performing system, and \secondbest{boxed} indicates the second-best.}
    \renewcommand{\arraystretch}{1.0} 
    \setlength{\tabcolsep}{4pt} 
    \begin{tabular}{lcccccc}
        \toprule
        \textbf{Model} & \multicolumn{2}{c}{\textbf{Chinese}} & \multicolumn{2}{c}{\textbf{Korean}} & \multicolumn{2}{c}{\textbf{Japanese}} \\
        \cmidrule(lr){2-3} \cmidrule(lr){4-5} \cmidrule(lr){6-7}
        & Lyrics & Music & Lyrics & Music & Lyrics & Music \\
        \midrule
        YuE & \secondbest{60} & \secondbest{62} & 55 & \secondbest{55} & \best{70} & 52 \\
        Udio & 36 & 46 & \secondbest{62} & \best{62} & 31 & 51 \\
        Suno V4 & \best{73} & \best{88} & \best{75} & 50 & \secondbest{60} & \best{80} \\
        Hailuo & 30 & 15 & 37 & 60 & 56 & 31 \\
        Tiangong & 51 & 39 & 20 & 22 & 32 & 35 \\
        \bottomrule
    \end{tabular}
    \label{tab:multilingual_results}
\end{table}

\section{Analysis and Ablations}
\label{sec:ablation}
\subsection{Comparison of Audio Tokenizers}
\label{sec: tokenizer comparison}

\begin{table}[htbp]
\centering

\caption{Qualitative comparison of different codec types based on reconstruction quality, LM convergence, and invalid probability. Invalid probability refers to the likelihood of generating noise or silence segments during LM token synthesis.}
\resizebox{\textwidth}{!}{
\begin{tabular}{lcccc}
\toprule
\textbf{Type} & \textbf{Codec} & \textbf{Reconstruction} & \textbf{LM Converge} & \textbf{Invalid Prob.} \\
\midrule
Acoustic & Encodec32k & Good & No & All \\
Acoustic & HiFiCodec & Good & No & All \\
Semantic + Acoustic & Semanticodec & Fair & Yes & High \\
Semantic + Acoustic & X-Codec & Fair & Yes & Low \\
\bottomrule
\end{tabular}}
\label{tab:codec_comparison}
\end{table}

In preliminary experiments on a 130k-hour subset of diverse music data, we conducted a qualitative analysis of four popular audio tokenizers, specifically focusing on acoustic tokens and fused semantic-acoustic tokens (see Table \ref{tab:codec_comparison}). Separate semantic and acoustic tokenizers would require retraining and thus were beyond the scope of this study, reserved for future work.

Acoustic tokenizers, including Encodec32k and HiFiCodec, exhibited decent reconstruction quality. However, their learned tokens proved challenging for LMs to converge due to the complexity and variability inherent in our in-the-wild dataset. Training a 0.5B LM with acoustic tokens consistently failed to converge, resulting primarily in invalid outputs characterized by noise or silence. Although prior studies indicated Encodec32k has been successfully applied to TTM~\citep{musicgen}, even scaling the LM to 7B and extending training up to 1 trillion tokens on our data yielded only intermittent success, with outputs still dominated by noise.

In contrast, tokenizers integrating semantic and acoustic features (Semanticodec, X-Codec) demonstrated significantly better convergence, largely due to the stable clustering provided by SSL encoders. This stability facilitated successful LM training at the 0.5B scale. However, the stable clustering slightly compromised acoustic dynamics, causing only fair reconstruction quality. We further identified a critical alignment flaw in Semanticodec related to AudioMAE's patch-based mechanism, where misalignment of one token propagated errors throughout reconstruction. X-Codec, using Hubert-derived semantics, avoided this issue and maintained lower invalid generation probability.

\subsection{Impact of Source Separation Prior and Dual-NTP}

We define a metric called the \textbf{Vocal-to-Accompaniment Ratio (VAR)}, to quantify the effect of track-wise energy distribution on linguistic information loss. Let \(v(n)\) denote the vocal signal and \(a(n)\) denote the accompaniment signal, over \(n = 1, 2, \ldots, N\). We compute \(\text{VAR}\) (in dB) as follows:
\begin{MiddleEquation}
\begin{equation}
\text{VAR} = 10 \log_{10} \left(
\frac{\sum_{n=1}^{N} \bigl(v(n)\bigr)^2}
     {\sum_{n=1}^{N} \bigl(a(n)\bigr)^2}
\right).
\end{equation}
\end{MiddleEquation}where higher VAR values indicate greater prominence of vocals relative to accompaniment, while lower VAR suggests accompaniment dominance.

\begin{figure}[htbp]
    \centering
    \begin{minipage}{0.5\linewidth}
        \centering
        \includegraphics[width=\linewidth]{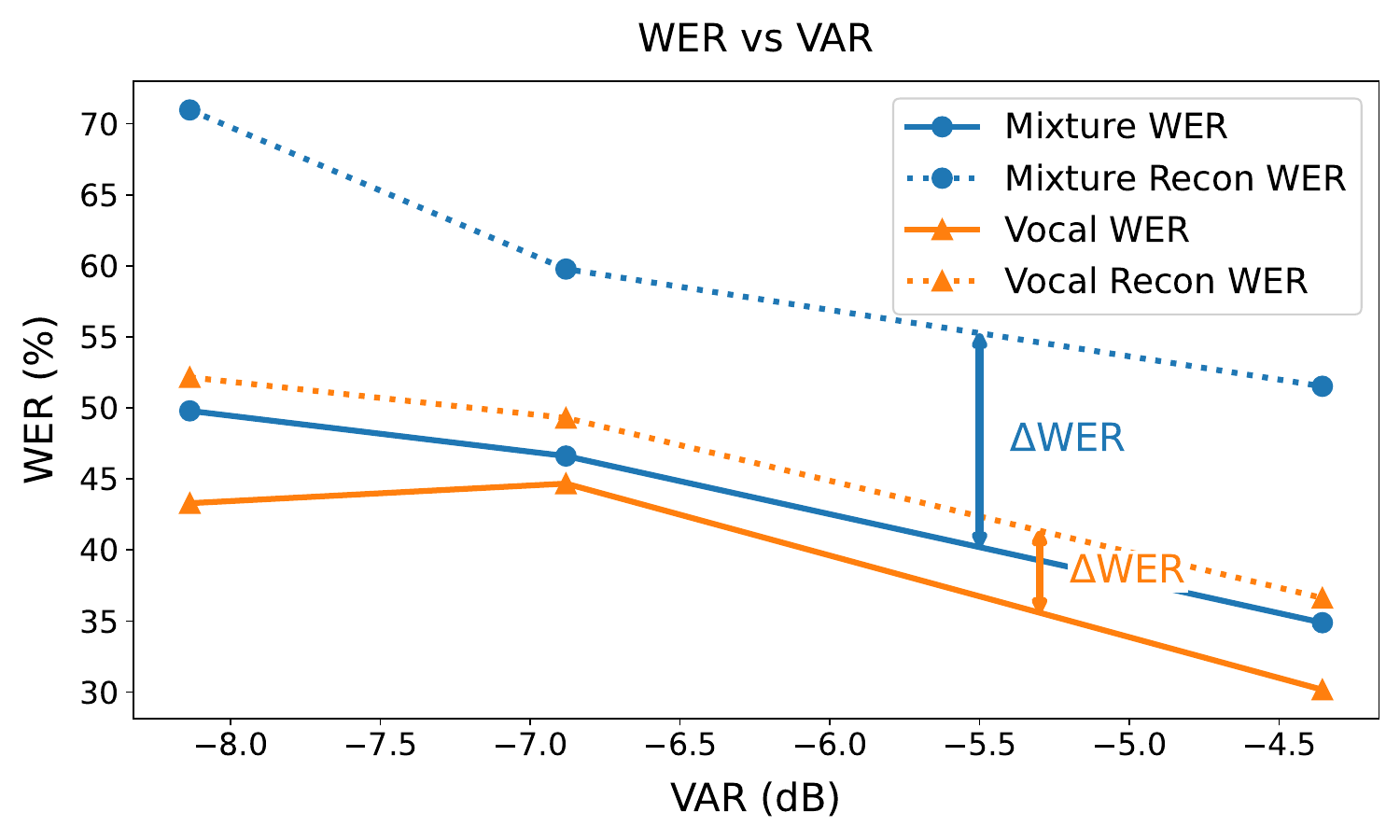}
        \caption{Comparison of WER-VAR plot for mixture and vocal tracks, including their tokenizer reconstructions, over 1K samples.}
        \label{fig:wer to var plot}
    \end{minipage}%
    \hfill
    \begin{minipage}{0.5\linewidth}
        \centering
        \includegraphics[width=\linewidth]{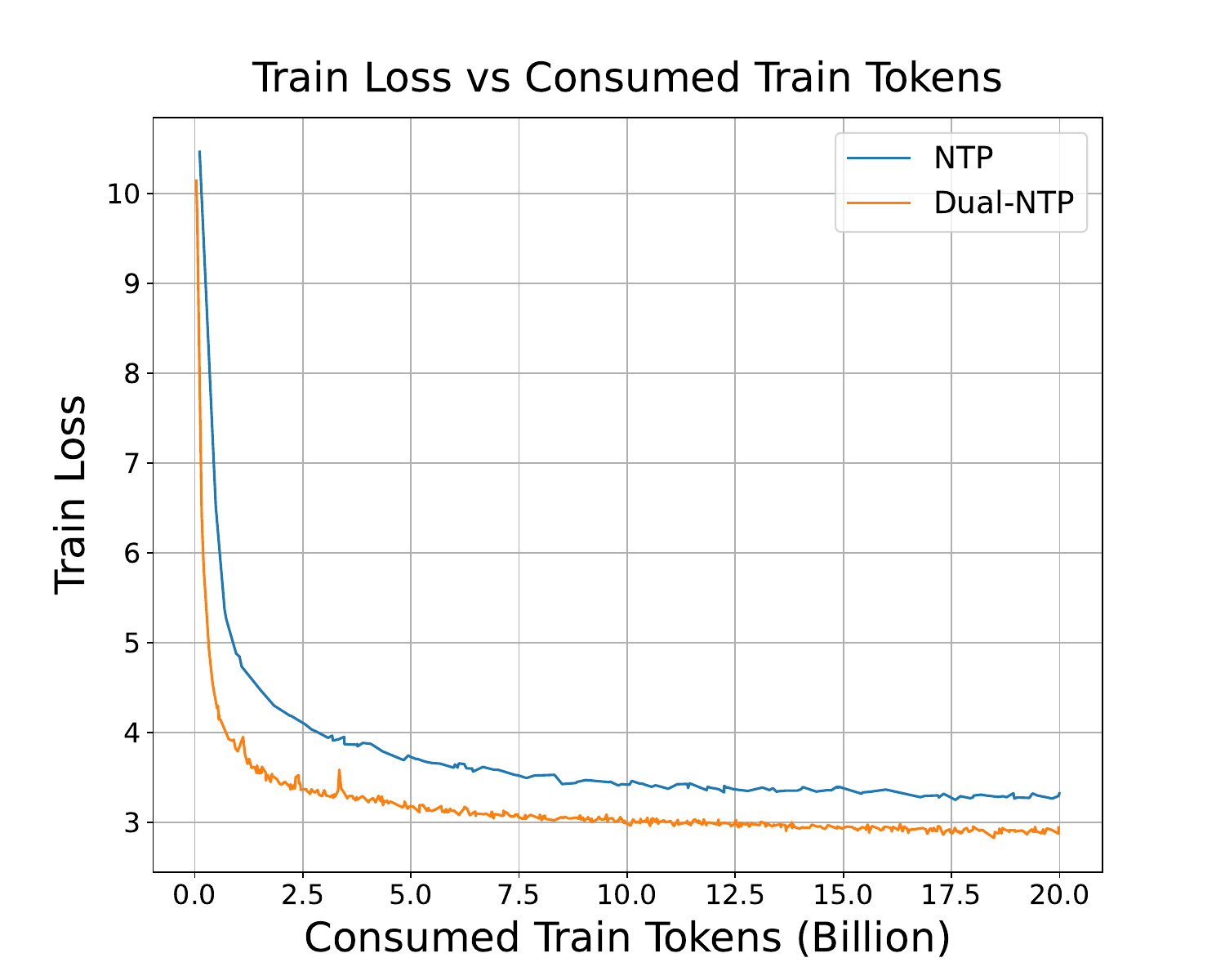}
        \caption{Training Loss over Consumed Train Tokens for NTP and Dual-NTP.}
        \label{fig:loss curve}
    \end{minipage}
\end{figure}

Similar to Figure~\ref{fig:delta_wer}, Figure~\ref{fig:wer to var plot} illustrates the WER-VAR relationship for mixture and vocal tracks across 1K samples, including tokenizer reconstructions. Although original vocal and mixture tracks exhibit similar absolute WER (solid blue and orange lines), mixture track reconstruction significantly increases WER (solid vs. dotted blue lines), especially as VAR declines, widening the gap ($\Delta\text{WER}$). A 20\%+ $\Delta\text{WER}$ is observed around -8.0 dB VAR.
In contrast, vocal tracks maintain low WER and smaller $\Delta\text{WER}$ (the worst case is 10\%- around -8.0dB VAR), indicating resilience of source separation priors to VAR degradation and reconstruction information loss.

Additionally, we perform an ablation study comparing Dual-NTP and standard NTP. Figure~\ref{fig:loss curve} presents training loss curves of two 0.5B LMs trained with identical data and computational budgets (20B tokens). Dual-NTP demonstrates a substantial reduction in loss (approximately 0.4 lower) compared to standard NTP, confirming its efficiency and robustness. Together, these analyses underscore the effectiveness of incorporating source separation priors with Dual-NTP into song modeling task.

\subsection{Ablation Analysis of Lyrics-following Capabilities with CoT}
\label{sec:cot_ablation}
\begin{figure}[htbp]
    \centering
    \includegraphics[width=0.5\textwidth]{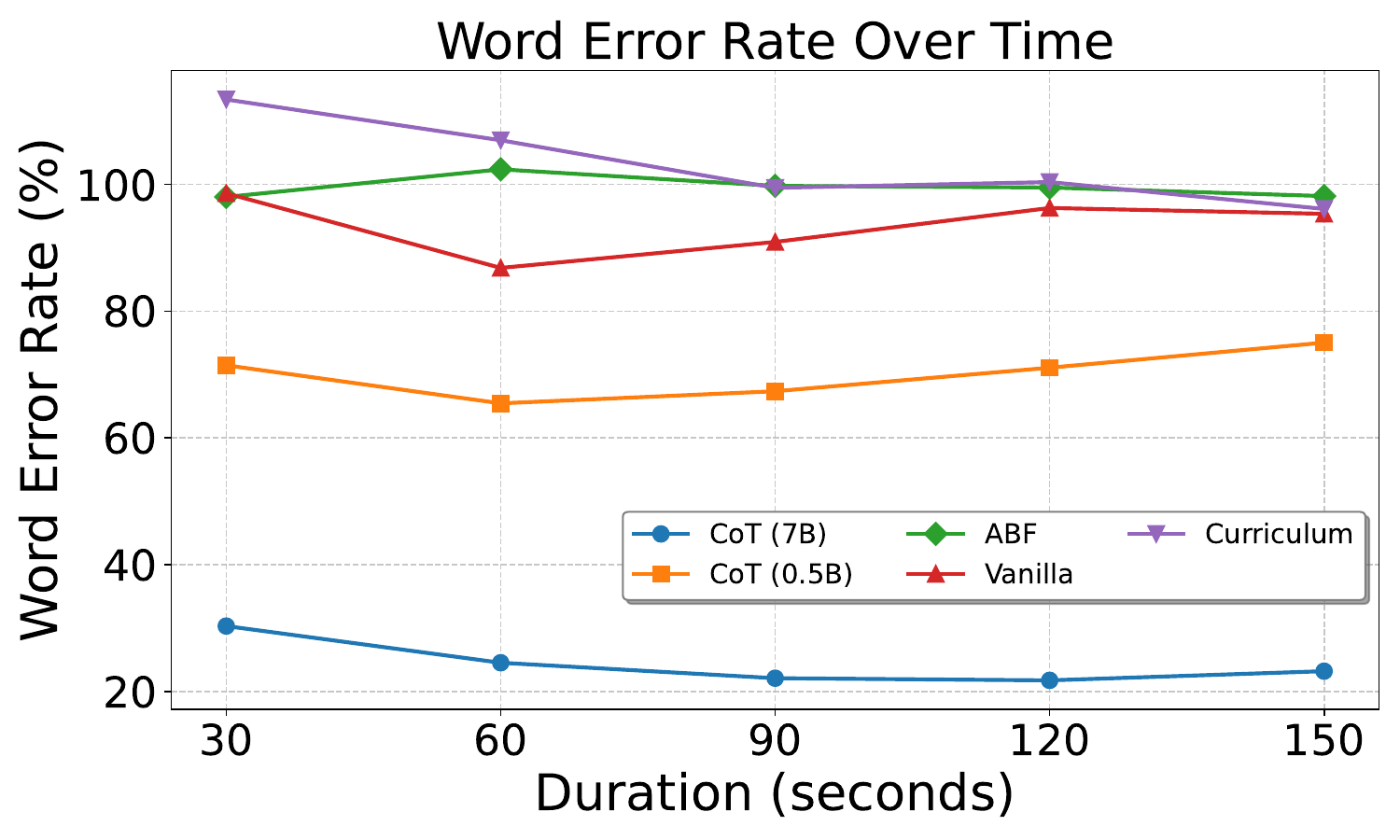}
    \caption{\footnotesize WER over time. Both CoT and model scaling significantly enhance lyrics-following capability.}
    \label{fig:lyricsfollowing}
\end{figure}

The analysis in Figure~\ref{fig:lyricsfollowing} examines an ablation setting involving a 0.5B LM, which was initially pretrained on a default mixture dataset\footnote{A mixture of speech and music. Text transcripts are in prepend format.} comprising 500B tokens and subsequently finetuned on the corresponding lyrics data for an additional 200B tokens using the specified methods: \textbf{Vanilla}, \textbf{Curriculum}, and \textbf{ABF}, and our proposed \textbf{CoT}. Additionally, we include results from the YuE-7B checkpoint to illustrate the performance gains achievable through scaling.

\textbf{Vanilla} refers to text prepend conditioning, where the model is trained with prepended lyrics as input for conditioning. \textbf{Curriculum} involves gradually increasing the text prepend data with progressively longer durations (e.g., 30s, 60s, 90s, etc.), aiming to improve the model's ability to follow lyrics over time. \textbf{ABF}~\citep{xiong2023effective} refers to adjusting the rope base frequency from 10k to 100k during finetuning to explore its effect on lyrics-following performance.

The WER over time is estimated using a fine-tuned Whisper model, with measurements recorded every 30 seconds up to 150 seconds. Overall, the proposed CoT method achieves consistently superior performance across all evaluated time intervals (30s to 150s). Scaling the model to 7B parameters demonstrates substantial improvements, reducing the WER from approximately 70\% at 0.5B parameters to around 20\%\footnote{Note that 20\% can be considered a relatively low number. Refer to the GT WER-to-VAR plot in Figure~\ref{fig:wer to var plot}.}.

In contrast, Vanilla, Curriculum, and ABF methods exhibit substantially worse WER, indicating a limited capability in maintaining lyrical coherence. Through manual inspection, we identified that the primary reason for failure in Vanilla and Curriculum was their tendency to generate instrumental preludes, causing the onset of singing to drift far from the original prepended lyrics condition, thus complicating accurate alignment.


\subsection{Effect of Scaling}
\begin{wrapfigure}{r}{0.45\textwidth}
    \centering
    \captionsetup{skip=0pt}
    \includegraphics[width=0.44\textwidth]{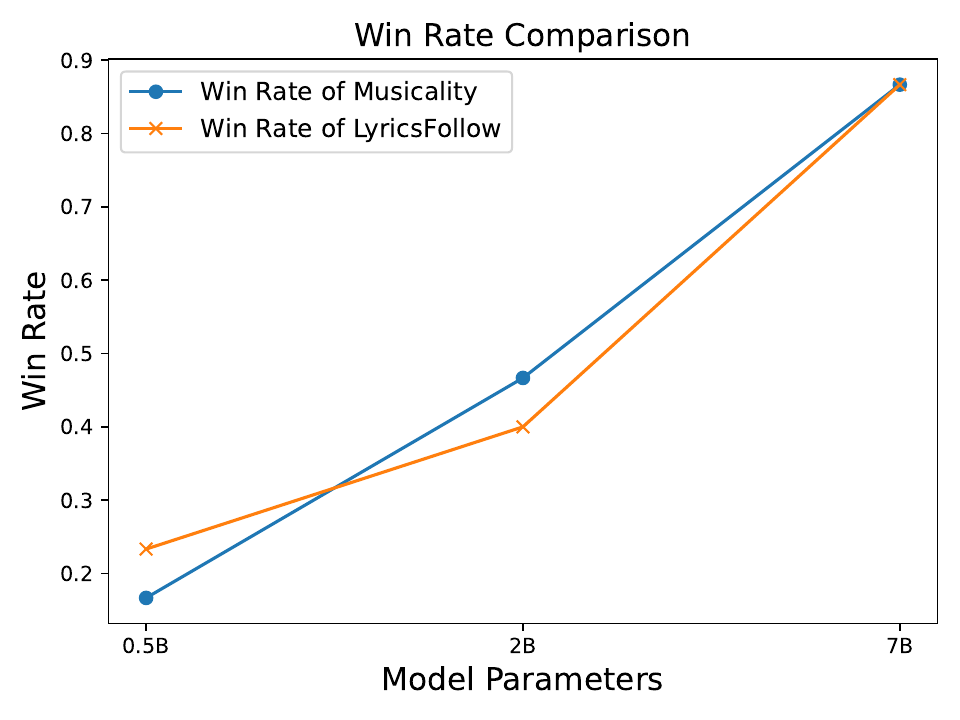}
    \caption{Human preference overall win rates for Musicality and Lyrics-following across model scales (0.5B, 2B, and 7B) in pairwise A/B tests. Larger models consistently achieve higher preferences.}
    \label{fig:scaling_win_rate}
    \vspace{-10pt}
\end{wrapfigure}

We investigate the impact of model scaling on musicality and lyrics-following capabilities. We compared checkpoints at 0.5B, 2B, and 7B scales. While the 0.5B and 2B models were trained with a limited budget of 500B tokens (in 16K context), the 7B model underwent complete scaling with a significantly larger 1.75T token budget using the full training dataset.

As illustrated in Figure \ref{fig:scaling_win_rate}, human evaluation demonstrates a clear improvement trend in both musicality and lyrics-following as model scale and training budget increase. Notably, the 7B model exhibits substantial enhancements, indicating that increased parameter counts and extensive training significantly boost the model's foundational creativity and compositional quality. These results confirm that scaling plays a crucial role in achieving higher musicality and improved lyric adherence.

\subsection{Analysis of Test-time Tricks}
\begin{figure}[htbp]
    \centering
    \includegraphics[width=0.6\textwidth]{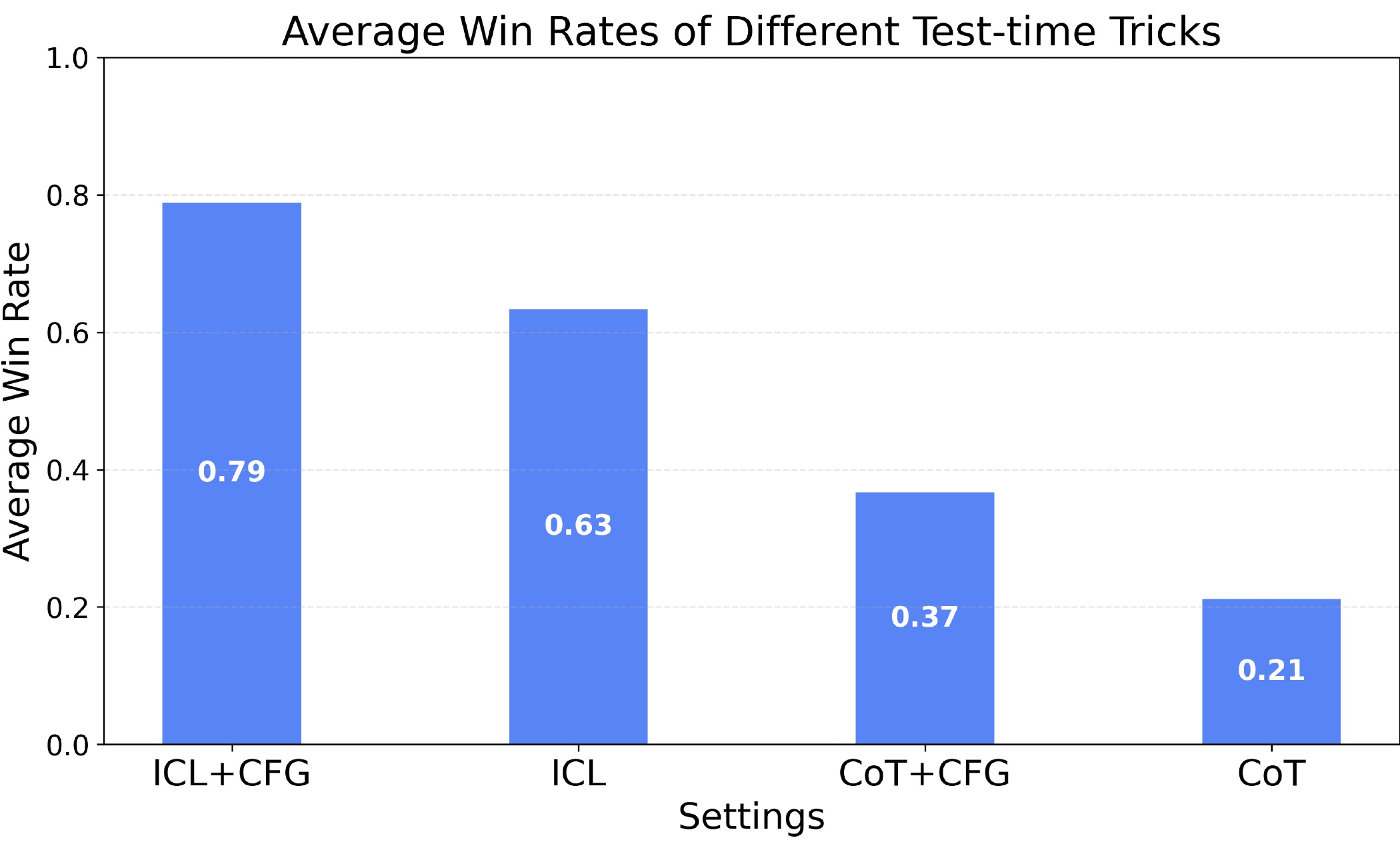}
    \caption{Human preference win rates for Musicality across different test-time tricks.}
    \label{fig:abtest_testtime_tricks}
    \vspace{-10pt}
\end{figure}

Figure~\ref{fig:abtest_testtime_tricks} presents human preference win rates for musicality obtained through A/B testing across different inference settings using YuE-7B checkpoints. Results clearly demonstrate that ICL-based methods outperform CoT-based methods significantly: ICL achieves a win rate of 0.63 compared to only 0.21 for CoT. Incorporating CFG further enhances these methods; specifically, ICL+CFG obtains the highest win rate (0.79), substantially exceeding both ICL alone and the CoT-based configurations.  

This performance advantage stems from the strong conditioning ability of ICL, which restricts the decoded token space to a musically favorable subspace guided by the provided human-generated music prompt. CFG similarly strengthens this conditioning by amplifying the influence of the text condition on next-token logits, making generated outputs more closely aligned with the intended prompt-guided subspace and thus further improving musicality.

\section{Representation Quality}
\label{sec:marble}

\begin{table*}[h]
    \centering
    \caption{Evaluation of YuE single-track unconditional mode on MARBLE. Including GTZAN genre classification, GS key recognition, MTG top 50 tagging, and EMO emotion regression.}
    \begin{tabular}{l ll ll ll}
    \toprule[1.5pt]
    \textbf{Dataset} & \textbf{GTZAN} & \textbf{GS} & \multicolumn{2}{c}{\textbf{MTG}} & \multicolumn{2}{c}{\textbf{EMO}} \\  
    \textbf{Task}    & Genre         & Key         & \multicolumn{2}{c}{Top50}      & \multicolumn{2}{c}{Emotion} \\  
    \midrule
    \textbf{Metrics} & Acc$\uparrow$ & Acc\textsuperscript{Refined}$\uparrow$ & AP$\uparrow$ & AUC$\uparrow$ & R2\textsuperscript{V}$\uparrow$ & R2\textsuperscript{A}$\uparrow$ \\ 
    \midrule
    MERT [\citeyear{li2023mert}]             & 78.6         & 65.6       & 29.9  & \textbf{83.4}  & 61.2  & 74.7 \\
    MusicFM [\citeyear{MusicFM}]          & 83.8         & 63.9       & -     & -     & 60.3  & \textbf{76.3} \\
    MuQ\textsubscript{iter} [\citeyear{zhu2025muq}] & 85.6 & 65.0  & -     & -     & \textbf{62.8}  & 76.1 \\
    CLAP [\citeyear{wu2023large-clap}]              & 82.1         & 16.0 & 27.7  & 82.0  & 54.1  & 70.3 \\
    CLaMP 3 [\citeyear{wu2025clamp3universalmusic}]              & \textbf{86.6}         & 53.8 & \textbf{30.2}  & 82.4  & 59.1  & 70.0 \\
    YuE              & 83.4         & \textbf{67.0} & 29.2  & 82.7  & 58.9  & 75.0 \\
    \bottomrule[1.5pt]
    \end{tabular}
    \label{tab:marble_res}
\end{table*}

YuE, fundamentally designed as a generative model rather than explicitly for representation learning, is evaluated with MARBLE~\citep{yuan2024marble} here using its Stage-1 LM in an unconditional single-track setting. Notably, this mode serves primarily as an auxiliary task and is disabled half way through the training. Moreover, it exclusively leverages discrete codes from codebook-0, implying a significant reduction in available information compared to dedicated representation learning models. 

Despite these inherent limitations, YuE achieves state-of-the-art performance on the GS key recognition task (Acc=67.0\%, see Table~\ref{tab:marble_res}), demonstrating a good sense of tonality and modality, which is essential for composing and singing in tune. Furthermore, its performance remains competitive with existing methods across other tasks, such as GTZAN genre classification, MTG tagging, and EMO emotion regression, underscoring YuE's robust general-purpose representation quality and learned musical skills.

\section{Emergent Abilities}
\label{sec:novel_abilities}
We strongly encourage readers to visit our demo page for audio examples illustrating the capabilities described.\footnote{\url{https://map-yue.github.io/}} Scaling up the model significantly enhances generation quality and unlocks novel abilities.

\paragraph{Advanced Vocal Techniques.} Beyond basic pop and rap vocals, our model spontaneously acquires diverse and expressive singing techniques, typically mastered only by gifted human vocalists through extensive training. These include vibrato, glissando, bel canto, death growl, mix voice, belting, riffs and runs, vocal fry, Beijing Opera, and Shanbei folk vocals. This indicates our Dual-NTP approach effectively captures subtle nuances in vocal performance.

\paragraph{Spontaneous Performance.}  
Our model spontaneously demonstrates musically expressive behaviors. For instance, in jazz performances, it naturally continues with scat singing after running out of lyrics; in a cappella, it simultaneously generates multi-part harmonies with distinct vocalists handling melody and accompaniment; in folk music, it inserts contextually appropriate instrumental solos, such as harmonica interludes, during vocal pauses.

\paragraph{World Music \& Pattern Mixing.}  
Our model effectively captures long-tail global music styles beyond mainstream western genres. For instance, it generates creative fusions such as Chinese gangsta rap accompanied by Japanese shamisen instrumentation and scales. It can also seamlessly blend distinct regional vocal styles, combining Chinese opera, Shanbei folk singing, and traditional Chinese vocals within a single cohesive performance.

\paragraph{Voice Cloning.}
Our model demonstrates high-fidelity voice cloning capabilities at inference time, successfully replicating distinct vocal identities. For example, we accurately reproduce the unique voices of Billie Eilish and Faye Wong (王菲) while generating entirely new lyrics and melodies. These cloned voices retain their signature timbral qualities, breathy textures, and emotional nuances, highlighting the model's ability to capture and reproduce subtle vocal characteristics from limited reference data provided only at inference.

\paragraph{Style Transfer.}
Our model shows versatile style transfer capabilities, enabling the generation of diverse and expressive vocal performances across different languages, genres, and timbres.
YuE enables cross-lingual and genre adaptation while preserving the original lyrical and melodic structure. In one example, a Japanese female J-pop vocal performance is transformed into an English male rap with the same city pop accompaniment. The model not only shifts the vocal characteristics but also adjusts prosody, phrasing, and expressiveness to ensure stylistic coherence, demonstrating its deep understanding of genre-specific vocal performance.

\paragraph{Code Switching.} The model naturally handles code-switching, smoothly transitioning between multiple languages or dialects within the same vocal performance, while preserving linguistic and stylistic consistency.

\section{Memorization Effect}
\label{sec:memorization}

Following previous literature~\citep{agostinelli2023musiclm, yuan2024chatmusician}, we investigate whether \textbf{YuE}, in its ICL mode—conditioned on a 30-second audio prompt and original lyrics—reproduces significant portions of its training data. ICL is generally more prone to memorization, making this evaluation critical. 

We employ ByteCover2~\citep{du2022bytecover2}, a state-of-the-art retrieval model optimized for melody-sensitive similarity across entire songs.\footnote{We do not use ByteCover3~\citep{du2023bytecover3} as it specializes in shorter segments.} Specifically, we create two sets of $N=1200$ music samples: $\mathcal{R}$ (\textbf{Ref}), comprising YuE’s training examples, and $\mathcal{G}$ (\textbf{Gen}), comprising corresponding samples generated by YuE in the ICL setting. We compute cosine similarity scores for each pair $(r,g)$ with $r \in \mathcal{R}$ and $g \in \mathcal{G}$, analyzing the top 1\% of scores since frequent high-similarity pairs would suggest substantial memorization.

\begin{figure}[htbp]
    \centering
    \includegraphics[width=0.7\linewidth]{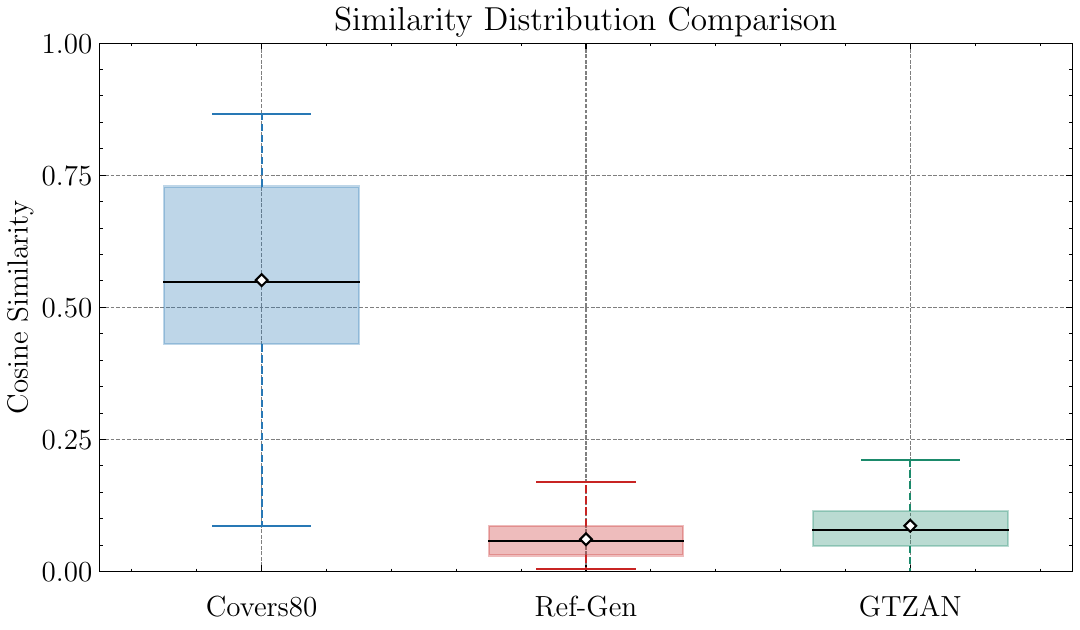}
    \caption{Box-plot comparison of cosine similarity across three scenarios: Covers80, \textit{Ref-Gen} (our training vs.\ generated sets), and GTZAN. The black bar denotes the median, and the diamond denotes the mean.}
    \label{fig:similarity_comparison}
\end{figure}

To contextualize these results, we compare them to real-world baselines from GTZAN (genre-level similarities) and Covers80 (known melodic duplicates). Results are shown in Figure~\ref{fig:similarity_comparison}. The similarity distribution for \textit{Ref-Gen} pairs is significantly lower than Covers80 and remains moderate even compared to GTZAN. While short repetitive motifs, particularly percussive loops, occasionally occur, overall results indicate that YuE’s ICL mode does not engage in extensive copying. Instead, YuE recombines learned musical patterns creatively, demonstrating that the ICL mode effectively generates original content rather than memorizing training samples.

\section{Unsuccessful Attempts}
\label{lab:unsuccessful_attempts}
During our initial scaling experiments, we encountered several challenges and setbacks. Here, we share these unsuccessful experiences to inform and inspire future research directions.

\paragraph{Acoustic Tokens.} As detailed in Section~\ref{sec: tokenizer comparison}, LMs trained on acoustic tokens consistently exhibited convergence difficulties and yielded higher losses compared to semantic-enhanced tokens. We attribute these challenges primarily to inherent limitations of current acoustic token representations, typically derived from RVQ-GANs. Such tokens often prioritize compression efficiency over representational quality and typically have limited capacity. Consequently, models trained on these tokens may tend to adopt shortcuts, frequently resorting to direct information copying. Even when scaled substantially, these models achieve only marginal improvements~\citep{hansen2025learnings, xin2024bigcodecpushinglimitslowbitrate, parker2024scaling}. We argue the lossy nature of discrete representations, limited semantic relevance~\citep{zhang2023speechtokenizer}, and excessive focus on reconstruction tasks collectively contribute to the difficulties observed in fitting acoustic tokens.

\paragraph{Unconditional Pre-train.}  
We initially pre-trained large models to learn general representations for cross-modal alignment (text-to-vocal) via fine-tuning. At smaller scales (e.g., sub-billion parameters), models showed moderate success in learning basic mappings. However, at 7B parameters, unconditional pre-training became counterproductive: fine-tuning failed to establish effective cross-modal alignment. We hypothesize that larger models internalize overly generic priors, overshadowing the specific conditional mappings needed for alignment. This ``catastrophic inertia'' prevents large models from adapting effectively to lyrics-to-song tasks.

\paragraph{Early Activation of ICL.}  
We observed that early activation of ICL data led to a poor musicality. Initially, the model began to excessively rely on the reference audio, resulting in overfitting and diminished musicality. After removing the reference audio later in the training process, the model continued to produce a significant number of invalid outputs, such as silence or noise. This problem became more pronounced with scaling, where larger models struggled even more to recover from this shortcut learning. These results highlight the importance of carefully managing the timing of ICL data activation to avoid overfitting and preserve the model’s creativity.

\section{Conclusion and Future Work}
We introduced YuE, an open-source foundation model family designed for long-form lyrics-to-song generation. By combining large-scale data, track-decoupled next-token prediction, a segment-wise conditioning strategy, and a redesigned in-context learning framework, YuE can generate coherent, full-length songs with expressive vocals and detailed musical structure. Experimental results show that YuE matches or exceeds several commercial systems in musicality, controllability, and cross-lingual lyrics following, and it also achieves competitive music understanding results on standard benchmarks. These findings highlight the promise of open, large-scale music models in enabling controllable, high-quality song generation and in advancing broader research into music-aware AI systems.

YuE’s approach can be extended by improving acoustic fidelity and mixing, incorporating musical knowledge such as chord progressions and instrumentation theory, and integrating deeper prosodic and emotional controls. Multilingual and cross-cultural expansions hold significant potential, especially for underrepresented musical traditions. Beyond music creation, YuE can benefit applications in music education, accessibility, and therapy, and can serve as an accessible platform for continued community-driven innovation in open music AI research.

\section{Ethics and Responsibility}
Ensuring ethical and responsible AI-generated music is crucial for fostering transparency, accessibility, and fair contribution to the music industry. As suggested by~\citet{ma2024foundation}, to promote accountability, we advocate for the inclusion of AI-generated / AI-assisted tags in generated content, increasing transparency for both musicians and audiences. Additionally, our memorization-effect experiments in Section~\ref{sec:memorization} demonstrate that our design maintains creativity without plagiarizing, even under strong training set conditioning.

In contrast to closed-source commercial systems, our model leverages an exceptionally diverse training dataset, explicitly enriched with culturally diverse music content. This enables the model to innovate and create within niche musical styles effectively (see Section~\ref{sec:novel_abilities}). As such, our model can serve as a \textbf{parameterized knowledge base}, contributing to the preservation and expansion of human musical artistry and cultural heritage.

This study has been reviewed and approved by the Human and Artefacts Research Ethics Committee under protocol HREP-2023-0230, titled Building Platform Technologies for Symbiotic Creativity in Hong Kong. The approval ensures that our research adheres to ethical guidelines in data usage, AI generation, and cultural representation. The approval remains effective until 30-Jan-2027.

\section{Contributions and Acknowledgments}
\begin{multicols}{2}

\textbf{Core Contributors~~~~~~~~~~~~~~~} \\[0.5ex]
\noindent
Ruibin Yuan, {\small\color{gray}Lead, Pre-train, Data, Eval}\\
\textit{\footnotesize HKUST, Moonshot.ai, MAP, ryuanab@connect.ust.hk}

Hanfeng Lin, {\small\color{gray}Pre-train, Data, Eval, Inference} \\
\textit{\footnotesize HKUST, MAP, hanfeng@ust.hk}

Shuyue Guo, {\small\color{gray}Pre-train, Demo} \\
\textit{\footnotesize MAP}

Ge Zhang, {\small\color{gray}Pre-train} \\
\textit{\footnotesize MAP, gezhang@umich.edu}

Jiahao Pan, {\small\color{gray}Pre-train, Eval, Data} \\
\textit{\footnotesize HKUST, MAP, fengshicherish@gmail.com}
\\[2ex]

\textbf{Contributors} \\[0.5ex]
Yongyi Zang, {\small\color{gray}Upsampler, Eval}\\
\textit{\footnotesize Independent} \\
Haohe Liu, {\small\color{gray}Upsampler, Tokenizer, Demo}\\
\textit{\footnotesize University Of Surrey, MAP} \\
Yiming Liang, {\small\color{gray}Eval Lead}\\
\textit{\footnotesize MAP} \\
Wenye Ma, {\small\color{gray}Representation Learning} \\
\textit{\footnotesize MBZUAI, MAP} \\
Xingjian Du, {\small\color{gray}Memorization Effect} \\
\textit{\footnotesize University of Rochester, MAP} \\
Xinrun Du, {\small\color{gray}Pre-train} \\
\textit{\footnotesize MAP} \\
Zhen Ye, {\small\color{gray}Tokenizer} \\
\textit{\footnotesize HKUST} \\
Tianyu Zheng, {\small\color{gray}Pre-train} \\
\textit{\footnotesize MAP} \\
Zhengxuan Jiang, {\small\color{gray}Inference} \\
\textit{\footnotesize MAP} \\
Yinghao Ma, {\small\color{gray}Eval} \\
\textit{\footnotesize MAP, Queen Mary University of London} \\
Minghao Liu, {\small\color{gray}Eval, Data} \\
\textit{\footnotesize 2077AI, MAP} \\
Zeyue Tian, {\small\color{gray}Eval} \\
\textit{\footnotesize HKUST, MAP} \\
Ziya Zhou, {\small\color{gray}Eval, Data} \\
\textit{\footnotesize HKUST, MAP} \\
Liumeng Xue, {\small\color{gray}Eval, Data} \\
\textit{\footnotesize HKUST, MAP} \\
Xingwei Qu, {\small\color{gray}Pre-train, Eval} \\
\textit{\footnotesize MAP} \\
Yizhi Li, {\small\color{gray}Eval} \\
\textit{\footnotesize MAP, University of Manchester} \\
Shangda Wu, {\small\color{gray}Eval} \\
\textit{\footnotesize Central Conservatory of Music, MAP} \\
Tianhao Shen, {\small\color{gray}Eval, Inference} \\
\textit{\footnotesize MAP} \\
Ziyang Ma, {\small\color{gray}Eval} \\ 
\textit{\footnotesize MAP, SJTU, NTU} \\
Jun Zhan, {\small\color{gray}Eval} \\
\textit{\footnotesize Fudan University} \\
Chunhui Wang, {\small\color{gray}Eval, Pre-train} \\
\textit{\footnotesize Geely} \\
Yatian Wang, {\small\color{gray}Eval} \\
\textit{\footnotesize HKUST} \\
Xiaowei Chi, {\small\color{gray}Eval} \\
\textit{\footnotesize HKUST} \\
Xinyue Zhang, {\small\color{gray}Eval} \\
\textit{\footnotesize HKUST} \\
Zhenzhu Yang, {\small\color{gray}Eval} \\
\textit{\footnotesize HKUST} \\
Xiangzhou Wang, {\small\color{gray}Eval} \\
\textit{\footnotesize MAP} \\
Shansong Liu, {\small\color{gray}Eval} \\
\textit{\footnotesize Meituan} \\
Lingrui Mei, {\small\color{gray}Eval} \\
\textit{\footnotesize Meituan} \\
Peng Li, {\small\color{gray}Eval} \\
\textit{\footnotesize HKUST} \\
Junjie Wang, {\small\color{gray}Eval} \\
\textit{\footnotesize Tsinghua University} \\
Jianwei Yu, {\small\color{gray}Data, Inference} \\
\textit{\footnotesize Moonshot.ai} \\
Guojian Pang, {\small\color{gray}Inference} \\
\textit{\footnotesize MAP} \\
Xu Li, {\small\color{gray}Eval} \\
\textit{\footnotesize Xiaohongshu} \\
Zihao Wang, {\small\color{gray}Data} \\
\textit{\footnotesize Zhejiang University, Carnegie Mellon University} \\[2ex]

\textbf{Academic Advisors} \\[0.5ex]
Xiaohuan Zhou\\ \textit{\footnotesize MAP} \\
Lijun Yu\\ \textit{\footnotesize Carnegie Mellon University} \\
Emmanouil Benetos\\ \textit{\footnotesize Queen Mary University of London, MAP} \\
Yong Chen\\ \textit{\footnotesize Geely}\\
Chenghua Lin\\ \textit{\footnotesize University of Manchester, MAP} \\
Xie Chen\\ \textit{\footnotesize Shanghai Jiao Tong University} \\
Gus Xia\\ \textit{\footnotesize MBZUAI, MAP} \\
Zhaoxiang Zhang\\ \textit{\footnotesize Chinese Academy of Sciences} \\
Chao Zhang\\ \textit{\footnotesize Tsinghua University} \\
Wenhu Chen\\ \textit{\footnotesize University of Waterloo, MAP} \\
Xinyu Zhou\\ \textit{\footnotesize Moonshot.ai} \\
Xipeng Qiu\\ \textit{\footnotesize Fudan University} \\
Roger Dannenberg\\ \textit{\footnotesize Carnegie Mellon University, MAP} \\[2ex]

\textbf{Correspondence (Alphabetical Order)} \\[0.5ex]
Jiaheng Liu \\
\textit{\footnotesize Nanjing University, MAP, 13121221227@163.com} \\

Jian Yang \\
\textit{\footnotesize MAP, jiaya@buaa.edu.cn} \\

Wenhao Huang \\
\textit{\footnotesize MAP, rubio8741@gmail.com} \\

Wei Xue \\
\textit{\footnotesize HKUST, weixue@ust.hk} \\ 

Xu Tan \\
\textit{\footnotesize Moonshot.ai, MAP, tanxu2012@gmail.com} \\

Yike Guo \\
\textit{\footnotesize HKUST, yikeguo@ust.hk} \\

\end{multicols}

\bibliography{iclr2025_conference.bib}


\clearpage
\appendix
\section{Subjective Evaluation}
\subsection{Evaluation Methods}
In this subjective evaluation experiment, annotators were required to perform pairwise comparative evaluations of music generation outputs from multiple models. Each test unit comprised two distinct musical pieces generated by different models. Following complete playback of both samples, annotators conducted binary comparative selections (options: Superiority of A, Superiority of B, or Equivalence between A and B) across predefined evaluation dimensions. Mandatory preference judgments were enforced for each dimensional criterion, with explicit instructions to minimize the frequency of selecting the equivalence option. The evaluation protocol incorporated a double-blind procedure with randomized presentation order of audio pairs to mitigate potential ordering effects.

\begin{figure}[htbp]
    \centering
    \includegraphics[width=0.85\linewidth]{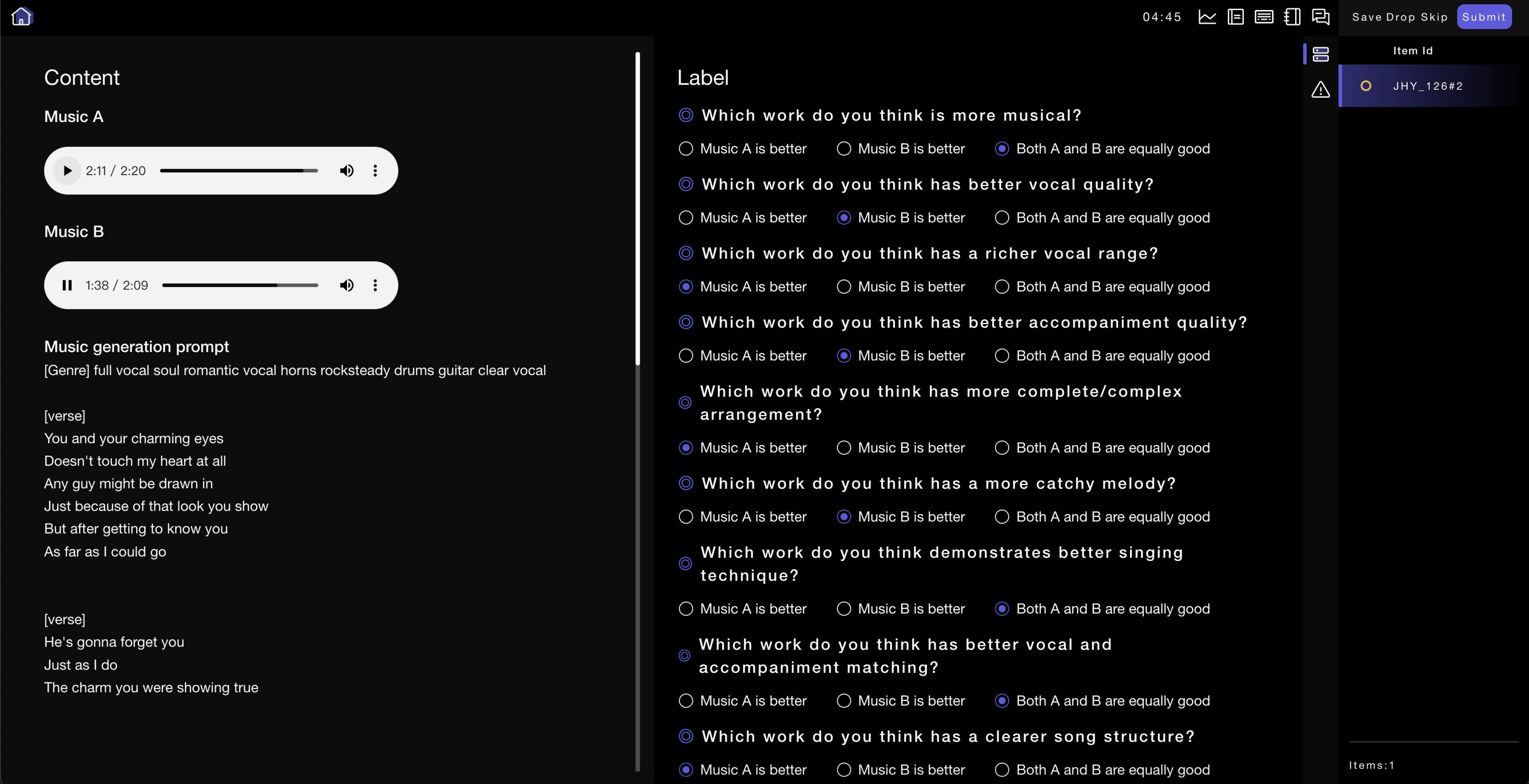}
    \caption{Subjective evaluation platform.}
    \label{fig:platform}
\end{figure}

\subsection{Evaluation Dimensions and Definitions}

\begin{enumerate}[label=\arabic*),topsep=1pt,itemsep=2pt,leftmargin=20pt]

\item \textbf{Overall Musicality} \\
\textbf{Definition:} The musical artistic value and professionalism demonstrated by the work as a whole, reflecting whether it approaches the creative level of professional musicians or composers. \\
\textbf{Evaluation Criteria:} Smoothness of the melody, complexity and rationality of the harmony, precision and rhythmic flow, and the artistic and creative qualities of the overall arrangement.

\item \textbf{Vocal Quality} \\
\textbf{Definition:} The acoustic quality of vocal performance in the work. \\
\textbf{Evaluation Criteria:} Pitch, rhythmic stability, naturalness of vocal timbre (resembling human singing), fullness and warmth of timbre, degree of mechanical or distorted sound, clarity of vocals, and richness in capturing delicate emotional expressions (e.g., variations in breath control, articulation precision, emotional conveyance).


\item \textbf{Accompaniment Quality} \\
\textbf{Definition:} The acoustic quality of the instrumental accompaniment in the work. \\
\textbf{Evaluation Criteria:} Realism and authenticity of instrumental timbres, dynamic variation and detail richness in instrumental expression (e.g., subtlety in guitar plucking or percussion dynamics).

\item \textbf{Arrangement Complexity} \\
\textbf{Definition:} The layering, coherence, balance, and creativity of the musical arrangement in the work. \\
\textbf{Evaluation Criteria:} Clarity of arrangement layers, coordination and interplay between instruments, balance of accompaniment within the overall audio track (e.g., appropriate volume and frequency distribution), fullness of low frequencies, brightness of high frequencies, diversity of arrangement elements (e.g., harmony, melodic lines, rhythm patterns across multiple dimensions), creativity, and variation and emotional progression between sections.

\item \textbf{Melodic Memorability and Catchiness} \\
\textbf{Definition:} The memorability, accessibility, and resonance-inducing capability of the melody. \\
\textbf{Evaluation Criteria:} Ease of memorization and singability, catchiness, emotional resonance, and repeated hooks or memorable elements, especially in the chorus.


\item \textbf{Vocal-Accompaniment Matching} \\
\textbf{Definition:} The consistency and compatibility between vocal melodies and instrumental accompaniment in terms of musical style, modality, harmony, and rhythm. \\
\textbf{Evaluation Criteria:} Compatibility of vocal melodies and accompaniment in modality, harmony, and rhythm, and absence of dissonance or conflict.

\item \textbf{Song Structure Clarity} \\
\textbf{Definition:} The logical coherence and sectional distinctiveness of the overall song structure. \\
\textbf{Evaluation Criteria:} Clarity of the song's structure (e.g., differentiation among verses, choruses, and interludes), naturalness of transitions between sections, and structural completeness.
  
\end{enumerate}

\subsection{Conditional Evaluation Dimension and Definitions}

\begin{enumerate}[resume,label=\arabic*),topsep=1pt,itemsep=2pt,leftmargin=20pt]

\item \textbf{Lyrics Following} \\
\textbf{Definition:} The accuracy of AI-generated vocals in performing the lyrics specified in the prompt. \\
\textbf{Evaluation Criteria:} Accuracy of lyric delivery (whether the specified lyrics are correctly performed), clarity of pronunciation (whether the lyrics are intelligible), alignment of lyrics rhythm with the musical beat, and naturalness and correctness of multilingual lyric transitions and pronunciations.

\item \textbf{Multilingual Lyrics Switching Naturalness and Correctness} \\
\textbf{Definition:} The fluency and accuracy of AI-generated vocals when performing lyrics in multiple languages, including the smoothness of transitions and the grammatical and pronunciation correctness of different languages. \\
\textbf{Evaluation Criteria:} Fluency and naturalness of multilingual transitions: whether transitions between languages are smooth and seamless without abrupt changes or noticeable interruptions; accuracy of pronunciation for multilingual lyrics: whether the pronunciation in different languages is precise, clear, and adheres to the phonetic norms of each language, avoiding mispronunciations or accent deviations that could hinder understanding.

\item \textbf{Genre Controllability} \\
\textbf{Definition:} The degree to which the generated music accurately reflects the musical genre specified in the prompt. \\
\textbf{Evaluation Criteria:} Accuracy of musical genre characteristics (whether the generated music aligns with the features of the genre specified in the prompt, such as jazz, pop, classical, rock, etc.).

\item \textbf{Instrument and Vocal Configuration Controllability} \\
\textbf{Definition:} The extent to which the generated music adheres to the instrument and vocal configuration specified in the prompt. \\
\textbf{Evaluation Criteria:} Matching of instrument and vocal configuration (whether the generated music follows the specifications in the prompt, such as piano, guitar, male or female vocals, choir, etc.).

\item \textbf{Emotional Expressiveness} \\
\textbf{Definition:} The accuracy and impact of emotional expression in the generated music, as specified in the prompt. \\
\textbf{Evaluation Criteria:} Alignment of musical emotions with the emotional description in the prompt (e.g., passionate, sorrowful, cheerful).

\item \textbf{Tempo and Rhythm} \\
\textbf{Definition:} The congruence of the music's tempo (BPM) and rhythm with the requirements specified in the prompt. \\
\textbf{Evaluation Criteria:} Consistency of generated music tempo (BPM) with the tempo specified in the prompt, and adherence to the required rhythmic patterns.
  
\end{enumerate}


\section{Qwen2Audio-Instruct Tagging Prompt}
\begin{tcolorbox}[colback=gray!10, colframe=black, title=Music Tagging Prompt]
Analyze the provided audio and describe its features in a valid JSON format 
with the following keys: \texttt{Music\_genre}, \texttt{Instrument}, and \texttt{Mood}. 
If there are multiple entries for any key, represent them as a list of strings. 

Example format:
\begin{lstlisting}[basicstyle=\ttfamily]
{
    "Music_genre": ["Jazz"],
    "Instrument": ["Saxophone", "Piano"],
    "Mood": ["Relaxed"]
}
\end{lstlisting}
\label{box:music tagging prompt}

\end{tcolorbox}

\begin{tcolorbox}[colback=gray!10, colframe=black, title=Vocal Tagging Prompt]
Analyze the provided audio and describe its vocal characteristics in a valid JSON format 
with the following keys: \texttt{gender}, \texttt{age}, and \texttt{vocal\_timbre}. 
If there are multiple entries for any key, represent them as a list of strings.

Example format:
\begin{lstlisting}[basicstyle=\ttfamily]
{
    "gender": ["female"],
    "age": ["adult"],
    "vocal_timbre": ["bright", "airy"]
}
\end{lstlisting}
\label{box:vocal tagging prompt}
\end{tcolorbox}

\clearpage
\section{Multilingual Subjective Evaluation}
\label{appendix:multilingual_subjective}
\begin{figure}[htbp]
    \centering
    \begin{subfigure}[b]{0.465\linewidth}
        \centering
        \includegraphics[width=\linewidth]{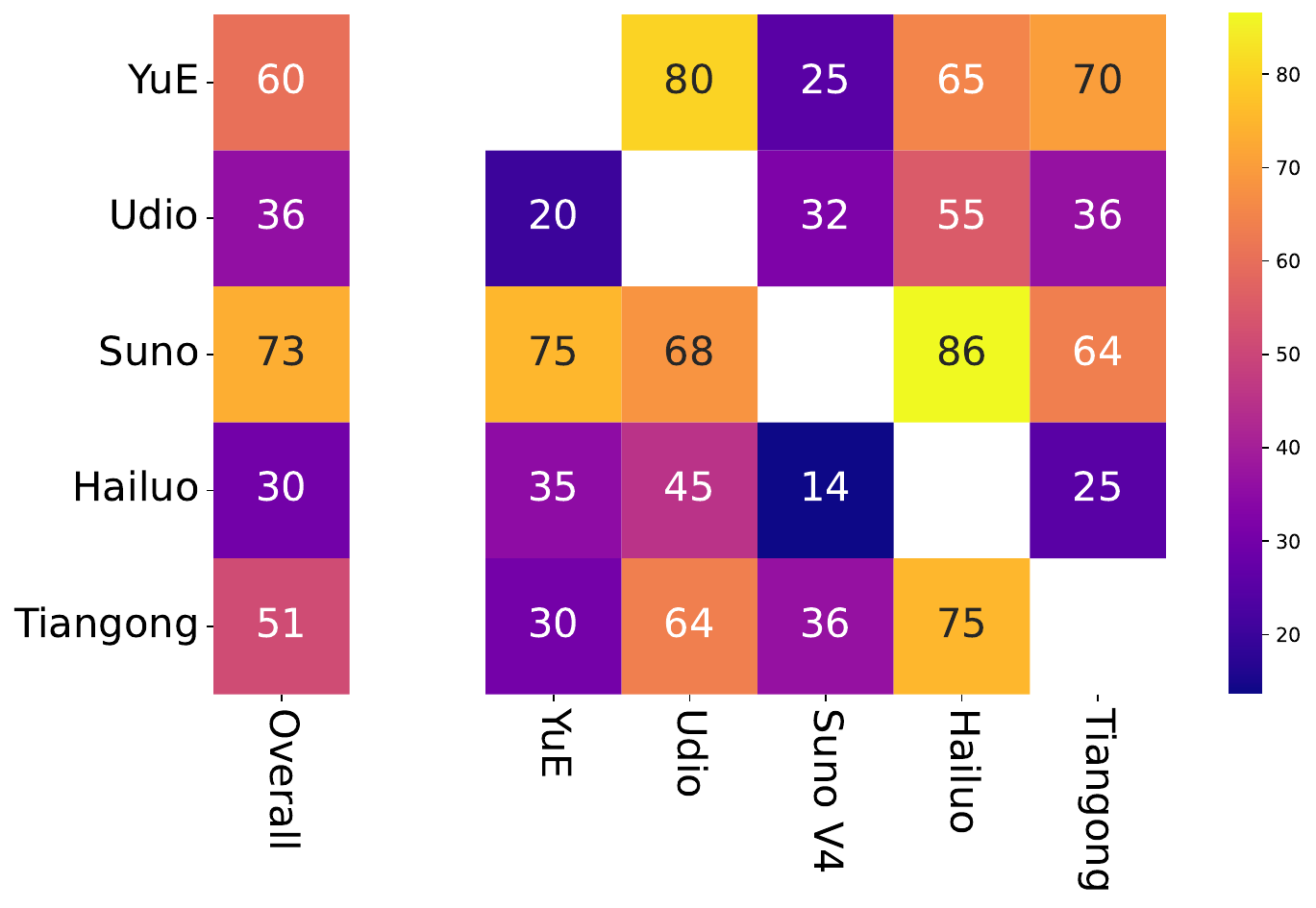}
        \caption{Chinese - Lyrics Following}
        \label{fig:zh_lrc}
    \end{subfigure}
    \hfill
    \begin{subfigure}[b]{0.48\linewidth}
        \centering
        \includegraphics[width=\linewidth]{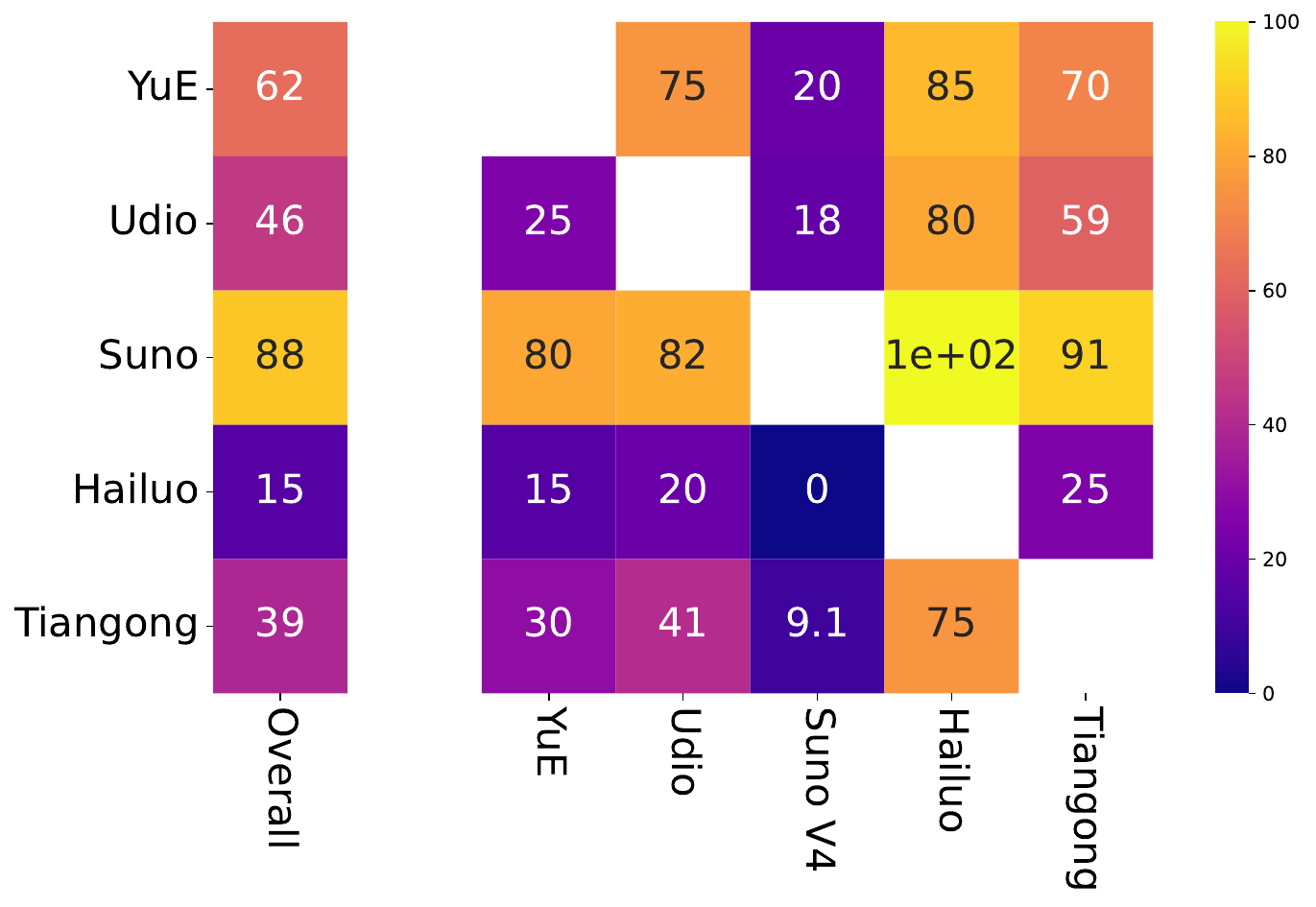}
        \caption{Chinese - Musicality}
        \label{fig:zh_mus}
    \end{subfigure}

    \vspace{0.5em} 

    \begin{subfigure}[b]{0.465\linewidth}
        \centering
        \includegraphics[width=\linewidth]{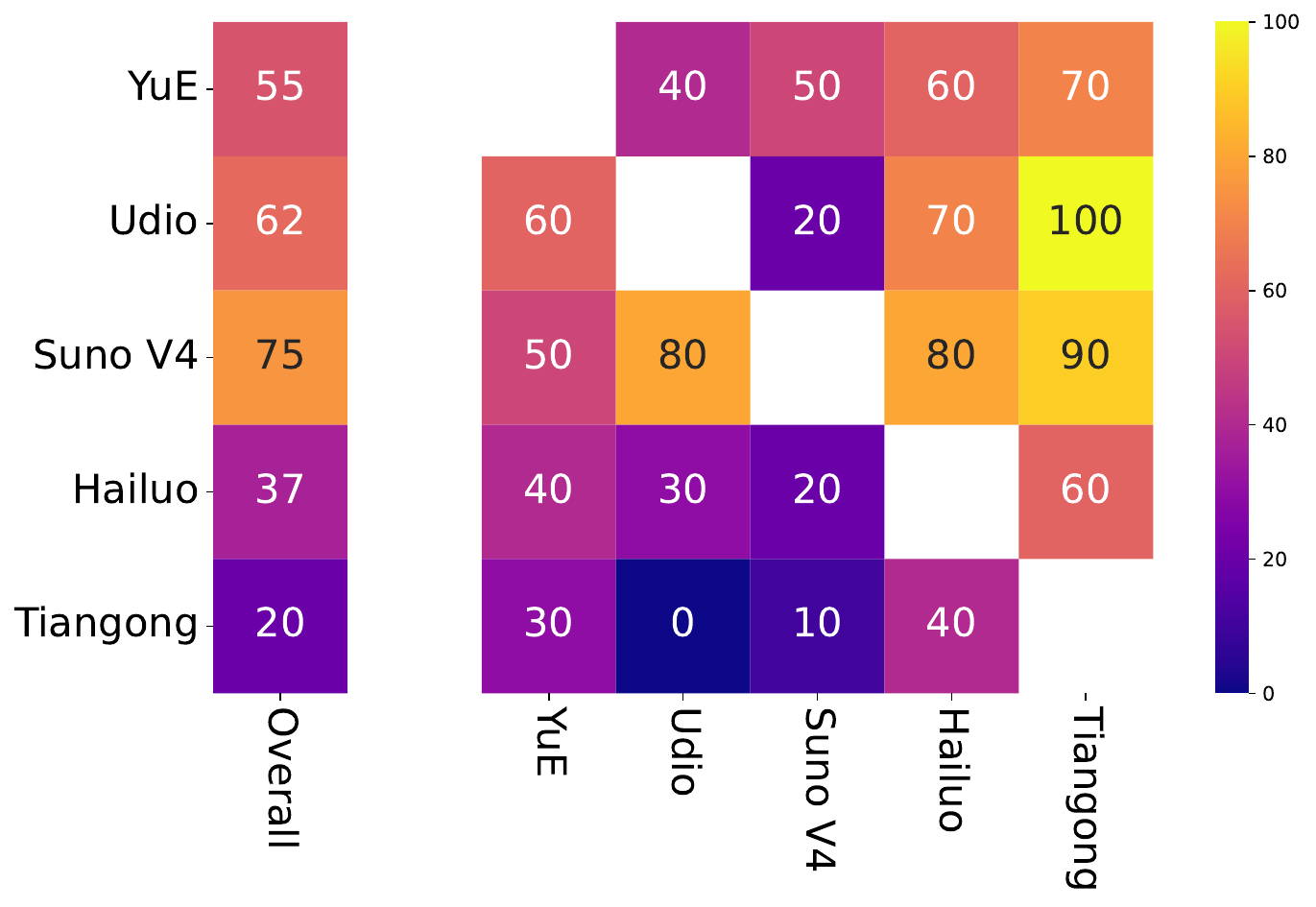}
        \caption{Korean - Lyrics Following}
        \label{fig:kr_lrc}
    \end{subfigure}
    \hfill
    \begin{subfigure}[b]{0.48\linewidth}
        \centering
        \includegraphics[width=\linewidth]{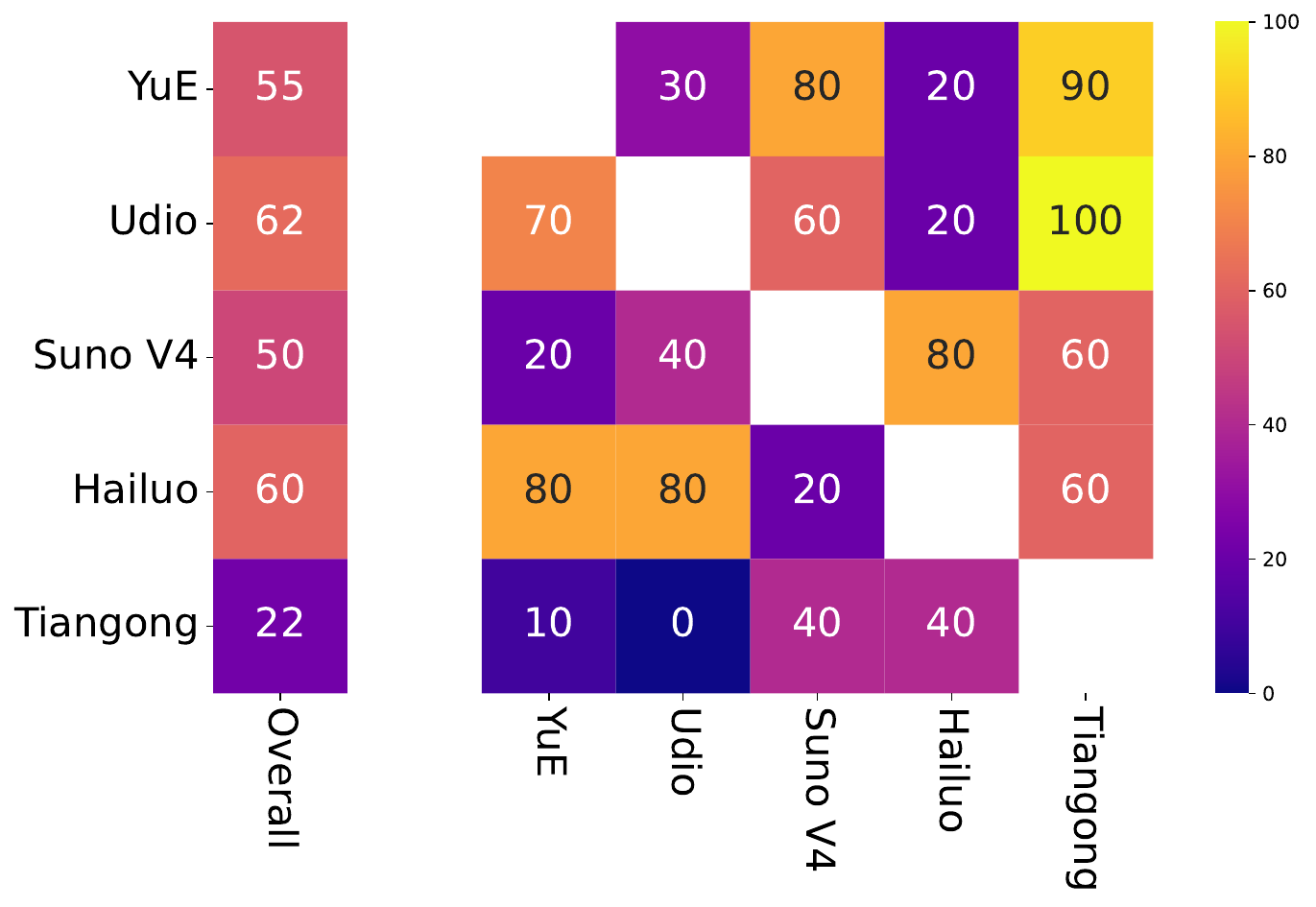}
        \caption{Korean - Musicality}
        \label{fig:kr_mus}
    \end{subfigure}

    \vspace{0.5em}

    \begin{subfigure}[b]{0.465\linewidth}
        \centering
        \includegraphics[width=\linewidth]{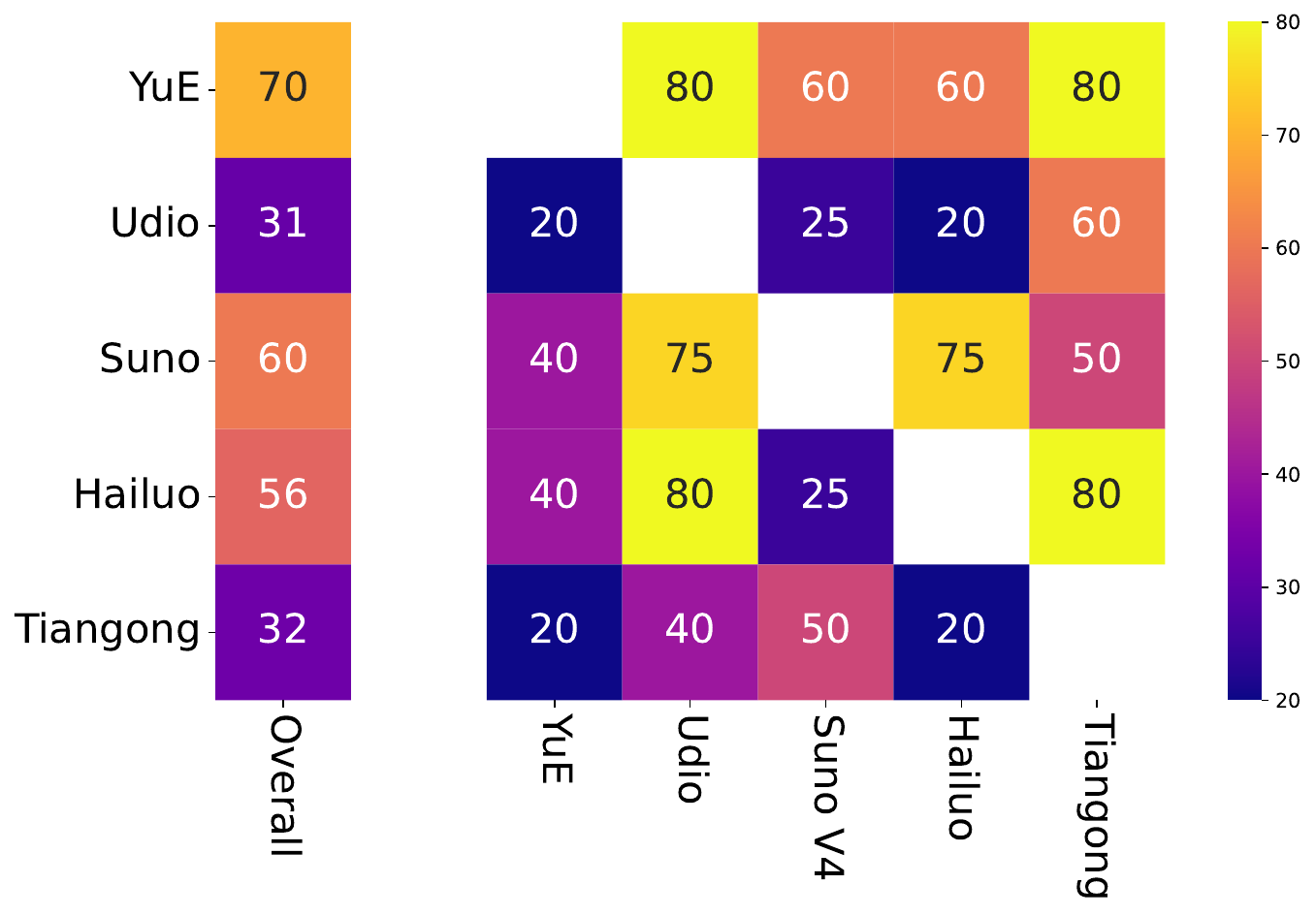}
        \caption{Japanese - Lyrics Following}
        \label{fig:jp_lrc}
    \end{subfigure}
    \hfill
    \begin{subfigure}[b]{0.48\linewidth}
        \centering
        \includegraphics[width=\linewidth]{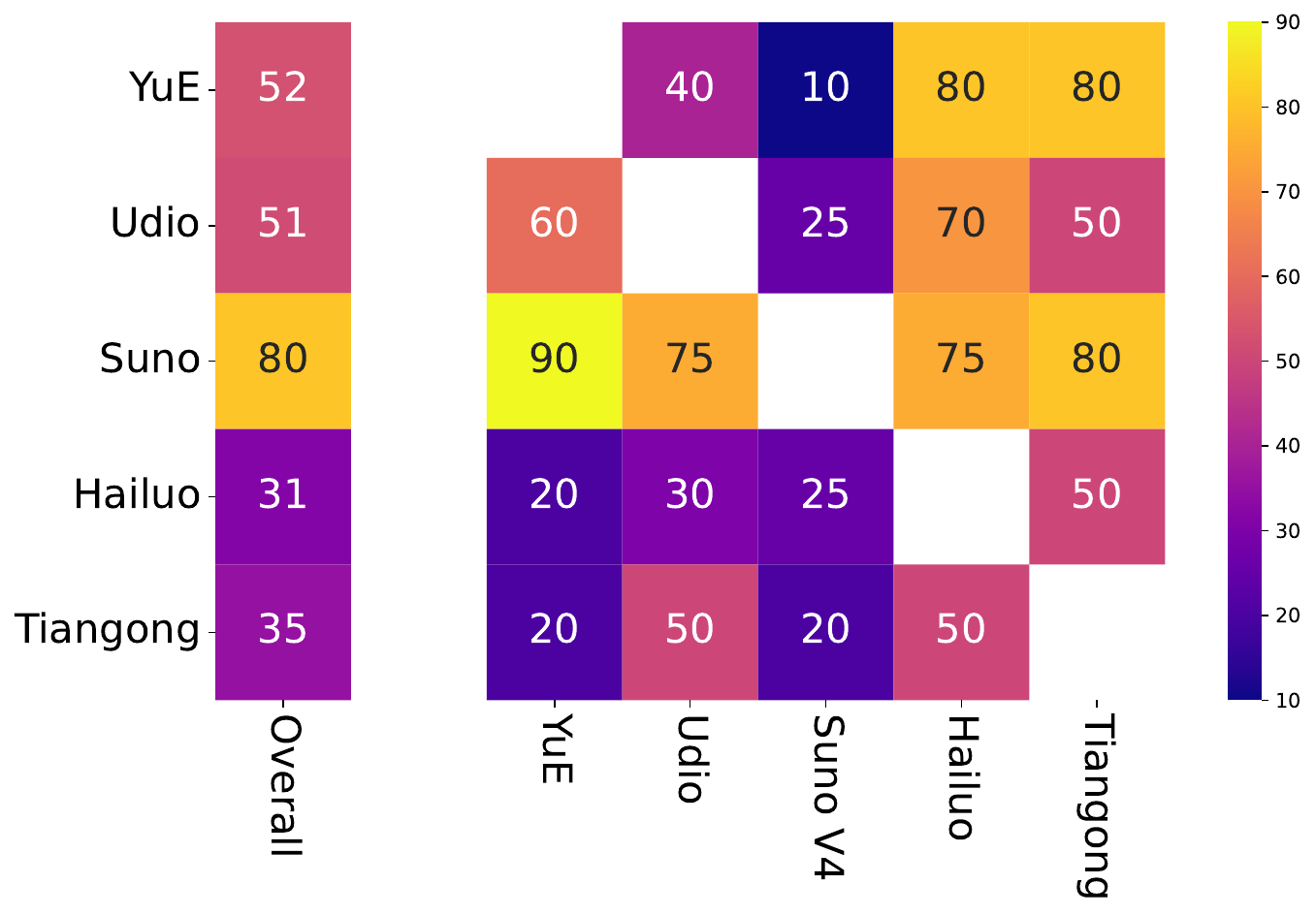}
        \caption{Japanese - Musicality}
        \label{fig:jp_mus}
    \end{subfigure}

    \caption{YuE vs. others across different languages on lyrics following and musicality.}
    \label{fig:multilingual_evaluation}
\end{figure}

\clearpage
\section{15 English Prompts From GPT}
\label{gpt_15_prompts}
\begin{tcolorbox}[colback=gray!10, colframe=black, title=ID: 1]
\footnotesize
[Genre] Rap  \\

[verse]  \\
Woke up in the morning, sun is shining bright \\  
Chasing all my dreams, gotta get my mind right \\  
City lights are fading, but my vision's clear \\  
Got my team beside me, no room for fear \\  
Walking through the streets, beats inside my head \\  
Every step I take, closer to the bread \\  
People passing by, they don't understand \\  
Building up my future with my own two hands   \\ 

[chorus]  \\
This is my life, and I'm aiming for the top \\  
Never gonna quit, no, I'm never gonna stop \\  
Through the highs and lows, I'mma keep it real \\  
Living out my dreams with this mic and a deal   \\ 

[verse]  \\
Late nights grinding, writing down these rhymes \\  
Clock is ticking fast, can't afford to waste time \\  
Haters gonna hate, but I brush it off \\  
Turn the negativity into something strong \\  
Mama working hard, wanna make her proud \\  
Echoes of her prayers cutting through the crowd \\  
Friends turned strangers, but it's all good \\  
Focused on my path like I always knew I would   \\ 

[chorus] \\
This is my journey, and I'm running this race \\  
Heart full of fire, you can see it in my face \\  
Obstacles ahead, but I got no fear \\  
Victory is close, yeah, it's almost here   \\ 

[bridge] \\
They said I couldn't do it, said I'd never rise \\  
But now I'm soaring high, reaching for the skies \\  
Lessons that I learned made me who I am \\  
Standing tall now, I don't give a damn   \\ 

[verse] \\
Echoes in the alley, music's getting loud \\  
Feeling the adrenaline pumping through the crowd \\  
Spotlights on me, it's my time to shine \\  
Living in the moment, everything's aligned \\  
Looking back now at the roads I've crossed \\  
Every single battle, every line I've tossed \\  
Made me stronger, wiser, ready for what's next \\  
Writing my own story, turning pages of the text   \\ 

[chorus]  \\  
This is my song, and I'm singing it proud \\  
Voices united, hear us shout out loud \\  
From the underground straight into the stars \\  
Carving out my name, leaving all these scars   \\ 

[outro]  \\ 
Yeah, this is for the dreamers, the ones who never quit \\  
Keep your head up high, and don't you ever submit \\  
Life is what you make it, so make it something great \\  
Step into your purpose, go and seize your fate  
\end{tcolorbox}

\begin{tcolorbox}[colback=gray!10, colframe=black, title=ID: 2]
\footnotesize
[Genre] Rock  \\

[verse]  \\
Standing on the corner, shadows in the night \\
The city's heartbeat echoes under lights \\
Hands deep in pockets, wandering alone \\
Footsteps tracing paths to the unknown \\
Suddenly he pauses, looks up to the sky \\
Eyes reflect the questions passing by \\
Whispers to the wind, words without a sound \\
Searching for the answers yet unfound \\

[chorus]  \\
Lost within the chaos, seeking out a sign \\
In a world of color, drawing blurred lines \\
Moving forward, looking back, unsure of the way \\
Trying to find a place where he can stay \\

[verse]  \\
He crosses empty streets, under neon glow \\
Faces in the crowd, stories left untold \\
Raises up his arms, reaching for the truth \\
Grasping at the fragments of his youth \\
Billboards and the banners flutter in the breeze \\
The rhythm of the city brings him to his knees \\
Heartbeat heavy, nowhere left to hide \\
Feeling like he's lost amidst the tide \\

[chorus]  \\
Lost within the chaos, seeking out a sign \\
In a world of color, drawing blurred lines \\
Moving forward, looking back, unsure of the way \\
Trying to find a place where he can stay \\

[bridge]  \\
Doesn't want to leave, doesn't want to fight \\
Caught between the darkness and the light \\
No need for reason, nothing to prove \\
Just a soul in transit, with nothing to lose \\

[outro]  \\
Doesn't want to leave, doesn't want to fight \\
Chasing after shadows in the night \\
He doesn't need the truth, doesn't need a name \\
Just looking for a spark to fan the flame
\end{tcolorbox}

\begin{tcolorbox}[colback=gray!10, colframe=black, title=ID: 3]
\footnotesize
[Genre] Pop  \\

[verse]  \\
Staring at the sunset, colors paint the sky \\
Thoughts of you keep swirling, can't deny \\
I know I let you down, I made mistakes \\
But I'm here to mend the heart I didn't break \\

[chorus]  \\
Every road you take, I'll be one step behind \\
Every dream you chase, I'm reaching for the light \\
You can't fight this feeling now \\
I won't back down \\
I'm the whisper in the wind, the shadow by your side \\
The warmth you feel within when you can't hide \\
You know you can't deny it now \\
I won't back down \\

[verse]  \\
They might say I'm foolish, chasing after you \\
But they don't feel this love the way we do \\
My heart beats only for you, can't you see? \\
I won't let you slip away from me \\

[chorus]  \\
Every road you take, I'll be one step behind \\
Every dream you chase, I'm reaching for the light \\
You can't fight this feeling now \\
I won't back down \\
I'm the whisper in the wind, the shadow by your side \\
The warmth you feel within when you can't hide \\
You know you can't deny it now \\
I won't back down \\

[bridge]  \\
No, I won't back down, won't turn around \\
Until you're back where you belong \\
I'll cross the oceans wide, stand by your side \\
Together we are strong \\

[outro]  \\
Every road you take, I'll be one step behind \\
Every dream you chase, love's the tie that binds \\
You can't fight this feeling now \\
I won't back down
\end{tcolorbox}

\begin{tcolorbox}[colback=gray!10, colframe=black, title=ID: 4]
\footnotesize
[Genre] Jazz  \\

[verse]  \\
In the quiet of the evening, shadows start to fall \\
Whispers of the night wind echo through the hall \\
Lost within the silence, I hear your gentle voice \\
Guiding me back homeward, making my heart rejoice \\

[chorus]  \\
Don't let this moment fade, hold me close tonight \\
With you here beside me, everything's alright \\
Can't imagine life alone, don't want to let you go \\
Stay with me forever, let our love just flow \\

[verse]  \\
Moonlight paints a picture upon your lovely face \\
Every glance between us fills the empty space \\
Time stands still around us when you're in my arms \\
Nothing else can matter, safe from any harm \\

[chorus]  \\
Don't let this moment fade, hold me close tonight \\
With you here beside me, everything's alright \\
Can't imagine life alone, don't want to let you go \\
Stay with me forever, let our love just flow \\

[bridge]  \\
Every touch ignites a fire, burning deep within \\
Every smile you give to me makes my head spin \\
Promise me you'll stay awhile, don't ever say goodbye \\
Together we'll chase every star across the sky \\

[chorus]  \\
Don't let this moment fade, hold me close tonight \\
With you here beside me, everything's alright \\
Can't imagine life alone, don't want to let you go \\
Stay with me forever, let our love just flow \\

[outro]  \\
Stay with me forever, let our love just flow
\end{tcolorbox}

\begin{tcolorbox}[colback=gray!10, colframe=black, title=ID: 5]
\footnotesize
[Genre] Blues  \\

[verse]  \\
Late last night, the rain was pouring down \\
Lonely footsteps echoed through the town \\
Thinking 'bout the love that slipped away \\
Wondering how I let you go that day \\

[chorus]  \\
Oh, my angel, where have you flown \\
Left me here to face this world alone \\
I'm just a fool, a fool in love with you \\
Can't deny this heartache's true \\

[verse]  \\
Streetlights flicker, shadows on the wall \\
Memories of you, I recall \\
Your laughter like a song inside my head \\
Without you here, my soul feels dead \\

[chorus]  \\
Oh, my angel, won't you return \\
In this fire of love, I still burn \\
I'm just a fool, a fool in love with you \\
Hoping someday you'll feel it too \\

[bridge]  \\
I fell for you, and I always knew \\
That my world revolves around you \\
I hope and I pray, both night and day \\
That you'll come back and choose to stay \\

[chorus]  \\
Oh, my angel, where have you flown \\
Left me here to face this world alone \\
I'm just a fool, a fool in love with you \\
Waiting here, what else can I do \\

[outro]  \\
I'm just a fool, a fool in love with you
\end{tcolorbox}

\begin{tcolorbox}[colback=gray!10, colframe=black, title=ID: 6]
\footnotesize
[Genre] RnB\_Soul  \\

[verse]  \\
Why don't we just find a place to hide \\
Leave all our worries and doubts behind \\
When nothing in this world is as it seems \\
Together we can live inside our dreams \\
There's no need to be afraid tonight \\
In the love we've made, we'll find the light \\
When we're living in a world of our own \\
It's you and me, we never feel alone \\

[chorus]  \\
They say it's hard for a man to let it show \\
But with you, I'm ready to let it all go \\
Whatever we try, we're gonna get there \\
You take control, baby, I don't care \\
I gotta keep on pushing when times get tough \\
We keep on making better love \\

[verse]  \\
Don't believe the things that others say \\
We've tried it all and found our way \\
They should take a look at you and me \\
Learning how to love and set it free \\
For every heartache, we take our time \\
You teach me yours and I'll show you mine \\
About the way that love is meant to be \\
Together we'll rewrite our history \\

[chorus]  \\
They say it's hard for a man to let it show \\
But with you, I'm ready to let it all go \\
Whatever we try, we're gonna get there \\
You take control, baby, I don't care \\
I gotta keep on pushing when times get tough \\
We keep on making better love \\

[bridge]  \\
Gotta take control and swallow my pride \\
Every man has feelings deep inside \\
You gotta find yourself before you can \\
Be ready to love and understand \\
Baby, I know what you're thinking of \\
We keep on making better love \\

[outro]  \\
I believe the love we're making's gonna last forevermore \\
Loving you feels so right, like never before \\
We'll be getting down tonight until the morning light \\
We keep on making better love \\
Better love, we'll be making \\
Better love, no more faking \\
They say it's hard for a man to let it show \\
But with you here, I'm ready to let go \\
Whatever we try, we're gonna get there \\
You take control, baby, I don't care \\
I gotta keep on pushing when times get tough \\
We keep on making better love \\
Better love (till fade out)
\end{tcolorbox}

\begin{tcolorbox}[colback=gray!10, colframe=black, title=ID: 7]
\footnotesize
[Genre] Ancient\_Chinese\_Style  \\

[verse]  \\
Beneath the moonlit sky so vast \\
A lone wanderer recalls the past \\
Whispers of the bamboo leaves \\
Echo tales the wind retrieves \\

[chorus]  \\
Oh, the rivers flow, mountains high \\
Journeying souls beneath the endless sky \\
Threads of fate entwine our way \\
Guiding us through night and day \\

[verse]  \\
Lanterns glow with softest light \\
Painting shadows in the night \\
Silken robes and ancient songs \\
Memories where hearts belong \\

[chorus]  \\
Oh, the rivers flow, mountains high \\
Journeying souls beneath the endless sky \\
Threads of fate entwine our way \\
Guiding us through night and day \\

[bridge]  \\
Stars reflect in tranquil ponds \\
Dreams unfold of times beyond \\
Lotus blooms and cranes in flight \\
Secrets held within the night \\

[outro]  \\
As the sunrise paints the east \\
Bringing hope and inner peace \\
Footprints fade upon the shore \\
But the spirit journeys evermore
\end{tcolorbox}

\begin{tcolorbox}[colback=gray!10, colframe=black, title=ID: 8]
\footnotesize
[Genre] Folk  \\

[verse]  \\
Underneath the open sky so clear, \\
We gather 'round with voices near. \\
Through trials faced and stories told, \\
Our spirits rise, our hearts unfold. \\

[chorus]  \\
So lift the lanterns to the sky, \\
Together we will soar and fly. \\
Though shadows loom and doubts appear, \\
We'll keep the flame forever here. \\

[verse]  \\
Remember all the paths we've crossed, \\
The battles won, the moments lost. \\
A banner of hope we hold up high, \\
A symbol shining in our eyes. \\

[chorus]  \\
So lift the lanterns to the sky, \\
Together we will soar and fly. \\
Though shadows loom and doubts appear, \\
We'll keep the flame forever here. \\

[bridge]  \\
With hands united, we stand tall, \\
Pledged to rise if we should fall. \\
Through darkest nights and stormy seas, \\
Our song will carry on the breeze. \\

[chorus]  \\
So lift the lanterns to the sky, \\
Together we will soar and fly. \\
Though shadows loom and doubts appear, \\
We'll keep the flame forever here. \\

[outro]  \\
We'll keep the flame forever here.
\end{tcolorbox}

\begin{tcolorbox}[colback=gray!10, colframe=black, title=ID: 9]
\footnotesize
[Genre] Dance  \\

[verse]  \\
Underneath the starlit sky, \\
We come alive, you and I. \\
City lights are shining bright, \\
Dancing through the endless night. \\

[chorus]  \\
Who are we? Let's break away, \\
Feel the beat and let it play. \\
Lost in music, hearts align, \\
In this moment, we define. \\

[verse]  \\
Shadows fade beneath the glow, \\
Rhythms guide us where to go. \\
Voices whisper in the crowd, \\
Turn it up, we'll sing aloud. \\

[chorus]  \\
Who are we? Let's break away, \\
Feel the beat and let it play. \\
Lost in music, hearts align, \\
In this moment, we define. \\

[bridge]  \\
Let the melody surround, \\
Lift us off the solid ground. \\
Every step and every move, \\
In this dance we find our groove. \\

[chorus]  \\
Who are we? Let's break away, \\
Feel the beat and let it play. \\
Lost in music, hearts align, \\
In this moment, we define. \\

[outro]  \\
Keep on dancing, feel the heat, \\
Moving to the pounding beat. \\
Who we are is here and now, \\
Take my hand, we'll show them how.
\end{tcolorbox}

\begin{tcolorbox}[colback=gray!10, colframe=black, title=ID: 10]
\footnotesize
[Genre] Country  \\

[verse]  \\
Da-dum, da-da-da-da-da-da-da \\
Da-dum, da-dum \\
Walking down this lonesome road \\
Thinking 'bout the love untold \\
Why haven't I told you \\
I've whispered to the midnight stars \\
Just how wonderful you are \\
Why haven't I told you \\

[chorus]  \\
Friends keep asking if I'm fine \\
I just smile and say you're mine \\
Might as well confess \\
Can't keep this inside \\
Maybe you feel the same way too \\
Oh darling, if you do \\
Why haven't you told me \\
Da-dum, da-da-da-da-da-da-da \\

[verse]  \\
I've sung it to the morning sun \\
That with you, my life's begun \\
Why haven't I told you \\
My heart's an open book today \\
Waiting for the words to say \\
Why haven't I told you \\

[chorus]  \\
Friends keep asking what's the news \\
I just grin and think of you \\
Time to take a chance \\
Let my feelings show \\
Maybe you feel the same way too \\
Oh darling, if you do \\
Why haven't you told me \\

[bridge]  \\
Da-dum, da-da-da-da-da-da-da \\
Da-dum, da-dum \\
No more holding back these words \\
Let them fly just like the birds \\

[chorus]  \\
Now I'm standing here tonight \\
Hoping that I got it right \\
Might as well confess \\
Can't keep this inside \\
Maybe you feel the same way too \\
Oh darling, if you do \\
Let's not waste another day \\
Why haven't we told us \\

[outro]  \\
Da-dum, da-da-da-da-da-da-da \\
Da-dum, da-dum \\
Now we've finally told us \\
Our new life's begun
\end{tcolorbox}

\begin{tcolorbox}[colback=gray!10, colframe=black, title=ID: 11]
\footnotesize
[Genre] Rap  \\

[verse]  \\
Woke up in the morning, sun is shining bright \\
Chasing all my dreams, gotta get my mind right \\
City lights are fading, but my vision's clear \\
Got my team beside me, no room for fear \\
Walking through the streets, beats inside my head \\
Every step I take, closer to the bread \\
People passing by, they don't understand \\
Building up my future with my own two hands \\

[chorus]  \\
This is my life, and I'm aiming for the top \\
Never gonna quit, no, I'm never gonna stop \\
Through the highs and lows, I'mma keep it real \\
Living out my dreams with this mic and a deal \\

[verse]  \\
Late nights grinding, writing down these rhymes \\
Clock is ticking fast, can't afford to waste time \\
Haters gonna hate, but I brush it off \\
Turn the negativity into something strong \\
Mama working hard, wanna make her proud \\
Echoes of her prayers cutting through the crowd \\
Friends turned strangers, but it's all good \\
Focused on my path like I always knew I would \\

[chorus]  \\
This is my journey, and I'm running this race \\
Heart full of fire, you can see it in my face \\
Obstacles ahead, but I got no fear \\
Victory is close, yeah, it's almost here \\

[bridge]  \\
They said I couldn't do it, said I'd never rise \\
But now I'm soaring high, reaching for the skies \\
Lessons that I learned made me who I am \\
Standing tall now, I don't give a damn \\

[verse]  \\
Echoes in the alley, music's getting loud \\
Feeling the adrenaline pumping through the crowd \\
Spotlights on me, it's my time to shine \\
Living in the moment, everything's aligned \\
Looking back now at the roads I've crossed \\
Every single battle, every line I've tossed \\
Made me stronger, wiser, ready for what's next \\
Writing my own story, turning pages of the text \\

[chorus]  \\
This is my song, and I'm singing it proud \\
Voices united, hear us shout out loud \\
From the underground straight into the stars \\
Carving out my name, leaving all these scars \\

[outro]  \\
Yeah, this is for the dreamers, the ones who never quit \\
Keep your head up high, and don't you ever submit \\
Life is what you make it, so make it something great \\
Step into your purpose, go and seize your fate
\end{tcolorbox}

\begin{tcolorbox}[colback=gray!10, colframe=black, title=ID: 12]
\footnotesize
[Genre] Rock  \\

[verse]  \\
Standing on the corner, shadows in the night \\
The city's heartbeat echoes under lights \\
Hands deep in pockets, wandering alone \\
Footsteps tracing paths to the unknown \\
Suddenly he pauses, looks up to the sky \\
Eyes reflect the questions passing by \\
Whispers to the wind, words without a sound \\
Searching for the answers yet unfound \\

[chorus]  \\
Lost within the chaos, seeking out a sign \\
In a world of color, drawing blurred lines \\
Moving forward, looking back, unsure of the way \\
Trying to find a place where he can stay \\

[verse]  \\
He crosses empty streets, under neon glow \\
Faces in the crowd, stories left untold \\
Raises up his arms, reaching for the truth \\
Grasping at the fragments of his youth \\
Billboards and the banners flutter in the breeze \\
The rhythm of the city brings him to his knees \\
Heartbeat heavy, nowhere left to hide \\
Feeling like he's lost amidst the tide \\

[chorus]  \\
Lost within the chaos, seeking out a sign \\
In a world of color, drawing blurred lines \\
Moving forward, looking back, unsure of the way \\
Trying to find a place where he can stay \\

[bridge]  \\
Doesn't want to leave, doesn't want to fight \\
Caught between the darkness and the light \\
No need for reason, nothing to prove \\
Just a soul in transit, with nothing to lose \\

[outro]  \\
Doesn't want to leave, doesn't want to fight \\
Chasing after shadows in the night \\
He doesn't need the truth, doesn't need a name \\
Just looking for a spark to fan the flame
\end{tcolorbox}

\begin{tcolorbox}[colback=gray!10, colframe=black, title=ID: 13]
\footnotesize
[Genre] Pop  \\

[verse]  \\
Staring at the sunset, colors paint the sky \\
Thoughts of you keep swirling, can't deny \\
I know I let you down, I made mistakes \\
But I'm here to mend the heart I didn't break \\

[chorus]  \\
Every road you take, I'll be one step behind \\
Every dream you chase, I'm reaching for the light \\
You can't fight this feeling now \\
I won't back down \\
I'm the whisper in the wind, the shadow by your side \\
The warmth you feel within when you can't hide \\
You know you can't deny it now \\
I won't back down \\

[verse]  \\
They might say I'm foolish, chasing after you \\
But they don't feel this love the way we do \\
My heart beats only for you, can't you see? \\
I won't let you slip away from me \\

[chorus]  \\
Every road you take, I'll be one step behind \\
Every dream you chase, I'm reaching for the light \\
You can't fight this feeling now \\
I won't back down \\
I'm the whisper in the wind, the shadow by your side \\
The warmth you feel within when you can't hide \\
You know you can't deny it now \\
I won't back down \\

[bridge]  \\
No, I won't back down, won't turn around \\
Until you're back where you belong \\
I'll cross the oceans wide, stand by your side \\
Together we are strong \\

[outro]  \\
Every road you take, I'll be one step behind \\
Every dream you chase, love's the tie that binds \\
You can't fight this feeling now \\
I won't back down
\end{tcolorbox}

\begin{tcolorbox}[colback=gray!10, colframe=black, title=ID: 14]
\footnotesize
[Genre] Jazz  \\

[verse]  \\
In the quiet of the evening, shadows start to fall \\
Whispers of the night wind echo through the hall \\
Lost within the silence, I hear your gentle voice \\
Guiding me back homeward, making my heart rejoice \\

[chorus]  \\
Don't let this moment fade, hold me close tonight \\
With you here beside me, everything's alright \\
Can't imagine life alone, don't want to let you go \\
Stay with me forever, let our love just flow \\

[verse]  \\
Moonlight paints a picture upon your lovely face \\
Every glance between us fills the empty space \\
Time stands still around us when you're in my arms \\
Nothing else can matter, safe from any harm \\

[chorus]  \\
Don't let this moment fade, hold me close tonight \\
With you here beside me, everything's alright \\
Can't imagine life alone, don't want to let you go \\
Stay with me forever, let our love just flow \\

[bridge]  \\
Every touch ignites a fire, burning deep within \\
Every smile you give to me makes my head spin \\
Promise me you'll stay awhile, don't ever say goodbye \\
Together we'll chase every star across the sky \\

[chorus]  \\
Don't let this moment fade, hold me close tonight \\
With you here beside me, everything's alright \\
Can't imagine life alone, don't want to let you go \\
Stay with me forever, let our love just flow \\

[outro]  \\
Stay with me forever, let our love just flow
\end{tcolorbox}

\begin{tcolorbox}[colback=gray!10, colframe=black, title=ID: 15]
\footnotesize
[Genre] Blues  \\

[verse]  \\
Late last night, the rain was pouring down \\
Lonely footsteps echoed through the town \\
Thinking 'bout the love that slipped away \\
Wondering how I let you go that day \\

[chorus]  \\
Oh, my angel, where have you flown \\
Left me here to face this world alone \\
I'm just a fool, a fool in love with you \\
Can't deny this heartache's true \\

[verse]  \\
Streetlights flicker, shadows on the wall \\
Memories of you, I recall \\
Your laughter like a song inside my head \\
Without you here, my soul feels dead \\

[chorus]  \\
Oh, my angel, won't you return \\
In this fire of love, I still burn \\
I'm just a fool, a fool in love with you \\
Hoping someday you'll feel it too \\

[bridge]  \\
I fell for you, and I always knew \\
That my world revolves around you \\
I hope and I pray, both night and day \\
That you'll come back and choose to stay \\

[chorus]  \\
Oh, my angel, where have you flown \\
Left me here to face this world alone \\
I'm just a fool, a fool in love with you \\
Waiting here, what else can I do \\

[outro]  \\
I'm just a fool, a fool in love with you
\end{tcolorbox}

\end{CJK*}
\end{document}